\documentclass[11pt,a4paper]{article}
\usepackage[a4paper,textwidth=160mm,textheight=240mm,centering,headheight=15pt,headsep=12pt]{geometry}
\usepackage[british]{babel}
\usepackage[T1]{fontenc}
\usepackage[utf8]{inputenc}
\usepackage{lmodern}
\usepackage{amsmath,amssymb,amsthm,mathtools}
\usepackage{graphicx,booktabs,longtable,array,float,placeins,needspace}
\usepackage{setspace,caption,etoolbox,accsupp}
\usepackage[authoryear,round]{natbib}
\usepackage{url}
\usepackage[hidelinks,breaklinks=true]{hyperref}
\usepackage{fancyhdr}
\newcommand{\accessiblefigure}[2]{%
  \BeginAccSupp{method=pdfstringdef,Alt={#1}}%
  \includegraphics[width=\textwidth]{#2}%
  \EndAccSupp{}}
\theoremstyle{plain}
\newtheorem{theorem}{Theorem}
\newtheorem{proposition}{Proposition}
\newtheorem{lemma}{Lemma}

\theoremstyle{definition}

\hypersetup{pdftitle={Minimax regret and monotone optimality in multiple testing},pdfauthor={Lei Li and Yuexiao Dong},bookmarksdepth=2}
\begin{document}
\begingroup
\setstretch{1}
\renewcommand{\thefootnote}{\fnsymbol{footnote}}
\begin{center}
{\Large\bfseries Minimax regret and monotone optimality in multiple testing\par}
\vspace{.8em}
Lei Li$^{1,}$\footnote{Corresponding author: \href{mailto:lei.li@lunarai.llc}{lei.li@lunarai.llc}.} and Yuexiao Dong$^2$\par
$^1$LunarAI LLC\qquad $^2$Temple University\par
\end{center}
\endgroup
\vspace{.7em}
\noindent\textbf{Abstract}\par
Confirmatory trials may seek efficacy claims in a prespecified subgroup and the
full population, although benefit outside the subgroup is uncertain. We measure
power loss relative to the marginal level-$\alpha$ test available if the population
to protect were known, using the same experiment. In a bivariate Gaussian model
with known covariance, we minimise the largest power loss, or regret, over
prespecified ranges of target and non-target effects, subject to strong familywise
error control and coordinatewise monotonicity. We optimise over monotone
closed-testing procedures with fixed marginal tests, allowing randomisation in
the shared intersection test. Monotonicity places the worst non-target mean at
its lower planning bound for each target effect. On the resulting reduced
planning set, every least-favourable distribution is supported on exactly two
efficacy scenarios, one for each target population. Under a condition on the
effect ranges and covariance, the minimax procedure is nonrandomised and unique
up to null sets. An integral criterion characterises exactly when a calibrated
step-up procedure attains the minimax value within this class. Bounds for the
continuous problem quantify possible improvements over simple procedures.
An asthma planning example illustrates power allocation and the compatibility
of target assignments with treatment benefits and burdens.
\medskip

\section{Introduction}\label{sec:introduction}

A targeted therapy may benefit a prespecified subgroup while its benefit
in the remaining patients is uncertain. A confirmatory trial may therefore
seek efficacy claims for both the subgroup and the full population.
Assessment of broader use also considers complement efficacy and treatment
burden. US Food and Drug Administration enrichment guidance discusses
evidence from marker-negative patients when considering broader use
\citep[Section~VII.B]{FDA2019}; European Medicines Agency guidance
emphasises efficacy and benefit--risk across relevant subpopulations
\citep[Sections~4.2 and~4.5]{EMA2019Subgroups}.
We study how a final testing procedure should allocate power between
the two prespecified claims.

Related work addresses uncertainty about treatment benefit
through enrichment designs, including interim population
selection, adaptive recruitment and utility-based optimisation
\citep{BrannathEtAl2009,JenkinsStoneJennison2011,
SimonSimon2013,RosenblumFangLiu2020,OndraEtAl2016,
BurnettJennison2021}.
With the trial populations fixed, attention turns to the
final testing procedure. Familiar choices include Holm's
step-down procedure \citep{Holm1979}, Hochberg's step-up
procedure \citep{Hochberg1988}, and the normal-theory
Dunnett--Tamhane (DT) step-up procedure \citep{dt1992}.
Optimal multiple-testing research has established maximin
results for stepwise procedures \citep{lrs2005} and developed
consonant procedures that relate intersection and elementary
rejections to maximin power
\citep{BittmanEtAl2009,RomanoShaikhWolf2011}.
For subgroup and overall-population claims,
\citet{BelitskayaLevyEtAl2018} account for possible harm
outside the subgroup; their COSST-O procedure maximises
the probability of at least one rejection at specified
alternatives. General state-dependent loss and minimax
formulations appear in \citet{rly2014}.
Further work develops optimal and maximin testing,
monotonicity-constrained optimisation, and optimal
intersection tests
\citep{rosset2022,RossetCorrection2024,hkr2023,khr2025}.
Related discovery criteria are considered by
\citet{Rosenblum2014UMP}.

These approaches optimise different power or loss criteria,
but do not directly characterise the worst-case power
shortfall for a designated population claim relative to
its marginal test over uncertain efficacy scenarios.
Giving priority to the full-population claim may reduce
subgroup power when benefit is concentrated there;
favouring the subgroup may reduce the chance of
establishing a broader claim. We consider prespecified
efficacy scenarios, each designating one claim for
protection. The procedure may reject either or both
hypotheses, but success means rejecting the designated
claim. Its benchmark is the marginal level-$\alpha$
test available if the target identity were known before
the trial, using the same experiment without knowing
the numerical treatment effects. Regret is the shortfall
in target power from this benchmark.

We minimise the largest regret over the planning scenarios, subject to
strong familywise error control and coordinatewise monotonicity:
each rejection probability is nondecreasing as either statistic increases.
Maximin power maximises the lowest absolute power; minimax regret
accounts for state-specific reference powers, following classical
most-stringent testing \citep{Wald1943,Hoeffding1951}.
A percentage point of lost power receives equal weight for either target.
We study a fixed experiment with bivariate normal subgroup and
full-population statistics and known correlation $0\le\rho<1$.
Planning families allow uncertainty in target efficacy and the non-target mean.
We optimise monotone closed-testing procedures with fixed marginal tests,
allowing randomisation in the shared intersection decision.
Adverse non-target effects can make a target likelihood ratio decrease
as the other statistic increases, linking power allocation to the
geometry of the monotone rejection region.

The analysis yields several structural results. First, monotonicity
reduces each planning family to its worst non-target offset, and
every least-favourable distribution on the reduced planning set
has one support point for each target claim, even when target
efficacy varies over an interval. Second, under a condition on
the effect ranges and covariance, the likelihood ratios increase
along a common direction, allowing conditional Gaussian
rearrangement to eliminate randomisation without reducing
target power or violating monotonicity and size constraints.
The resulting minimax rule is unique up to null sets.
Third, an integral criterion together with regret equalisation
characterises exactly when a calibrated step-up procedure is
minimax over the randomised monotone class $\mathcal U$.
Under the common-direction condition, the same characterisation
holds among nonrandomised monotone procedures with strong
familywise error control. The zero-offset case gives an explicit
step-up optimality criterion and a cubic expansion for excess
regret immediately beyond the symmetric transition at fixed
correlation below one. Bounds for the continuous problem
quantify possible improvements over computed procedures.
Numerical examples illustrate changes in rejection boundaries
and cutoff reoptimisation, while an asthma planning example
relates target power to treatment benefits and burdens.

The remainder of the paper is organised as follows. Section~\ref{sec:model} formulates the problem;
Section~\ref{sec:theory} establishes the structural results. Section~\ref{sec:computation} presents computation and bounds,
Sections~\ref{sec:numerical}--\ref{sec:clinical-example} give illustrations,
and Section~\ref{sec:discussion} discusses implications.
Proofs and further results appear in the supplement.

\section{Statistical formulation and minimax-regret objective}\label{sec:model}

We seek a testing procedure that limits the power lost for the population claim designated for protection under each efficacy scenario. Let $(X,Y)\sim N_2(\mu,\Sigma_\rho)$, where
$\mu=(\mu_S,\mu_F)$, both variances are one, and the known correlation
satisfies $0\le\rho<1$. The efficacy hypotheses are
$H_S:\mu_S\le0$ and $H_F:\mu_F\le0$.
Write $P_\mu,E_\mu$ for probability and expectation, $P_0$ for the
law at $\mu=0$, $f_0$ for its density, and $\Phi,\varphi$ for the
standard normal distribution function and density.
A planning state $\theta=(j,\mu)$ specifies the mean and the designated
claim $j\in\{S,F\}$; $\Theta$ is the set defined below.
With $c=\Phi^{-1}(1-\alpha)$, $0<\alpha<1$, define regret by
$R(\theta;D)=\Phi(\mu_j-c)-P_\mu(D_j=1)$ for rejection indicators
$D=(D_S,D_F)$. Our objective is
\begin{equation}\label{eq:decisionminimax}
 v_{\rm det}
 =\inf_{D\in\mathcal D_{\rm det}}
   \sup_{\theta\in\Theta}R(\theta;D).
\end{equation}
Here $\mathcal D_{\rm det}$ comprises nonrandomised, coordinatewise
nondecreasing decision pairs with strong familywise error control:
$P_{\mu}\{D_j=1\text{ for some }j\text{ with }\mu_j\le0\}\le\alpha$
for every $\mu\in\mathbb R^2$.
The validity domain includes unrestricted nuisance means under partial
nulls; $\Theta$ specifies the states whose power is protected.
Success means rejecting the designated claim, including when both
hypotheses are rejected. Proposition~\ref{prop:gates} establishes
$\Phi(\mu_j-c)$ as a separately attainable power envelope over
$\mathcal D_{\rm det}$.
If the target were known, applying its marginal test and never rejecting
the other hypothesis would attain this benchmark.
The nulls concern average efficacy: a full-population rejection alone
does not establish complement efficacy \citep{LiIvanova2024} or net benefit. For a fixed distribution over planning states, minimising average regret is equivalent to maximising weighted target power, since the reference-power term does not depend on the procedure. Under the minimax criterion, however, the state-specific benchmarks affect which efficacy scenarios are least favourable.

\subsection{Planning families and their interpretation}\label{sec:planningfamilies}

Uncertainty in target efficacy determines marginal reference power,
whereas uncertainty in the other mean affects the power of the joint
procedure. We specify these separately.
Let $I_j=[\underline\mu_j,\overline\mu_j]\subset(0,\infty)$,
$j\in\{S,F\}$, be nonempty intervals of plausible target
noncentralities. These are means of unit-variance statistics;
equal endpoints allow a fixed target effect.
Set $s=\sqrt{1-\rho^2}$, the conditional standard deviation of either
statistic given the other. Let $\delta$ and $\eta$ denote non-target mean offsets: $\delta$ describes variation in the full-population mean
when the subgroup is the target, while $\eta$ describes
variation in the subgroup mean when the full population
is the target. The finite bounds $e_S,e_F\ge0$ limit their magnitudes. The subgroup-target and full-population-target planning states
are denoted by $\theta_S(\mu_S,\delta)$ and
$\theta_F(\mu_F,\eta)$, respectively, and defined as
\begin{equation}\label{eq:alternatives}
\begin{split}
 \theta_S(\mu_S,\delta)
 &=\bigl(S,(\mu_S,\rho\mu_S+s\delta)\bigr),
 \quad \mu_S\in I_S,\ |\delta|\le e_S,\\
 \theta_F(\mu_F,\eta)
 &=\bigl(F,(\rho\mu_F+s\eta,\mu_F)\bigr),
 \quad \mu_F\in I_F,\ |\eta|\le e_F.
\end{split}
\end{equation}
The complete planning set is defined as $\Theta=
\{\theta_S(\mu_S,\delta):\mu_S\in I_S,\ |\delta|\le e_S\}
\cup
\{\theta_F(\mu_F,\eta):\mu_F\in I_F,\ |\eta|\le e_F\}.$ For each planning state $\theta=(j,\mu)\in\Theta$, the target designation $j\in\{S,F\}$ is prespecified,
although the true state $\theta$ is unknown at analysis.

The offsets $\delta$ and $\eta$ measure departures from $\mu_F=\rho\mu_S$ in subgroup-target
states and $\mu_S=\rho\mu_F$ in full-population-target states, in units
of $s$. Thus $\delta=(\mu_F-\rho\mu_S)/s$ in subgroup-target states and
$\eta=(\mu_S-\rho\mu_F)/s$ in full-population-target states. Each offset
varies the non-target mean while holding target efficacy fixed; a zero
bound fixes the corresponding reference relation. Negative offsets lower the non-target mean relative to its
zero-offset reference, without requiring that mean to be negative.
At fixed $\delta$,
$Y\mid X=x\sim N(\rho x+s\delta,s^2)$ is unchanged as $\mu_S$ varies;
likewise, $X\mid Y=y\sim N(\rho y+s\eta,s^2)$ is unchanged as $\mu_F$
varies with $\eta$ fixed.
This conditional invariance drives the regret-shape analysis.
Section~\ref{sec:clinical-example} illustrates the assignments and their
benefit--burden interpretation.
Under proportional recruitment and common per-observation variance,
zero offsets represent zero complement effect in subgroup-target states
and homogeneous benefit in full-population-target states.

\subsection{Reduction to a shared intersection test}\label{sec:intersectionreduction}
The original problem involves two rejection decisions under strong
familywise error control. The following proposition shows that
optimisation over nonrandomised monotone procedures can be reduced
to a single shared intersection test with fixed marginal tests,
without increasing regret. The reduction uses coordinatewise upper rejection regions, which contain every point larger than a rejected point in both coordinates. Specifically, let $M_S=1\{X>c\}$, $M_F=1\{Y>c\}$, and
$\mathcal G=\{x>c\}\cup\{y>c\}$.
Closed testing rejects an elementary hypothesis when both its marginal
test and the shared intersection test reject. The reduction builds on
the marginal rejection requirements of \citet{hkr2023} and leaves the
shared test as the design choice \citep{khr2025}.
\begin{proposition}[Reduction to closed testing]\label{prop:gates}
Every nonrandomised monotone decision pair with strong familywise
error control at level $\alpha$ satisfies $D_j\le M_j$ almost
everywhere. It is dominated by a closed-testing procedure
$D^\phi=(\phi M_S,\phi M_F)$, where $\phi$ is the indicator of an
upper set, vanishes outside $\mathcal G$, and satisfies
$E_0\phi\le\alpha$. Conversely, every such closed-testing
procedure strongly controls the familywise error rate.
\end{proposition}
Under a partial null, letting the non-null mean tend to infinity
forces each rejection decision to satisfy its marginal test
requirement. Taking $\phi=\max(D_S,D_F)$ then gives a shared
intersection test that preserves all original rejections.
The resulting closed-testing procedure can only increase
target power. Its regret is therefore the probability that
the target's marginal test rejects but the shared intersection
test does not, leading to the representation below.

Proposition~\ref{prop:gates} allows us to optimise the shared
intersection test rather than the two rejection decisions.
We first express regret in terms of this test and then
introduce a randomised relaxation. For an intersection test $\phi$, let
$D^\phi=(\phi M_S,\phi M_F)$ denote the induced
closed-testing decision pair. Its regret at $\theta=(j,\mu)\in\Theta$ is
\begin{equation}\label{eq:regret}
 R(\theta;\phi):=R(\theta;D^\phi)
 =\Phi(\mu_j-c)-E_{\mu}[\phi M_j],
 \qquad \theta=(j,\mu)\in\Theta.
\end{equation}
Within the two planning families, use
$R_S(\mu_S,\delta;\phi)=R(\theta_S(\mu_S,\delta);\phi)$ and
$R_F(\mu_F,\eta;\phi)=R(\theta_F(\mu_F,\eta);\phi)$. Additionally, let $\mathcal U$ be the class of coordinatewise nondecreasing
functions $0\le\phi\le1_{\mathcal G}$ with $E_0\phi\le\alpha$,
and let $\mathcal U_{\rm det}$ denote its indicator-valued
members. Every $D^\phi$ with $\phi\in\mathcal U_{\rm det}$
satisfies the monotonicity and strong familywise
error requirements. Conversely, Proposition~\ref{prop:gates}
gives every $D\in\mathcal D_{\rm det}$ a dominating
$D^\phi$ with no larger regret at any state.
Thus the deterministic minimax value is preserved:
\begin{equation}\label{eq:minimax}
v_{\rm ran}:=\inf_{\phi\in\mathcal U}
\sup_{\theta\in\Theta}R(\theta;\phi)
\le
v_{\rm det}=\inf_{\phi\in\mathcal U_{\rm det}}
\sup_{\theta\in\Theta}R(\theta;\phi).
\end{equation}
Allowing $\phi$ to take values in $[0,1]$ gives a convex
relaxation with fixed marginal tests. For a randomised rule, $\phi(X,Y)$ is the conditional probability of rejecting the shared intersection hypothesis; a single random decision is then combined with both marginal tests.

\section{Minimax-regret optimality and the form of the test}\label{sec:theory}

We first identify the efficacy scenarios that determine worst-case regret.
This reduction leads to an optimal shared intersection test and a way to
replace randomised decisions by a nonrandomised boundary. We then use the
boundary characterisation to determine when a familiar step-up procedure
already attains the minimax value, and quantify the possible loss when it
does not.

\subsection{Least-favourable scenarios and minimax optimality}\label{sec:saddle}
Increasing $\delta$ or $\eta$ raises only the non-target mean,
which cannot reduce target power by coordinatewise monotonicity,
while marginal reference power remains unchanged.
Hence the largest regret occurs at $\delta=-e_S$ and
$\eta=-e_F$, respectively, which are referred to as the worst-case offsets.
Restricting the planning families to these offsets gives two
one-parameter subfamilies, with states
$\theta_S^-(\mu_S)=\theta_S(\mu_S,-e_S)$ and
$\theta_F^-(\mu_F)=\theta_F(\mu_F,-e_F)$.
Their union forms the reduced planning set $\Theta_{\rm red}$;
write $R_S(\mu_S;\phi)$ and $R_F(\mu_F;\phi)$ for their regrets.
Within the subgroup-target subfamily,
$Y\mid X=x\sim N(\rho x-se_S,s^2)$ does not depend on $\mu_S$.
This allows target efficacy to vary while the conditional
probability of missing the claim, given $X=x>c$, remains fixed.
We use this property to characterise regret as a function of
target efficacy and identify its maximum within each
planning interval.

\begin{lemma}[Regret shape]\label{lem:shape}
For every $\phi\in\mathcal U$, each target-specific regret at the
worst-case offset either vanishes identically or has a unique maximiser
on every compact noncentrality interval. Its nonzero extension to real
noncentrality parameters is strictly increasing, or increases strictly
to one finite mode and then decreases strictly.
\end{lemma}
Increasing target efficacy raises both marginal reference power and the
chance of passing the shared test. Their difference can therefore be
largest at an intermediate effect. For the subgroup target,
\begin{equation}\label{eq:miss}
\begin{split}
R_S(\mu_S;\phi)
&=\int_c^\infty g(x)\varphi(x-\mu_S)\,dx,\\
g(x)
&=E[1-\phi(x,\rho x-se_S+sW)],
\quad W\sim N(0,1).
\end{split}
\end{equation}
Here $g(x)$ is the conditional probability of failing the shared test
given $X=x>c$. It is independent of $\mu_S$, nonnegative and nonincreasing
by coordinatewise monotonicity and $\rho\ge0$. Varying target efficacy
changes only the normal weighting density in \eqref{eq:miss}.

The normal convolution gives a short explanation of the shape result.
After division by the positive density $\varphi(c-\mu_S)$, the regret
derivative is a constant minus an increasing exponential integral.
When $g$ is nonconstant, this expression crosses zero once, from positive
to negative; when $g$ is constant, nonzero regret is strictly increasing.
The same argument applies to the full-population target. Thus a nonzero shortfall
has a unique maximiser on each planning interval, although it can lie in
the interior. The Stieltjes argument extends classical shortcoming analysis
\citep{VanZwetOosterhoff1967} by allowing $g(x)$ to have a positive limit as $x\to\infty$.

Lemma~\ref{lem:shape} identifies one worst-case target effect per family.
The next theorem shows how these two states jointly determine the minimax
solution and when that solution is nonrandomised and unique.

\begin{theorem}[Minimax solution and two least-favourable scenarios]
\label{thm:saddle}
For every finite pair $(e_S,e_F)$, the minimax problem over
$\mathcal U$ has an attained saddle point and $v_{\rm ran}>0$.
Every least-favourable prior on $\Theta_{\rm red}$ has the form
\begin{equation}\label{eq:prior}
\Pi_*
=q_*\delta_{\theta_S^-(\mu_{S,*})}
 +(1-q_*)\delta_{\theta_F^-(\mu_{F,*})},
\qquad 0<q_*<1,
\end{equation}
where $\mu_{S,*}$ and $\mu_{F,*}$ uniquely maximise the
corresponding regrets and both regrets equal $v_{\rm ran}$. If
\begin{equation}\label{eq:J}
J_0
=\underline{\mu}_S\underline{\mu}_F
+\frac{\rho(\underline{\mu}_S e_F+\underline{\mu}_F e_S)}s
-e_Se_F>0,
\end{equation}
then $v_{\rm det}=v_{\rm ran}$, and the minimax monotone rule is
nonrandomised, unique up to null sets, and has size $\alpha$
under $P_0$.
\end{theorem}
Theorem~\ref{thm:saddle} reduces the continuous planning
problem to two least-favourable efficacy scenarios, one
from each target family. Their noncentrality parameters
may lie in the interiors of the planning intervals, so
the most adverse effects need not be the smallest plausible
effects. Both families receive positive weight and attain
the same maximum regret at the minimax solution.
The measure $\Pi_*$ is conventionally called a
least-favourable prior; its weights are determined by
the optimisation rather than by clinical frequencies.
A prior supported on only one target family has zero
minimum average regret, whereas achieving zero regret
for both families would require rejection throughout
$\mathcal G$, whose null probability exceeds $\alpha$. The condition $J_0>0$ provides a further conclusion:
the minimax solution can be chosen nonrandomised and
is unique up to Gaussian null sets. It ensures a common
increasing direction for the target likelihood ratios,
allowing conditional Gaussian rearrangement to preserve
size and monotonicity without reducing target power.
The construction is developed in the next section.

The two-point characterisation follows from minimax
attainment and Lemma~\ref{lem:shape}, which gives a
unique regret maximiser within each target family.
The saddle point pairs these two scenarios with a
minimax testing procedure (Supplementary Section~S2). The remaining question is whether randomisation can improve the minimax procedure. Condition~\eqref{eq:J}
provides a positive direction $(d_x,d_y)$ along which
both target likelihood ratios increase, together with
the covariance restriction needed for conditional
Gaussian rearrangement \citep{Ehrhard1983}.
Although adverse offsets can make a likelihood ratio
decrease in one coordinate, both increase along this
shared direction. Rearrangement then replaces a
randomised rule by a nonrandomised monotone rule of
the same size without reducing any target power.
The condition evaluated at $\mu_{S,*},\mu_{F,*}$
also suffices for the nonrandomised minimax conclusion.
We next construct this replacement and characterise
its boundary.

\subsection{A convex characterisation of the optimal boundary}
\label{sec:boundary}

Theorem~\ref{thm:saddle} reduces the minimax problem to two
least-favourable efficacy scenarios. For fixed target
noncentrality parameters $(\mu_S,\mu_F)$ and prior weight
$q\in(0,1)$, minimising average regret is equivalent to
maximising weighted target power. We therefore consider
the corresponding Bayes problem of maximising
$E_0[\phi H]$, where
$H=qM_SL_S+(1-q)M_FL_F$.
Here $L_S$ and $L_F$ are the likelihood ratios at the
worst-case offsets:
\begin{equation}\label{eq:likelihoods}
\begin{split}
L_S(x,y)
&=\exp\{\mu_S x-\mu_S^2/2
-e_S(y-\rho x)/s-e_S^2/2\},\\
L_F(x,y)
&=\exp\{\mu_F y-\mu_F^2/2
-e_F(x-\rho y)/s-e_F^2/2\}.
\end{split}
\end{equation}
Under the size constraint alone, rejecting at the largest values of $H$
maximises weighted power. However, $\partial_y\log L_S=-e_S/s$ and
$\partial_x\log L_F=-e_F/s$: raising one statistic can lower the other
target's likelihood ratio. A threshold on $H$ may therefore reject at one
point and accept at a larger point. We enforce monotonicity through
rearrangement along the common increasing direction.

Under \eqref{eq:J}, choose $d_x,d_y>0$ so that both likelihood ratios
increase along $(d_x,d_y)$ and $\rho<d_x/d_y<1/\rho$, with
infinite-endpoint conventions at $\rho=0$. Choose the direction at the
lower target noncentrality endpoints.
The ratio restriction makes $d_y-\rho d_x$ and $d_x-\rho d_y$ positive,
so each directional log-likelihood-ratio derivative is smallest at the
worst offset for a fixed target noncentrality. The same direction therefore
works throughout the full planning set $\Theta$.

To exploit this common increasing direction, we transform the observations into coordinates $(T,Z)$, so that optimisation can proceed along lines parallel to
$(d_x,d_y)$.
Set $X=d_x(T+Z)$ and $Y=d_y(T-Z)$. The contrast $Z$ indexes parallel
lines, and $T$ locates an observation along each line. Increasing $T$
raises both likelihood ratios. Conditional on $Z$, allocating null
rejection probability to the largest $T$ values thus improves both target
powers. The challenge is to make these conditional decisions jointly
coordinatewise monotone. Under $P_0$, $Z\sim N(0,\tau^2)$ and
$T\mid Z=z\sim N(\beta z,\sigma^2)$; the ratio restriction gives
$|\beta|<1$.

\begin{figure}[!htbp]
\normalsize\setstretch{1}\centering
\accessiblefigure{The same upper rejection region is shown in original and transformed coordinates. Lines of fixed contrast become vertical; the highlighted line enters the rejection region above its boundary threshold.}
{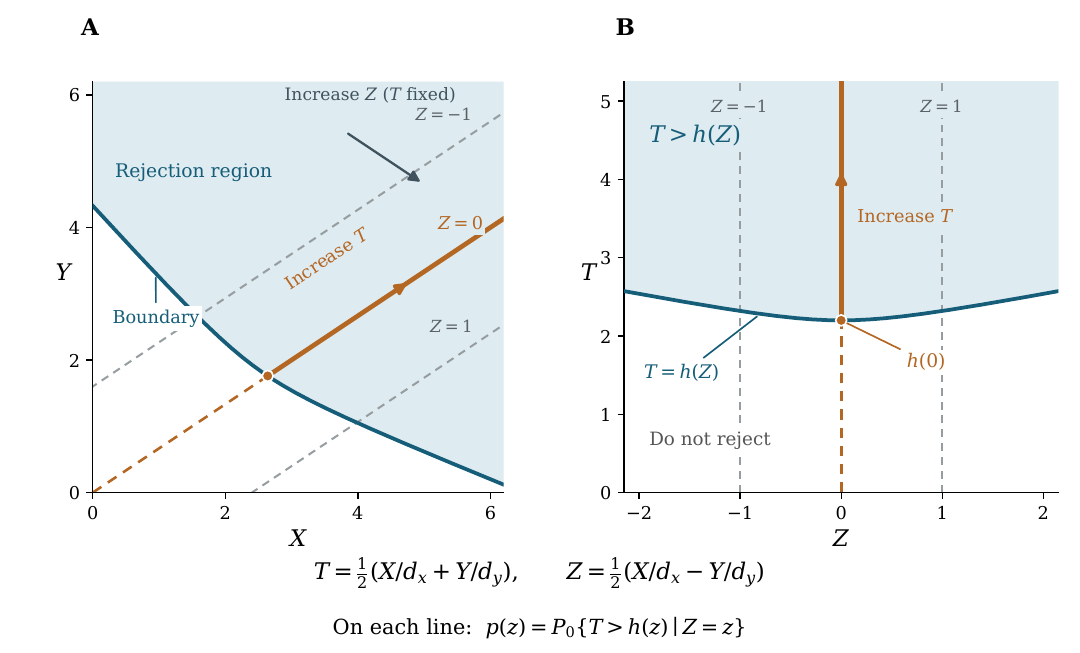}
\caption{Coordinate transformation. Lines of constant $Z$ follow the common increasing direction (A) and become vertical in $(Z,T)$ coordinates (B). Shading marks rejection above $h(Z)$; orange highlights $Z=0$.}
\label{fig:coordinate-transformation}
\end{figure}

Every coordinatewise upper rejection region $A$ with $0<P_0(A)<1$
has, up to $P_0$-null sets, the representation $A=\{T>h(Z)\}$, where
$h:\mathbb R\to\mathbb R$ satisfies
$|h(z_1)-h(z_2)|\le |z_1-z_2|$ for all $z_1,z_2\in\mathbb R$.
To see why the threshold is finite, move far enough along a conditional
line to place both coordinates above a point in $A$, or far enough
backwards to place them below a point outside $A$. Comparing adjacent
lines gives the Lipschitz bound. The empty region has $h\equiv+\infty$.
Writing $\overline\Phi=1-\Phi$, the conditional null rejection probability
is $p(z)=\overline\Phi\{(h(z)-\beta z)/\sigma\}$.
Figure~\ref{fig:coordinate-transformation} illustrates the
change of coordinates: lines of constant $Z$ become vertical,
and the rejection region is an upper tail in $T$ on each line.

Let $I(p)=\varphi(\Phi^{-1}(p))$, continuously
extended to zero at both endpoints. The coordinatewise
monotonicity constraint becomes
\begin{equation}\label{eq:shapeconstraints}
-\gamma_-I(p)\le p'\le\gamma_+I(p),
\qquad
\gamma_-=(1-\beta)/\sigma,\quad
\gamma_+=(1+\beta)/\sigma.
\end{equation}
Since $I$ is concave, these inequalities define a convex class.
Containment in $\mathcal G$ requires
$p(z)\le p_{\mathcal G}(z)=\overline\Phi\{(h_{\mathcal G}(z)-\beta z)/\sigma\}$,
where $h_{\mathcal G}(z)=\min\{c/d_x-z,c/d_y+z\}$.
Let $\mathcal K_{\mathcal G}$ be the convex class of locally absolutely
continuous functions $p:\mathbb R\to[0,1]$ with
$0\le p(z)\le p_{\mathcal G}(z)$ for every $z$, satisfying
\eqref{eq:shapeconstraints} almost everywhere; $p\equiv0$ is included.
The conditional probabilities cannot be chosen independently: the slope
bounds tie neighbouring lines together to preserve coordinatewise
monotonicity, while the upper bound restricts rejection to $\mathcal G$.
Equivalent distributional and finite-difference formulations, including
endpoints, and the conditional-normal parameters are given in
Supplementary Section~S3.

Put $w(z)=\varphi(z/\tau)/\tau$,
$\widetilde H(t,z)=H(d_x(t+z),d_y(t-z))$, and
\begin{equation}\label{eq:F}
F(z,p)=
\int_{\overline\Phi^{-1}(p)}^\infty
\widetilde H(\beta z+\sigma u,z)\varphi(u)\,du.
\end{equation}
For fixed $z$, $F(z,p)$ is the conditional weighted target
power obtained by rejecting an upper tail with conditional
null probability $p$, with $F(z,0)=0$.
Both mixture weights are positive. Wherever at least one marginal indicator
is active, its weighted likelihood-ratio contribution increases strictly
along $T$, while the other contribution is nondecreasing. Increasing $p$
therefore includes observations with strictly smaller values of $H$ and
gives decreasing marginal power gains. Consequently, $F(z,p)$ is continuous
and strictly concave in $p$ for $0\le p\le p_{\mathcal G}(z)$.
Upward jumps in $H$ at the marginal rejection thresholds
give one-sided slope changes consistent with concavity.

\begin{theorem}[Conditional rearrangement]\label{thm:boundary}
Under \eqref{eq:J}, every $\phi\in\mathcal U$ is dominated
in all target powers by a nonrandomised monotone rule
of the same intersection-test size. For the fixed two-state
prior with $q\in(0,1)$, the unique Bayes rule in
$\mathcal U$ is obtained from
\begin{equation}\label{eq:convexproblem}
\max_{p\in\mathcal K_{\mathcal G}}
\int w(z)F(z,p(z))\,dz
\quad\text{subject to}\quad
\int w(z)p(z)\,dz=\alpha.
\end{equation}
A finite positive Lagrange multiplier for the size
constraint exists. At a proposed feasible solution $p_*$,
global optimality is equivalent to
\begin{equation}\label{eq:VI}
\int w(z)
F'_+(z,p_*(z);p(z)-p_*(z))\,dz\le0
\end{equation}
for every feasible $p$ of size $\alpha$. Here $F'_+(z,p_*(z);p(z)-p_*(z))$ denotes the right derivative
at $t=0$ of
$F(z,p_*(z)+t\{p(z)-p_*(z)\})$.
\end{theorem}
The theorem expresses the choice of a monotone boundary as an allocation
of conditional null rejection probabilities, with a fixed total type~I
error budget. Its convex formulation permits global optimality assessment
over $\mathcal U$.
To construct the replacement, write
$\phi=\int_0^1 1\{\phi>u\}\,du$. Each level set is upper and lies inside
$\mathcal G$, so its conditional rejection probabilities belong to
$\mathcal K_{\mathcal G}$. Their average
$p_\phi(z)=E_0(\phi\mid Z=z)$ belongs to this convex class, although
individual levels need not have size at most $\alpha$. Replace the rule
on each line by the upper Gaussian tail with probability $p_\phi(z)$.
Each target likelihood ratio multiplied by its marginal indicator is
nondecreasing along that line, so this replacement improves or maintains
every target power. Membership in $\mathcal K_{\mathcal G}$ preserves
coordinatewise upperness and containment in $\mathcal G$; averaging over
$Z$ preserves the intersection-test size.

For $q\in(0,1)$, strict increase of $H$ inside $\mathcal G$ makes equality
in the conditional power comparison possible only for this upper-tail
allocation. Thus every Bayes optimiser has the same tail form. Strict
concavity of $F$ gives a unique optimal probability function and hence a
unique Bayes rule, up to null sets. Fixing one least-favourable prior then
gives minimax uniqueness: every minimax rule must attain its Bayes value
and agree with that rule.

The variational inequality \eqref{eq:VI} states that
no feasible redistribution of conditional rejection
probability has a positive directional gain in weighted
target power. Concavity makes this a global optimality
condition. On a smooth boundary portion where neither
slope constraint nor the upper-bound constraint is active,
$H$ equals the Lagrange multiplier for the size constraint
on the boundary. Active slope constraints produce vertical
or horizontal boundary segments where the unconstrained
level curve would violate coordinatewise monotonicity.
Combining \eqref{eq:VI} with the two maxima over
noncentrality parameters and regret equalisation gives
a necessary and sufficient condition for a saddle point
under \eqref{eq:J}.

\subsection{When a step-up procedure is optimal}\label{sec:rectangles}

Section~\ref{sec:boundary} characterises the optimal monotone
intersection test without restricting its rejection boundary
to a prespecified shape. We now ask when a familiar step-up
procedure is already minimax optimal, so that allowing more
flexible rejection regions cannot improve worst-case regret.
We study the two-hypothesis step-up family with intersection
rejection region $R=\{x>x_0\}\cup\{y>y_0\}\cup Q,~~
 Q=(c,\infty)^2, ~ x_0,y_0>c.$ The procedure rejects both hypotheses when both marginal tests reject. When only one rejects, its statistic must also exceed $x_0$ for the subgroup or $y_0$ for the full population. The single-rejection arms are $\{x>x_0,y\le c\}$ and $\{y>y_0,x\le c\}$. Equal arm cutoffs give the
two-test Gaussian Dunnett--Tamhane procedure \citep{dt1992}.

We first determine the optimal cutoffs within the step-up
family. Exact calibration requires the intersection test
to have null rejection probability $P_0(R)=\alpha$.
This constraint links $x_0$ and $y_0$, leaving one
free parameter to balance the worst-case regrets of
the two target claims. 

Exact calibration creates a trade-off between the two
target claims. Increasing $x_0$ raises subgroup regret
but requires a decrease in $y_0$, thereby reducing
full-population regret. Thus the optimal calibrated
step-up procedure is the unique rule with finite cutoffs
that equalises the two interval maximum regrets.
This identifies the best rule within the step-up family,
but does not establish optimality over all monotone
procedures.

For any calibrated rule $R$ in this family, let $\mu_{S,R},\mu_{F,R}$
maximise its two regrets over $I_S,I_F$, respectively.
Write $z(x)=(c-\rho x)/s$ and $w_S(x)=\varphi(x)\Phi(z(x))$.
The conditional average likelihood ratios in the two arms are $C_S(x)
 =e^{\mu_{S,R}x-\mu_{S,R}^2/2}
 \frac{\Phi(z(x)+e_S)}{\Phi(z(x))}$ and $C_F(y)
 =e^{\mu_{F,R}y-\mu_{F,R}^2/2}
 \frac{\Phi(z(y)+e_F)}{\Phi(z(y))}$, respectively.
These scores are strictly increasing. Their cutoff values measure marginal
target-power gain per unit null rejection probability. Balance the weighted
gains and fit a size penalty $k$ by
\begin{equation}\label{eq:fit}
 qC_S(x_0)=(1-q)C_F(y_0)=k,
 \quad
 q=\frac{C_F(y_0)}{C_S(x_0)+C_F(y_0)}.
\end{equation}
Equation~\eqref{eq:fit} balances these gains after
weighting the two target families and determines
a candidate multiplier $k$ for the size constraint.
This local balance does not establish global optimality,
which also requires considering changes within $Q$.

\begin{figure}[!htbp]
\normalsize\setstretch{1}\centering
\accessiblefigure{A step-up rejection region is compared with the region after a lower subset of the joint quadrant is deleted. The deleted subset and the two required arm deletions are red; retained rejection regions are blue.}
{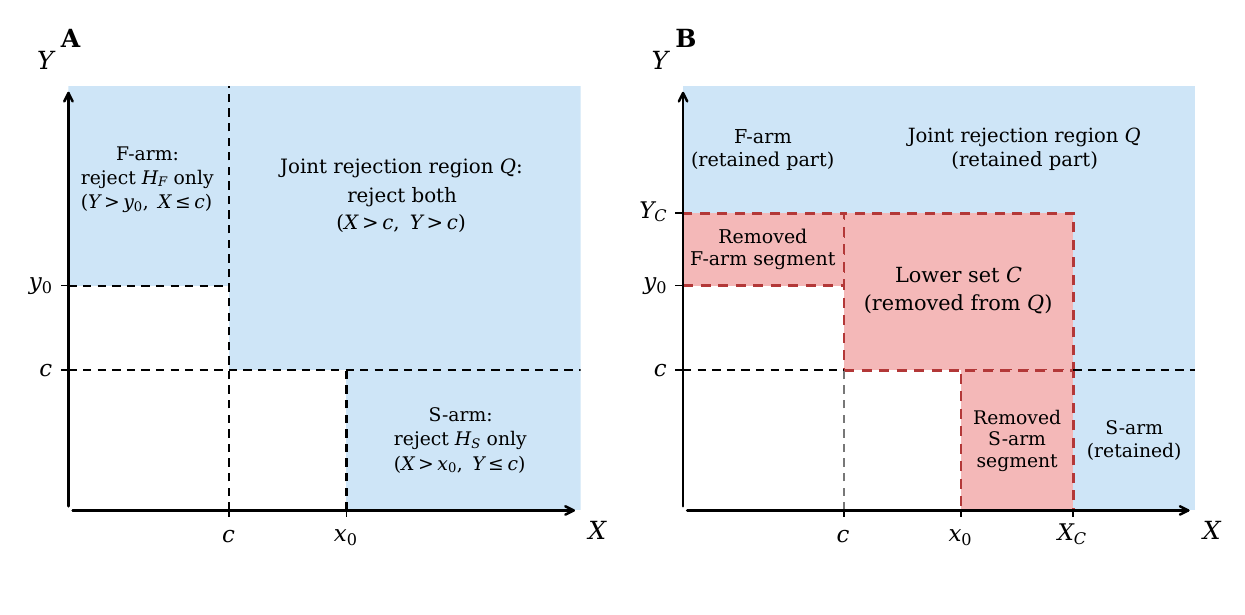}
\caption{Coupled deletions under the integral criterion. Removing a lower set $C$ from $Q$ also shortens both rejection arms. The three red regions correspond to the terms in $\Psi_R(C)$.}
\label{fig:stepup_deletion}
\end{figure}

Using \eqref{eq:likelihoods} at $\mu_{S,R},\mu_{F,R}$,
define $H_Q=qL_S+(1-q)L_F$ on $Q$.
Removing points where $H_Q<k$ can increase penalised
weighted power by releasing null rejection probability.
However, monotonicity may also require shortening the
two single-rejection arms. The integral criterion measures
the combined effect of these changes, as illustrated in
Figure~\ref{fig:stepup_deletion}.

To assess these modifications under monotonicity,
consider deleting a lower set $C\subset Q$, which contains
every coordinatewise smaller point of $Q$ whenever it
contains a point. Write $X_C,Y_C$ for its coordinate
suprema, taking both to be $c$ for the empty set and
allowing infinity. Define
\begin{equation}\label{eq:Psi}
\begin{aligned}
\Psi_R(C)
&=\int_C(H_Q-k)f_0(x,y)\,dx\,dy\\
&+\int_{x_0}^{\max(x_0,X_C)}[qC_S(x)-k]w_S(x)\,dx
+\int_{y_0}^{\max(y_0,Y_C)}[(1-q)C_F(y)-k]w_S(y)\,dy.
\end{aligned}
\end{equation}
The first term is the loss in penalised weighted power
from deleting $C$; it can be negative where $H_Q<k$.
The other two terms are the additional arm costs,
finite and nonnegative even at infinite endpoints.
Thus $\Psi_R(C)<0$ identifies an improvement for
this deletion. Requiring nonnegative loss for every
lower $C$, together with regret equalisation, gives
the following exact criterion.

\begin{theorem}[Integral criterion for step-up optimality]
\label{thm:rectangle}
An exactly calibrated step-up procedure is minimax over
$\mathcal U$ if and only if its two interval worst regrets
are equal and $\Psi_R(C)\ge0$ for every lower $C\subset Q$,
with the fit \eqref{eq:fit} at its own worst noncentrality
parameters. Under \eqref{eq:J}, the same statement is
necessary and sufficient among nonrandomised monotone
procedures with strong familywise error control.
Without \eqref{eq:J}, the condition remains sufficient
for nonrandomised minimax optimality.
\end{theorem}
Equalisation balances the two worst regrets within the step-up family;
the integrated criterion checks more general monotone boundaries. For a
fixed missed set $C$, conditional averaging replaces each single-rejection
arm by a complete tail with no smaller power. An arm cannot start below
its corresponding coordinate supremum without forcing rejection of a
missed quadrant point. The best cutoffs for $E_0[\phi(H-k)]$ are therefore
$\max(x_0,X_C)$ and $\max(y_0,Y_C)$, giving exactly $\Psi_R(C)$.
Calibration details and the proof are given in Supplementary Section~S4.

We call $H_Q\ge k$ throughout $Q$ the pointwise
sufficient condition. It implies the integrated
criterion, but is not necessary: a step-up procedure can be minimax with
$H_Q<k$ away from the corner, yet nonnegative integrated loss for every
permissible deletion (Supplementary Section~S7). Thus, global step-up
optimality depends on the combined consequences
of admissible boundary changes, rather than on
the likelihood ratio at individual observations.
The corner condition $H_Q(c,c)\ge k$ itself is
necessary, since a small corner deletion has no
arm cost. When the integrated criterion holds
and the regrets are equal, the fitted $q$ and
the step-up procedure's own worst noncentrality
parameters form a least-favourable prior.

\subsection{The zero-offset case and the cost of a simple rule}\label{sec:zero-offset}
The zero-offset case provides a benchmark in which the
non-target means follow their reference relations,
while target efficacy may still vary over the planning
intervals. At $e_S=e_F=0$, the likelihood ratios are
coordinatewise nondecreasing, and
Theorem~\ref{thm:rectangle} reduces to the scalar condition
\begin{equation}\label{eq:nominalcriterion}
 \exp\{-\mu_{S,R}(x_0-c)\}
 +\exp\{-\mu_{F,R}(y_0-c)\}\ge1,
\end{equation}
in addition to regret equalisation; \citet{hkr2023}
obtain the symmetric independent likelihood-ratio
condition. For identical intervals $[\ell,u]$,
$0<\alpha\le1/2$, and symmetric calibrated cutoff
$t$, the rule is minimax precisely when
$\ell\le\log(2)/(t-c)$. The upper endpoint
does not enter. Thus, the calibrated step-up rule
is minimax over $\mathcal U$ when the lower effect
bound is at or below this threshold; above it,
a more flexible monotone rejection boundary can
reduce worst-case regret.

Failure of the exact optimality criterion does not indicate
how much worst-case regret could be reduced by a more
flexible procedure. We therefore bound the possible
improvement over a calibrated step-up rule. Let
$r_S,r_F$ be its interval maximum regrets,
$v_R=\max(r_S,r_F)$, and
$\epsilon_R=v_R-qr_S-(1-q)r_F$, using its fitted
two-point prior. The term $\epsilon_R$ measures
regret imbalance and vanishes when the two
maximum regrets are equal. In addition, write $W_R$ for the step-up rule's weighted target
power under this prior and $W^*$ for the maximum
weighted target power over $\mathcal U$.
The fixed-prior lower bound on minimax regret gives
$v_R-v_{\rm ran}\le\epsilon_R+(W^*-W_R)$.
The fitted arm scores ensure that conditional averaging
leaves no positive penalised-power gain from changing
the two arms alone. Any remaining gain is bounded by
the negative part of $H_Q-k$ within the joint
rejection quadrant. Consequently, $0\le v_R-v_{\rm ran}
\le \epsilon_R+E_0[1_Q(k-H_Q)_+].$
The same bound applies to $v_R-v_{\rm det}$.
With strict inequality in
\eqref{eq:nominalcriterion}, the equal-regret
step-up procedure with reoptimised cutoffs stays
minimax in a positive neighbourhood of zero
offset bounds. Retaining the zero-offset cutoffs
under asymmetric uncertainty can instead
create a first-order loss. The bound quantifies possible improvement over a step-up rule,
but does not describe how the actual regret loss emerges
when exact optimality fails. The following proposition
characterises this transition in the symmetric zero-offset case.

\begin{proposition}[Sharp transition with zero offsets]
\label{prop:cubic}
Fix $\rho<1$ and $\alpha$, let $t$ be the symmetric
step-up cutoff and $h_0=\log(2)/(t-c)$.
For the two zero-offset simple alternatives with
common noncentrality parameter $h>h_0$, put
$d(h)=1-2e^{-h(t-c)}$ and
$k_0(h)=\tfrac12e^{ht-h^2/2}$.
The actual minimax regret loss of the step-up
rule satisfies
\begin{equation}\label{eq:cubic}
 R_S(h;1_R)-v_{\rm det}(h)
 =\frac{2k_0(h_0)f_0(c,c)}{3h_0^2}\,d(h)^3
 +O(d(h)^4)
 \quad\text{as }h\downarrow h_0.
\end{equation}
For $\alpha\le1/2$, the same expansion holds
locally for identical intervals $[h,h+w]$,
with any fixed finite $w\ge0$.
\end{proposition}

Proposition~\ref{prop:cubic} shows that the loss
from using the step-up procedure grows only at
cubic order immediately beyond the optimality
threshold. Thus, failure of exact minimax
optimality need not imply a substantial increase
in worst-case regret. The cubic order has a geometric explanation.
Here the deficit of the mixture likelihood
ratio below its critical value is linear
near the corner, and the deleted region
has area of order $d^2$. Restoring its
probability mass moves the critical value
by order $d^2$ and costs only order $d^4$.
The expansion fixes a nonsingular correlation
and is not uniform as $\rho$ approaches one.

\section{Computation and bounds for the continuous problem}
\label{sec:computation}

Computation has two tasks: construct feasible testing procedures
and bound how much any better procedure could reduce worst-case
regret. We construct feasible nonrandomised candidates using
monotone staircase dynamic programming, calibrated step-up
and closed max-$Z$ procedures, and conditionally averaged
likelihood-ratio rules when their sufficient monotonicity
check holds numerically. A feasible rule provides an upper
bound on the minimax value, while a lower bound limits
the possible improvement. The bounds below apply to the
full class $\mathcal U$, including rules that vary within
computational grid cells.

To bound further improvement, follow \citet[Section~6.3]{rly2014}
and fix any two-point prior $(q,\mu_S,\mu_F)$ on the worst-offset
families, with $0<q<1$, $\mu_j\in I_j$, and a penalty $k\ge0$.
Write $H=qM_SL_S+(1-q)M_FL_F$ at these states and let
$O=q\Phi(\mu_S-c)+(1-q)\Phi(\mu_F-c)$ be its weighted
reference power. Partition a finite box $K$ into rectangles,
including the marginal critical lines. Monotonicity bounds
a rule within each cell by its lower-left and upper-right
vertex values. Maximising cellwise upper bounds on penalised
power over ordered vertex values in $[0,1]$, vanishing outside
$\mathcal G$, gives $M_K(k)$. This relaxation covers arbitrary
monotone behaviour within cells. A tail allowance
$T_K\ge\int_{K^c}Hf_0$ then gives
$E_0[\phi H]\le k\alpha+M_K(k)+T_K$ for every
$\phi\in\mathcal U$. Subtracting this bound from $O$
bounds prior-average regret from below, and hence
the maximum regret:

\begin{equation}\label{eq:continuumbracket}
 O-k\alpha-M_K(k)-T_K
 \le v_{\rm ran}\le v_{\rm det}
 \le\sup_{\theta\in\Theta}R(\theta;\widehat\phi)
\end{equation}
for every feasible nonrandomised rule $\widehat\phi$,
at all finite offset bounds.

The endpoint difference bounds possible improvement over the candidate;
the prior and penalty may be selected to tighten it. For a fixed prior
and penalty, the vertex relaxation converges to the continuous penalised-power
optimum under mesh refinement and expanding boxes.
Table~\ref{tab:continuum-certificate} reports floating-point evaluations.
Displayed gaps exclude rigorous integration and rounding allowances;
mesh and quadrature diagnostics assess numerical sensitivity separately.
Construction, calibration, convergence and integration-error details are
given in Supplementary Section~S10.

\begin{table}[!htbp]
\centering\normalsize\setstretch{1}
\caption{Continuous-problem regret bounds in percentage points. The interval rows use $e=e_S=e_F$. Feasible regret is the best evaluated nonrandomised value; gaps use unrounded values in $10^{-3}$ percentage points. QUEST A/B use (complement-effect, stratum-difference) bounds $(20,80)$ and $(5,125)$ mL. Numerical error allowances are excluded; Section~S10 gives full refinement diagnostics.}
\label{tab:continuum-certificate}
\begin{tabular}{lrrr}
\toprule
Scenario & Lower bound & Feasible regret & \shortstack{Gap\\($10^{-3}$ pp)} \\
\midrule
$e=0.1$ & 8.122 & 8.122 & 0.472 \\
$e=0.25$ & 8.469 & 8.471 & 1.324 \\
$e=0.4$ & 8.704 & 8.707 & 2.812 \\
$e=0.8$ & 9.036 & 9.041 & 5.174 \\
QUEST A & 6.181 & 6.188 & 6.438 \\
QUEST B & 5.984 & 5.990 & 5.449 \\
\bottomrule
\end{tabular}
\end{table}

\FloatBarrier
\section{Numerical illustration}\label{sec:numerical}

We examine how non-target uncertainty changes the preferred rejection
boundary and the allocation of power between claims. We use
$\alpha=0.025$, $\rho=\sqrt{0.4}$, $I_S=[1.5,4]$ and $I_F=[2,4.5]$,
and also consider fixed target noncentrality parameters of $3.2$.
Regrets are reported in percentage points and maximised over the planning
intervals at the worst-case offsets. We compare optimised step-up
procedures, symmetric closed max-$Z$, monotone staircase boundaries and
conditionally averaged likelihood-ratio rules.

For the stated intervals and equal offset bounds, the optimised step-up
procedure is symmetric Dunnett--Tamhane (DT), with arm cutoff
approximately $2.208$. Symmetric closed max-$Z$ rejects the intersection
when $\max(x,y)>t_{\max Z}\approx2.193$. Relative to DT, it removes
rejection points near the joint marginal corner and extends the arms.
For a fixed target effect, wider offset bounds lower the non-target mean
in the worst-case states, making joint marginal rejection less likely.
In these examples, reallocating null rejection probability towards the
arms protects the designated claim when the other statistic is small.

\begin{figure}[!htbp]
\normalsize\setstretch{1}\centering
\accessiblefigure{Three rejection boundaries as uncertainty increases. The first coincides with DT, the second removes a small corner, and the third coincides with closed max-Z. Shading marks rejection regions.}
{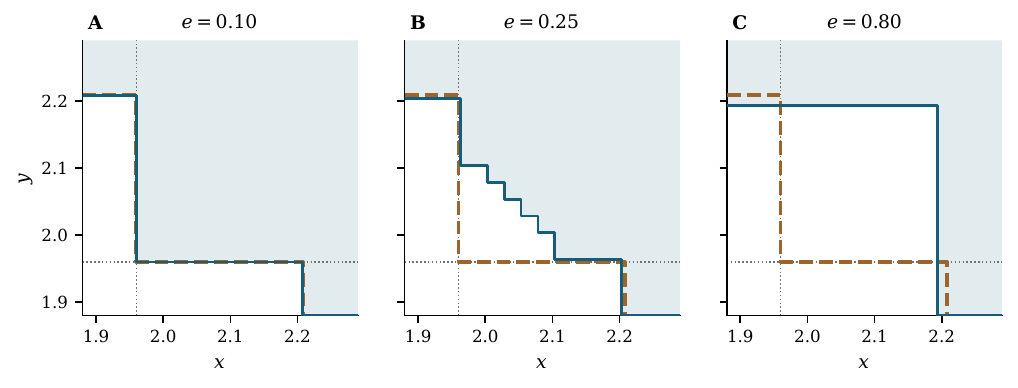}
\caption{Intersection rejection boundaries at common offset bounds $0.10$, $0.25$ and $0.80$ (A--C). Dashed brown: DT; solid blue: calibrated grid rule; dotted: marginal critical values. Shading marks rejection; Section~\ref{sec:numerical} gives the regrets.}
\label{fig:robust-boundaries}
\end{figure}

Figure~\ref{fig:robust-boundaries} shows how the preferred boundary changes.
At common offset bound $0.10$, DT passes the numerical pointwise
sufficient check and has worst regret $8.122$. At $0.25$, a modest
change in geometry helps: the conditionally averaged likelihood-ratio
rule reduces regret from $8.479$ to $8.471$. The grid candidate has the
same displayed regret; panel B shows its staircase boundary. At $0.80$,
symmetric closed max-$Z$ reduces DT's regret from $9.355$ to $9.041$,
a gain of $0.314$ points, and the refined staircase coincides with it.
Thus a familiar procedure with longer arms captures the improvement at
this larger offset bound. When both target noncentralities are fixed at
$3.2$, the conditionally averaged rule improves DT even at $0.10$:
optimality depends on the target ranges as well as non-target uncertainty.

Table~\ref{tab:continuum-certificate} assesses the remaining scope for
improvement. At $0.80$, the evaluated gap is small relative to the gain
over DT. At $0.10$, the sufficient check supports DT independently of
the lower-bound calculation.

Asymmetric uncertainty also makes cutoff reoptimisation valuable. With
$e_S=0.25$ and $e_F=0$, the subgroup and full-population cutoffs change
to approximately $2.198$ and $2.218$, respectively, reducing worst regret
from $8.479$ to $8.157$. Lowering the subgroup cutoff and raising the
full-population cutoff reallocates power towards the adversely affected
target. This gain comes from changing the allocation within the step-up
family, with its boundary form fixed. Correlation misspecification changes
intersection-null size while preserving the DT versus closed max-$Z$
regret ordering in the examined scenarios. The next example connects
these power allocations to clinically motivated population targets.

\FloatBarrier
\section{An asthma planning example}
\label{sec:clinical-example}\label{sec:priority}\label{sec:asthma-example}

We use planning inputs from LIBERTY ASTHMA QUEST for change in
pre-bronchodilator forced expiratory volume in one second (FEV$_1$) at
week 12. The prespecified subgroup has baseline blood eosinophils at least
300 cells/$\mu$L and anticipated prevalence $\pi=0.42$
\citep{BusseEtAl2018}. For 200 mg dupilumab versus matched placebo,
reported mean differences were 0.21 L in the subgroup and 0.14 L overall
\citep{CastroEtAl2018,DupixentLabel2026}. For future planning, proportional
recruitment and a common per-observation variance give $\rho=\sqrt{0.42}$.
We use a working full-population standard error of approximately 28.1 mL
and construct sensitivity ranges from the published marginal confidence
intervals.

To interpret the target assignments, compare treating nobody, the subgroup
$S$ and the full population $F$. Let $\Delta_S,\Delta_C$ be treatment
effects in the subgroup and its complement, and let $\tau_S\ge0$ and
$\tau_C>0$ be prespecified treatment burdens on the same additive scale.
Their population-weighted net-benefit values are
\begin{equation}\label{eq:actionvalue}
U_0=0,\qquad
U_S=\pi(\Delta_S-\tau_S),\qquad
U_F=U_S+(1-\pi)(\Delta_C-\tau_C).
\end{equation}
Provided subgroup net benefit is positive,
$U_F-U_S=(1-\pi)(\Delta_C-\tau_C)$ shows why the complement's net
benefit determines whether extending treatment is preferred. A common
burden has $\tau_S=\tau_C$. Compatibility requires fixed burdens favouring $S$ throughout
subgroup-target states and $F$ throughout full-population-target states. Requiring
positive net benefit in every treated stratum gives the stronger
conditions that imply $J_0>0$ (Supplementary Section~S8).

Table~\ref{tab:clinical-calibration} compares reoptimised procedures across
bounds on absolute complement effects in subgroup-target states and
absolute between-stratum effect differences in full-population-target
states. At respective bounds of 20 and 80 mL, reoptimising closed max-$Z$
reduces worst regret from
$6.760$ to $6.188$ percentage points. It is the best evaluated rule and
improves on reoptimised step-up by $0.279$ points; differences use
unrounded values. At the respective lower-endpoint states and worst-case offsets, reoptimised closed max-$Z$
has subgroup and full-population target powers of $78.949\%$ and
$75.163\%$, compared with marginal reference powers of $85.137\%$ and
$81.351\%$. Each target-specific shortfall is $6.188$ percentage points.

\begin{table}[!htbp]
\centering\normalsize\setstretch{1}
\setlength{\tabcolsep}{4pt}
\caption{QUEST-informed worst-case regrets in percentage points, for complement-effect and stratum-difference bounds in mL. Step-up and closed max-$Z$ cutoffs are reoptimised by row. Symmetric closed max-$Z$ uses equal cutoffs; DT retains its original cutoffs. Bold marks the smallest displayed regret.}
\label{tab:clinical-calibration}
\begin{tabular}{rrrrrr}
\toprule
\shortstack{Complement\\effect\\bound} & \shortstack{Stratum\\difference\\bound} & \shortstack{Equal-regret\\step-up} & \multicolumn{2}{c}{Closed max-$Z$} & DT \\
\cmidrule(lr){4-5}
& & & Reoptimised & Symmetric & \\
\midrule
10 & 40 & 6.105 & \textbf{5.983} & 6.561 & 6.798 \\
20 & 80 & 6.466 & \textbf{6.188} & 6.760 & 7.185 \\
40 & 80 & 6.686 & \textbf{6.312} & 6.760 & 7.185 \\
5 & 125 & 6.173 & \textbf{5.990} & 6.798 & 7.276 \\
60 & 80 & 6.764 & \textbf{6.350} & 6.760 & 7.185 \\
\bottomrule
\end{tabular}
\end{table}

\begin{figure}[!htbp]
\normalsize\setstretch{1}\centering
\accessiblefigure{Target powers approach one as effects increase. Difference panels show that closed max-Z reoptimisation lowers subgroup power and raises full-population power relative to the symmetric rule, with both differences tending to zero.}
{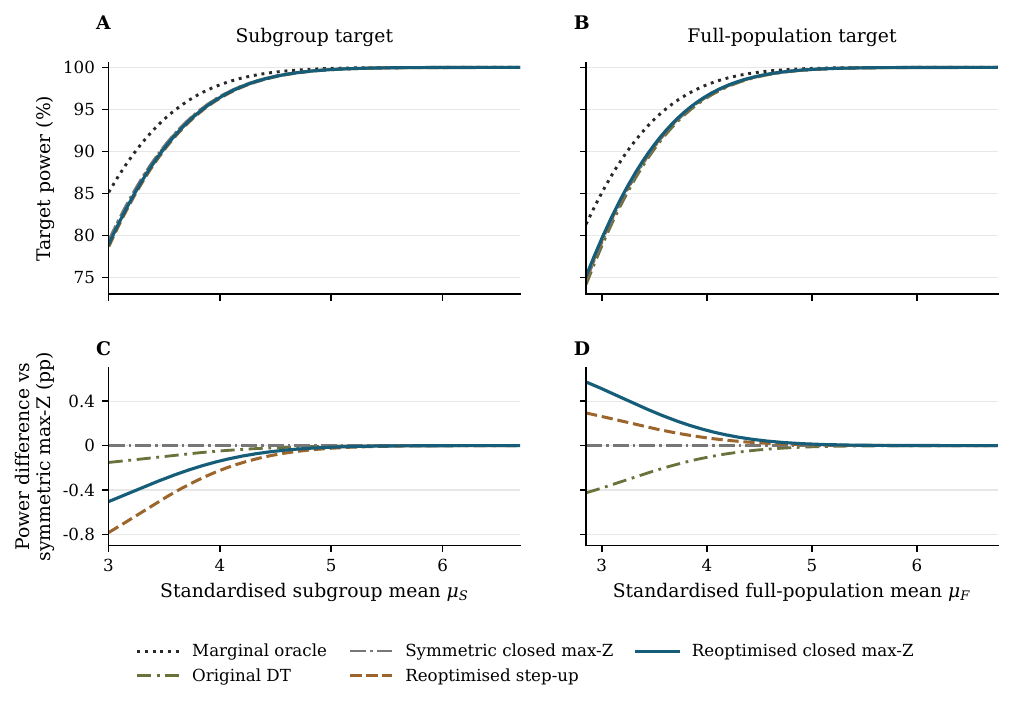}
\caption{Power allocation for QUEST bounds $(20,80)$ mL at $\alpha=0.025$. Top: target powers (\%); bottom: differences from symmetric closed max-$Z$ (percentage points), at worst offsets.}
\label{fig:power-tradeoff}
\end{figure}

Figure~\ref{fig:power-tradeoff} shows the allocation producing this balance.
At the lower endpoints, reoptimisation gains $0.573$ points of
full-population target power and gives up $0.503$ points of subgroup power
relative to symmetric closed max-$Z$. Reducing the larger shortfall
therefore involves accepting a loss of power for the other target.
Clinical calibration
and sensitivity analyses are given in Supplementary Sections~S11 and S13.

\FloatBarrier
\section{Discussion}\label{sec:discussion}

Simple step-up procedures can be minimax optimal even when non-target
effects are uncertain. Their optimality depends on the likelihood ratios
within the joint rejection quadrant and on the changes to the two arms
required by monotonicity. The integral criterion accounts for these
coupled changes and can establish optimality when the pointwise
sufficient condition fails. When a more flexible boundary improves the
step-up rule, the gain need not be large. Immediately beyond the symmetric
zero-offset transition at fixed nonsingular correlation, excess regret
has a cubic leading term.

On the reduced planning set, every least-favourable distribution has one
support point per target claim, despite the intervals of possible target
effects. These states determine the optimal compromise. Under the common-direction condition,
conditional Gaussian rearrangement makes the minimax rule nonrandomised
and unique up to $P_0$-null sets. The numerical examples show how effect uncertainty changes the preferred
boundary and how reoptimising familiar cutoffs can matter more than
changing rejection-region geometry.

The asthma example provides an interpretation of the target assignments
through treatment benefits and burdens. Their compatibility depends on
the assumed effect ranges and on whether positive net benefit is required
in each treated stratum. The minimax-regret objective protects designated
efficacy claims; a welfare-based formulation would directly incorporate
the consequences of treatment decisions.

The exact characterisation uses the two-hypothesis structure, in which
fixed marginal tests leave one intersection-size constraint and a joint
quadrant with two rejection arms. Conditional Gaussian rearrangement
extends to higher dimensions under a common-direction condition while
preserving one null rejection-probability constraint. Additional
intersection constraints arise with more hypotheses, and extending the
complete minimax characterisation to that setting remains open. The
analysis also assumes known covariance and a fixed experiment. Further
work could allow estimated covariance or adaptive recruitment, relax the
common-direction condition, and develop finite checks for the integral
criterion beyond the pointwise sufficient condition.

\section*{Data and code availability}
The studies use specified Gaussian distributions and published aggregate
QUEST summaries. The accompanying source-code archive (version 0.2.0)
contains inputs, saved results, analysis and figure scripts, and
reproduction instructions. A public repository record is in preparation.

\section*{Acknowledgements}
OpenAI Codex was used as an auxiliary tool for literature searches, mathematical exploration, coding and editorial assistance. The authors independently verified the results and take full responsibility for the manuscript. The authors received no funding for this work and declare no conflicts of interest.

\clearpage
\section*{Supplementary material}
\noindent This supplement gives the complete mathematical arguments and reproducible numerical calculations. It uses the notation of the main paper and states its assumptions where each extension is introduced.
\renewcommand{\theHsection}{S\arabic{section}}
\renewcommand{\theHequation}{S\arabic{equation}}
\renewcommand{\theHtable}{S\arabic{table}}
\renewcommand{\theHfigure}{S\arabic{figure}}
\renewcommand{\thetable}{S\arabic{table}}
\setcounter{table}{0}
\renewcommand{\thefigure}{S\arabic{figure}}
\setcounter{figure}{0}

\theoremstyle{plain}
\newtheorem{suppthm}{Theorem}
\renewcommand{\thesuppthm}{S\arabic{suppthm}}
\newtheorem{supplemma}{Lemma}
\renewcommand{\thesupplemma}{S\arabic{supplemma}}
\newtheorem{suppprop}{Proposition}
\renewcommand{\thesuppprop}{S\arabic{suppprop}}
\newtheorem{suppcor}{Corollary}
\renewcommand{\thesuppcor}{S\arabic{suppcor}}
\setcounter{equation}{0}
\renewcommand{\theequation}{S\arabic{equation}}
\setcounter{section}{0}
\renewcommand{\thesection}{S\arabic{section}}

\section{Experiment, decision class and validity}
\label{supp:validity}

Throughout this supplement,
\[
 (X,Y)^{\mathsf T}\sim N_2(\mu,\Sigma),\qquad
 \Sigma=\begin{pmatrix}1&\rho\\\rho&1\end{pmatrix},\qquad
 0\le\rho<1,
\]
with known covariance. Let $0<\alpha<1$, $c=\Phi^{-1}(1-\alpha)$ and
$s=(1-\rho^2)^{1/2}$. The standard normal density and distribution function
are $\varphi$ and $\Phi$. Write $P_{\mu}$ and $E_{\mu}$
for probability and expectation under mean $\mu$; $P_0$, $E_0$
and $f_0$ refer to the centred bivariate law. Define
\[
 M_S=1\{X>c\},\qquad M_F=1\{Y>c\},\qquad
 \mathcal G=\{x>c\}\cup\{y>c\},\qquad Q=(c,\infty)^2.
\]
A measurable test function is called upper if it is coordinatewise
nondecreasing. A set is upper if its indicator is coordinatewise
nondecreasing; a lower set is closed under coordinatewise decreases.
Our randomised monotone class with fixed marginal tests is
\[
 \mathcal U=\{\phi:0\le\phi\le1_{\mathcal G},\ \phi\text{ upper},\ E_0\phi\le\alpha\}.
\]
The subset of indicator-valued functions is $\mathcal U_{\rm det}$.
All identifications of functions and sets are modulo Gaussian null sets.
Every mean has a strictly positive density, so the null sets are common to
all distributions in the experiment.

For $\phi\in\mathcal U$, use one independent uniform variable $V$ to form
the induced decision pair $D^\phi=(D_S^\phi,D_F^\phi)$, where
$D_j^\phi=1\{V\le\phi(X,Y)\}M_j$, $j=S,F$.
This is a monotone closed-testing procedure: closed testing combines the fixed marginal
tests with one shared, coordinatewise nondecreasing intersection test.
Its conditional rejection probabilities are $\phi(X,Y)M_j$.
For indicator-valued $\phi$, $D_j^\phi=\phi M_j$. The
elementary null hypotheses are $\mu_S\le0$ and $\mu_F\le0$; strong
familywise error control is required over all $\mu\in\mathbb R^2$.

\begin{suppprop}[Validity and marginal-test restrictions for nonrandomised procedures]
\label{supp:gate}
Every member of $\mathcal U$ gives strong familywise error control at level
$\alpha$. Conversely, every nonrandomised pair of monotone elementary
decisions with strong familywise error control is dominated, componentwise, by a member of
$\mathcal U_{\rm det}$ followed by the marginal tests.
\end{suppprop}
\begin{proof}
For an upper $\phi$, the coupling $Z_{\mu}=Z_0+\mu$ gives
$E_{\mu}\phi\le E_0\phi$ whenever both coordinates of $\mu$ are nonpositive.
Under this intersection null, any final rejection requires the common
intersection rejection and hence has probability at most $\alpha$. If
only the $S$ null is true, an erroneous rejection requires $X>c$, whose
probability is at most $\alpha$. The $F$ case is identical.

Now consider any nonrandomised upper decisions $D_S,D_F$. If
$D_S(x_0,y_0)=1$ at a point with $x_0<c$, coordinatewise monotonicity forces rejection on
$\{x\ge x_0,y\ge y_0\}$. Along the partial-null sequence
$\mu_S=0$, $\mu_F\to\infty$, the probability of this rectangle tends to
$1-\Phi(x_0)>\alpha$: the difference from $P(X\ge x_0)$ is bounded by
$P(Y<y_0)\to0$. This contradicts strong validity. Thus $D_S\le M_S$,
apart from the null boundary line, and similarly $D_F\le M_F$.
The set $R=\{D_S=1\text{ or }D_F=1\}$ is upper, contained in $\mathcal G$, and
$P_0(R)\le\alpha$ by global-null control. The decisions $1_RM_j$ dominate
$D_j$ and are valid by the first part of the proof.
\end{proof}

The marginal-test reduction applies to nonrandomised decisions. For example,
rejecting with a constant probability $\alpha$ can be valid and monotone
while rejecting outside the marginal rejection regions. We optimise over the randomised
monotone closed-testing class $\mathcal U$, which retains those restrictions. See also
\citet{hkr2023} for the multiple-testing antecedent of the marginal-test argument.
The class $\mathcal D_{\rm det}$ in the main paper consists of the
nonrandomised monotone decision pairs in the converse. Every
$\phi\in\mathcal U_{\rm det}$ induces a pair in $\mathcal D_{\rm det}$.
The proposition supplies a dominating induced pair for every original pair.
The replacement gives at least as much target power and no larger regret.
Thus optimisation over $\phi\in\mathcal U_{\rm det}$ preserves the minimum
worst-case regret over $\mathcal D_{\rm det}$.

As in the main paper, let $I_j=[\underline{\mu}_j,\overline{\mu}_j]$,
$j\in\{S,F\}$, be nonempty compact intervals in $(0,\infty)$;
singleton intervals are allowed. The labelled planning states are
\[
\begin{aligned}
 \theta_S(\mu_S,\delta)&=\bigl(S,(\mu_S,\rho\mu_S+s\delta)\bigr),
 \quad \mu_S\in I_S,\ |\delta|\le e_S,\\
 \theta_F(\mu_F,\eta)&=\bigl(F,(\rho\mu_F+s\eta,\mu_F)\bigr),
 \quad \mu_F\in I_F,\ |\eta|\le e_F,
\end{aligned}
\]
where $e_S,e_F$ are finite and nonnegative. Their complete planning set is
\[
 \Theta=\{\theta_S(\mu_S,\delta):\mu_S\in I_S,\ |\delta|\le e_S\}\cup\{\theta_F(\mu_F,\eta):\mu_F\in I_F,\ |\eta|\le e_F\}.
\]
For $\theta=(j,\mu)\in\Theta$, define
\[
\begin{aligned}
 R(\theta;D)&=\Phi(\mu_j-c)-P_{\mu}(D_j=1),\\
 R(\theta;\phi)&:=R(\theta;D^\phi)
 =\Phi(\mu_j-c)-E_{\mu}[\phi M_j].
\end{aligned}
\]
Thus $R(\theta;\phi)$ is the regret of the induced pair for both
nonrandomised and randomised $\phi$. Write $R_S(\mu_S,\delta;\phi)=R(\theta_S(\mu_S,\delta);\phi)$
and $R_F(\mu_F,\eta;\phi)=R(\theta_F(\mu_F,\eta);\phi)$.
The interval parameters are
the target noncentrality parameters, since both marginal variances equal one.
The reference powers are
$\Phi(\mu_S-c)$ and $\Phi(\mu_F-c)$, respectively: they use the same experiment
and the designated marginal level-$\alpha$ test. By the marginal-test reduction,
these powers are upper bounds for the respective target powers in the
nonrandomised class considered here.

\begin{suppprop}[Worst-case offset reduction]
\label{supp:adverse}
For every monotone final decision, target regret is maximised over the
non-target mean offsets at $\delta=-e_S$ and $\eta=-e_F$.
\end{suppprop}
\begin{proof}
Increasing $\delta$ adds a nonnegative constant to $Y$ while leaving $X$,
the covariance and the subgroup reference power unchanged. A coupling using
the same centred Gaussian observation therefore increases the upper
$S$ decision pointwise. Its power cannot decrease and its regret cannot
increase. Interchange the coordinates for the full-population-target subfamily.
\end{proof}

Henceforth write $\theta_S^-(\mu_S)=\theta_S(\mu_S,-e_S)$ and
$\theta_F^-(\mu_F)=\theta_F(\mu_F,-e_F)$ for the states at the worst-case offsets.
Their regrets are
\[
\begin{aligned}
 R_S(\mu_S;\phi)&=E_{(\mu_S,\rho\mu_S-se_S)}[M_S(1-\phi)],\\
 R_F(\mu_F;\phi)&=E_{(\rho\mu_F-se_F,\mu_F)}[M_F(1-\phi)].
\end{aligned}
\]
These expressions also fix the benchmark marginal power in the randomised
monotone closed-testing class.

\section{Shape of the regret function and a least-favourable prior}
\label{supp:saddle}

The argument for power loss in the normal location family below is related to the
radial regret analysis of \citet{VanZwetOosterhoff1967} and
\citet{Oosterhoff1969}. We give the complete Stieltjes calculation
because a nonzero tail limit is permitted here.

\begin{supplemma}[Unimodality of the regret function]
\label{supp:shape}
For any $\phi\in\mathcal U$, extend each subfamily's regret formula to all real
noncentrality parameters. This extension either vanishes identically, is strictly
increasing, or increases strictly to one finite maximiser and then decreases
strictly. A nonzero regret function consequently has a unique maximum on
its compact noncentrality interval.
\end{supplemma}
\begin{proof}
For a subgroup-target state at the worst-case offset,
\[
 Y\mid X=x\ \overset{d}=\ \rho x-se_S+sW,\qquad W\sim N(0,1),
\]
independently of $\mu_S$. Put
\[
 g(x)=E[1-\phi(x,\rho x-se_S+sW)],\qquad x>c.
\]
The same $W$ can be used at two values of $x$. Since $\rho\ge0$ and
$\phi$ is upper, $g$ is bounded, nonnegative and nonincreasing. Taking a
right-continuous version, write $g_0=g(c+)$ and
$\ell=\lim_{x\to\infty}g(x)$. There is a finite positive Stieltjes
measure $\nu$ on $(c,\infty)$ satisfying
$g(x)=\ell+\nu((x,\infty))$ and $\nu((c,\infty))=g_0-\ell$.
Tonelli's theorem gives
\begin{equation}
 R_S(\mu_S)=\ell\Phi(\mu_S-c)+\int_{(c,\infty)}
       [\Phi(u-\mu_S)-\Phi(c-\mu_S)]\,\nu(du).
 \label{eq:supp:stieltjes}
\end{equation}
Boundedness of the normal density justifies differentiation under this
finite-measure integral. Dividing by $\varphi(c-\mu_S)>0$ yields
\begin{equation}
 \frac{R_S'(\mu_S)}{\varphi(c-\mu_S)}
 =g_0-\int_{(c,\infty)}
   \exp\{(u-c)\mu_S-(u^2-c^2)/2\}\,\nu(du)=K(\mu_S).
 \label{eq:supp:shape}
\end{equation}
Writing $t=u-c>0$, the exponent is $t(\mu_S-c)-t^2/2$, bounded above by
$(\mu_S-c)_+^2/2$. Thus the integral is finite and locally dominated in $\mu_S$,
so $K$ is continuous. If $\nu\ne0$, its integral increases strictly
with $\mu_S$, tends to zero as $\mu_S\to-\infty$ by dominated convergence, and
tends to infinity as $\mu_S\to\infty$ by monotone convergence. Consequently
$K$ decreases strictly from $g_0>0$ to $-\infty$. Moreover,
\[
 K(c)=\ell+\int_{(c,\infty)}
       [1-e^{-(u-c)^2/2}]\,\nu(du)>0.
\]
There is therefore one maximiser, above $c$. If $\nu=0$ and $g_0>0$,
regret is $g_0\Phi(\mu_S-c)$ and is strictly increasing. If $g_0=0$,
regret is zero. Interchanging the coordinates proves the result for the other subfamily.
The argument does not require $\ell=0$, integrability or eventual vanishing of $g$.
\end{proof}

The reduced, target-labelled planning set is
\[
 \Theta_{\rm red}
 =\{\theta_S^-(\mu_S):\mu_S\in I_S\}
 \cup\{\theta_F^-(\mu_F):\mu_F\in I_F\}.
\]
Let $\mathcal P(\Theta_{\rm red})$ denote its probability
measures. For $\Pi\in\mathcal P(\Theta_{\rm red})$ set
$\mathcal R(\phi,\Pi)=\int R(\theta;\phi)\,\Pi(d\theta)$.

\begin{suppthm}[A minimax rule and a least-favourable prior for all finite offset bounds]
\label{supp:randomsaddle}
The minimax problem over $\mathcal U$ has an attained, strictly positive
value $v_{\rm ran}$, and an attained saddle point with probability measures
on $\Theta_{\rm red}$. Every least-favourable prior has exactly one support point
on each labelled subfamily, with positive weights. At a saddle point,
\[
 \Pi_*=q_*\delta_{\theta_S^-(\mu_{S,*})}+(1-q_*)\delta_{\theta_F^-(\mu_{F,*})},\qquad 0<q_*<1,
\]
where $\mu_{S,*},\mu_{F,*}$ uniquely maximise their respective regret functions and
$R_S(\mu_{S,*};\phi_*)=R_F(\mu_{F,*};\phi_*)=v_{\rm ran}$.
\end{suppthm}
\begin{proof}
Regard $\mathcal U$ as a subset of $L^\infty(P_0)$ with its weak-star
topology. The unit ball is compact, and the range, support and size
constraints are weak-star closed. Monotonicity is closed as well: up
to a null-set version it is equivalent to nonnegative distributional
coordinate derivatives, or
\[
 \int_{\mathbb R^2}\phi(z)\partial_j\psi(z)\,dz\le0
 \quad(j=1,2)
\]
for every nonnegative smooth compactly supported $\psi$.
The test functional is weak-star continuous because
$(\partial_j\psi)/f_0\in L^1(P_0)$. Conversely, nonnegative
distributional coordinate derivatives yield a monotone version by
mollification and almost-everywhere passage to a subsequence. Thus
$\mathcal U$ is compact and convex.

For each alternative, the likelihood ratio restricted to its target's
marginal rejection region lies in $L^1(P_0)$, so
regret is affine and weak-star continuous in $\phi$. Gaussian location
densities vary continuously in total variation, uniformly on compact
mean sets; hence the dependence on the noncentrality parameter is continuous
uniformly in $0\le\phi\le1$. The space of priors is compact and convex
in its weak topology. Average regret is affine and separately
continuous in both variables. The compact-convex minimax theorem of
\citet{Sion1958} gives
\[
 \min_{\phi\in\mathcal U}\max_{\Pi\in\mathcal P(\Theta_{\rm red})}
  \mathcal R(\phi,\Pi)
 =\max_{\Pi\in\mathcal P(\Theta_{\rm red})}\min_{\phi\in\mathcal U}
  \mathcal R(\phi,\Pi),
\]
with both extrema attained. A minimax rule paired with a least-favourable
prior forms a saddle point.

All regrets are nonnegative. If the attained worst regret were zero,
zero regret at any one subgroup-target noncentrality parameter would force $\phi=1$ almost
everywhere on $\{x>c\}$, because its alternative density is positive.
A full-population-target noncentrality parameter similarly forces $\phi=1$ on $\{y>c\}$. This is
impossible since
\[
 P_0(\mathcal G)=2\alpha-P_0(Q)>\alpha.
\]
The strict inequality follows from the positive probability of each
strip where only one marginal test rejects. Therefore $v_{\rm ran}>0$.
A prior supported on only one subfamily has minimum prior-average regret
zero, attained by the corresponding marginal test. Every least-favourable
prior must assign positive probability to both labelled subfamilies.

At a saddle point, regret is at most $v_{\rm ran}$ at every parameter and
its prior average equals this value. The prior is supported on the
set of alternatives where regret equals the minimax value. Each subfamily
assigned positive prior probability has positive regret at such a point, and
Lemma~\ref{supp:shape} gives a unique maximising noncentrality parameter there.
This proves the support and equalisation assertions. No uniqueness
of $q_*$ is implied.
\end{proof}

\section{A convex boundary representation and elimination of randomisation}
\label{supp:purification}

For fixed target noncentrality parameters $\mu_S,\mu_F>0$, the likelihood ratios
at the worst-case offsets are
\begin{align*}
 L_S(x,y)&=\exp\{\mu_S x-\mu_S^2/2-e_S(y-\rho x)/s-e_S^2/2\},\\
 L_F(x,y)&=\exp\{\mu_F y-\mu_F^2/2-e_F(x-\rho y)/s-e_F^2/2\}.
\end{align*}
For $0<q<1$, put $H=qM_SL_S+(1-q)M_FL_F$.
This is a weighted likelihood-ratio statistic. The prior-weighted target
power is $E_0(\phi H)$; the reference-power terms do
not depend on the decision rule.

\begin{supplemma}[Common increasing direction]
\label{supp:direction}
Put
\[
 A_0=\mu_S+\rho e_S/s,\quad u=e_S/s,\qquad
 B_0=\mu_F+\rho e_F/s,\quad v=e_F/s,\qquad J=A_0B_0-uv.
\]
If $J>0$, there are $d_x,d_y>0$ along whose direction both likelihood
ratios strictly increase. The ratio $r=d_x/d_y$ may also be chosen in
$(\rho,1/\rho)$ when $\rho>0$. For $\rho=0$ this extra interval is
$(0,\infty)$. If
\begin{equation}
 J_0=(\underline{\mu}_S+\rho e_S/s)(\underline{\mu}_F+\rho e_F/s)-e_Se_F/s^2>0,
 \label{eq:supp:J0}
\end{equation}
the same direction works for all target noncentrality parameters in both intervals.
\end{supplemma}
\begin{proof}
The directional log derivatives are positive precisely when
$u/A_0<r<B_0/v$, with $B_0/v=\infty$ if $v=0$. This interval is nonempty
if and only if $J>0$. When $\rho>0$, the identities $A_0=\mu_S+\rho u$ and
$B_0=\mu_F+\rho v$ imply $u/A_0<1/\rho$ and $B_0/v>\rho$, with the same
endpoint conventions. The two open intervals therefore intersect.
Choosing the direction at the lower endpoints of the noncentrality intervals suffices
for all larger noncentrality parameters, since their directional derivatives
increase.
\end{proof}

For a labelled state $\theta=(j,\mu)$, let $L_\theta=dP_\mu/dP_0$.
Along $d=(d_x,d_y)$, the directional log-likelihood-ratio derivatives
at general offsets are
\begin{equation}\label{eq:supp:full-direction}
\begin{aligned}
 \partial_d\log L_{\theta_S(\mu_S,\delta)}
 &=\mu_Sd_x+\delta(d_y-\rho d_x)/s,\\
 \partial_d\log L_{\theta_F(\mu_F,\eta)}
 &=\mu_Fd_y+\eta(d_x-\rho d_y)/s.
\end{aligned}
\end{equation}
The ratio restriction makes both offset coefficients in
\eqref{eq:supp:full-direction} positive. Each derivative is therefore
smallest at the adverse offset and lower target noncentrality endpoint.
Positivity there extends to every state in $\Theta$, so the same
conditional rearrangement improves or maintains all target powers
throughout the full planning set.

The construction preserves a rule's conditional null rejection probability
on each line parallel to this direction and allocates that probability to an
upper tail, where the target likelihood ratios are largest. The remaining
issue is compatibility across lines: the constraints on the conditional
probabilities below ensure that the resulting rejection region remains
coordinatewise monotone and contained in $\mathcal G$.

Fix such a direction and set
\[
 T=\tfrac12(X/d_x+Y/d_y),\qquad
 Z=\tfrac12(X/d_x-Y/d_y).
\]
Then $X=d_x(T+Z)$ and $Y=d_y(T-Z)$. Under $P_0$,
\begin{align*}
 \tau^2&=\operatorname{Var}_0(Z)
 =\tfrac14(d_x^{-2}+d_y^{-2}-2\rho/(d_xd_y)),\\
 \beta&=(d_x^{-2}-d_y^{-2})/(4\tau^2),\qquad
 \sigma^2=s^2/(4d_x^2d_y^2\tau^2),
\end{align*}
and $Z\sim N(0,\tau^2)$, $T\mid Z=z\sim N(\beta z,\sigma^2)$.
The restriction on $r$ gives $|\beta|<1$: the inequalities
$\beta<1$ and $\beta>-1$ reduce to $r>\rho$ and $r<1/\rho$.

Every nonempty proper upper set has a representation $T>h(Z)$ with
$h$ finite and globally 1-Lipschitz. Indeed, any line along the chosen direction translated
sufficiently far in the positive direction lies above a fixed point
in the upper set, and translated sufficiently far in the negative
direction lies below a point outside it. Its threshold is finite.
Comparing lines indexed by $z,z'$ after translating $T$ by $|z-z'|$ gives
$|h(z)-h(z')|\le|z-z'|$. The same comparisons prove the converse.
The union of marginal rejection regions has the boundary
\[
 h_{\mathcal G}(z)=\min(c/d_x-z,c/d_y+z).
\]

Parameterise a boundary by its conditional null rejection probability
\[
 p(z)=\overline\Phi\{(h(z)-\beta z)/\sigma\},\qquad
 p_{\mathcal G}(z)=\overline\Phi\{(h_{\mathcal G}(z)-\beta z)/\sigma\}.
\]
Let $I(p)=\varphi(\Phi^{-1}(p))$ for $0<p<1$, extended by zero at
the endpoints, and put $\gamma_-=(1-\beta)/\sigma$,
$\gamma_+=(1+\beta)/\sigma$. The boundary constraints become
\begin{equation}
 -\gamma_-I(p)\le p'\le\gamma_+I(p),\qquad 0\le p\le p_{\mathcal G}.
 \label{eq:supp:boundary}
\end{equation}
The derivative bounds are interpreted in the distributional sense.
Equivalently, $\mathcal K_{\mathcal G}$ consists of locally absolutely
continuous functions $p:\mathbb R\to[0,1]$ with
$0\le p\le p_{\mathcal G}$, whose derivative bounds in
\eqref{eq:supp:boundary} hold almost everywhere.
The identically-zero function is included.

\begin{supplemma}[Convexity and exact representation]
\label{supp:convex}
The class $\mathcal K_{\mathcal G}$ is convex and consists exactly of the
conditional null rejection probabilities of upper subsets
of $\mathcal G$, including the empty set.
\end{supplemma}
\begin{proof}
The identities $I'(p)=-\Phi^{-1}(p)$ and $I''(p)=-1/I(p)$ show that
$I$ is concave. Thus both differential inequalities in
\eqref{eq:supp:boundary} are preserved under convex mixtures.
Alternatively, let $G_t(p)=\Phi(\Phi^{-1}(p)-t)$, extended
continuously at $0,1$. For $t\ge0$,
\[
 G_t'(p)=\exp\{t\Phi^{-1}(p)-t^2/2\},
\]
so $G_t$ is convex. For $z_2-z_1=d\ge0$, the exact finite-difference
form of the boundary constraint is
\begin{equation}
 p(z_2)\ge G_{\gamma_-d}(p(z_1)),\qquad
 p(z_1)\ge G_{\gamma_+d}(p(z_2)).
 \label{eq:supp:finite}
\end{equation}
These are closed convex inequalities. A finite Lipschitz boundary gives
them by writing $h(z)=\beta z-\sigma\Phi^{-1}(p(z))$, and
differentiation gives the differential form.

Conversely, the distributional derivative bounds imply local absolute
continuity and a uniform Lipschitz bound, since $I\le1/\sqrt{2\pi}$.
On every interval where $0<p<1$, the chain rule for
$u=\Phi^{-1}(p)$ gives $-\gamma_-\le u'\le\gamma_+$.
Thus $u$ cannot diverge at a finite endpoint of such an interval.
By continuity, either $p$ is identically zero or one, or $0<p<1$
everywhere; $p\le p_{\mathcal G}<1$ excludes the identically-one case.
In the interior case, integrating the bounds on $u$ gives
\eqref{eq:supp:finite} and reconstructs a finite 1-Lipschitz $h$.
The identically-zero case represents the empty set. Finally,
$p\le p_{\mathcal G}$ is exactly $h\ge h_{\mathcal G}$.
\end{proof}

The Gaussian tail maps in \eqref{eq:supp:finite} also arise in
constrained testing \citep{DongRothSu2022,AwanVadhan2023}.

Let $w_Z(z)=\varphi(z/\tau)/\tau$,
$\widetilde H(t,z)=H(d_x(t+z),d_y(t-z))$, and define, for
$0\le p\le p_{\mathcal G}(z)$,
\begin{equation}
 F(z,p)=\int_{\overline\Phi^{-1}(p)}^\infty
       \widetilde H(\beta z+\sigma u,z)\varphi(u)\,du,
 \qquad F(z,0)=0.
 \label{eq:supp:F}
\end{equation}
For $0<q<1$, both mixture weights are positive. Under the common-direction
condition, each likelihood ratio increases strictly along $T$, while its
marginal rejection indicator is nondecreasing. At every point with
$t>h_{\mathcal G}(z)$, at least one indicator is active. Its positively
weighted contribution increases strictly, and the other contribution is
nondecreasing. Consequently, $\widetilde H(t,z)$ increases strictly
throughout this portion of the conditional line, including where only one
marginal indicator is active. Increasing $p$ lowers the conditional cutoff
and therefore gives strictly decreasing marginal weighted-power gains.
Thus $F(z,\cdot)$ is continuous and strictly concave on
$[0,p_{\mathcal G}(z)]$. Away from values
corresponding to marginal rejection boundaries, its derivative is the test statistic on the boundary
\[
 F_p(z,p)=\widetilde H(\beta z+\sigma\overline\Phi^{-1}(p),z).
\]
At a marginal rejection boundary this formula is replaced by the corresponding one-sided
derivatives.

\begin{suppthm}[Simultaneous elimination of randomisation]
\label{supp:purify}
If one direction satisfies Lemma~\ref{supp:direction} for all
alternatives being considered, every $\phi\in\mathcal U$ is dominated
in all target powers by a nonrandomised monotone rule of the same null
size. In particular, \eqref{eq:supp:J0} suffices for simultaneous
elimination of randomisation throughout both noncentrality intervals.
\end{suppthm}
\begin{proof}
The construction is a conditional Gaussian tail rearrangement,
in the symmetrisation tradition of \citet{Ehrhard1983}. We verify
both the order constraint and the target-power comparison directly.
Use the layer representation
$\phi=\int_0^1 1\{\phi>u\}\,du$. Each level set is upper and
contained in $\mathcal G$, so its conditional null probabilities are in
$\mathcal K_{\mathcal G}$. Convexity, or Jensen's inequality applied to
\eqref{eq:supp:finite}, shows that
$p_\phi(z)=E_0(\phi\mid Z=z)$ also belongs to $\mathcal K_{\mathcal G}$.
Recover its nonrandomised upper tail in $T$ by
Lemma~\ref{supp:convex}. This preserves the conditional null rejection probability.

For a target alternative let $S(t,z)$ be its likelihood ratio multiplied
by its marginal rejection indicator, nondecreasing in $t$. If $\psi$ is any
conditional rejection function with rejection probability
$p$ and $1\{t>h\}$ the upper tail with that probability, then
\[
 \int [1\{t>h\}-\psi(t)]\,[S(t,z)-S(h,z)]\,P_0(dt\mid z)\ge0.
\]
Each integrand has nonnegative sign on either side of $h$; the
constant term vanishes because the rejection probabilities are equal. Points
on marginal rejection boundaries have zero conditional probability, so one-sided values can
be used when required. Apply this inequality to $\phi$ on each
conditional line and integrate over $z$. The same upper-tail rearrangement
works for every alternative sharing the specified increasing direction.
\end{proof}

\begin{suppthm}[Bayes optimality and a unique nonrandomised minimax rule]
\label{supp:deterministic}
For $0<q<1$ and $J>0$, the fixed-prior Bayes rule is unique modulo
null sets, nonrandomised, and has size exactly $\alpha$. It solves
\begin{equation}
 \max_{p\in\mathcal K_{\mathcal G}}\int w_Z(z)F(z,p(z))\,dz,
 \qquad \int w_Z(z)p(z)\,dz=\alpha.
 \label{eq:supp:program}
\end{equation}
A finite multiplier $k>0$ makes this same $p$ a global maximiser of
$\int w_Z(F-kp)$. Under \eqref{eq:supp:J0}, the nonrandomised and
randomised minimax values coincide, and the minimax monotone closed-testing
rule is unique and nonrandomised. Its least-favourable prior has
the two-point form in Theorem~\ref{supp:randomsaddle}.
\end{suppthm}
\begin{proof}
The feasible functions are uniformly bounded and Lipschitz.
Arzel\`a--Ascoli on compact intervals followed by a diagonal argument
gives local uniform subsequential convergence. Testing
\eqref{eq:supp:boundary} against nonnegative compactly supported
smooth functions proves that the inequalities are closed under
this convergence. The upper-bound constraint is closed as well. Size is
continuous by bounded convergence against $w_Z$. Power is continuous
because $F$ is pointwise continuous in $p$ and bounded above by the
integrable function $E_0(H\mid Z=z)$.

The exact-size class is nonempty: $h=h_{\mathcal G}+\delta$ remains 1-Lipschitz
and its size decreases continuously from $P_0(\mathcal G)>\alpha$ to zero as
$\delta$ increases from zero to infinity. Compactness and continuity
give existence. Strict concavity of $F$ gives uniqueness of $p$
up to $w_Z$-null sets. The theorem on elimination of randomisation equates the
nonrandomised and randomised optima. Moreover, its conditional
comparison for $H$ is strict if a rejection function inside $\mathcal G$ differs from
the upper-tail indicator on a set of positive conditional probability, because $H$ is strictly
increasing on that conditional line within $\mathcal G$. Every randomised Bayes
optimiser must therefore agree almost everywhere with the unique
nonrandomised rule.

If a candidate has size smaller than $\alpha$, mix its conditional
probabilities toward $p_{\mathcal G}$ until the size constraint is attained. Convexity
preserves feasibility, and $H>0$ in $\mathcal G$ makes power strictly increase.
Hence every optimum has size exactly equal to the allowed level. For $0<A<P_0(\mathcal G)$, let
$V(A)$ be the optimal power at exact size $A$. Convexity of
$\mathcal K_{\mathcal G}$ and concavity of power make $V$ finite and concave;
the same enlargement makes it strictly increasing. At the interior
point $A=\alpha$, a finite supporting slope $k$ exists.
Strict increase with the allowed size forces $k>0$. The inequality
$V(A)\le V(\alpha)+k(A-\alpha)$ proves global penalised optimality.

Under \eqref{eq:supp:J0}, eliminating randomisation from the attained randomised
minimax rule yields a nonrandomised rule with the same minimax value.
Choose any least-favourable prior supplied by
Theorem~\ref{supp:randomsaddle}. Every minimax rule is Bayes for
that prior. Both weights are positive, and its fixed-pair determinant
is positive by \eqref{eq:supp:J0}; fixed-prior uniqueness therefore
makes every minimax rule agree with one nonrandomised rule. Equality
in the size constraint and the support assertions follow as above.
\end{proof}

The determinant condition is sufficient for a nonrandomised optimum.
When it fails, Theorem~\ref{supp:randomsaddle} still supplies a randomised
saddle point; equality of the nonrandomised and randomised values requires
further analysis.

\begin{suppprop}[Complete variational condition]
\label{supp:variational}
A feasible $p_*$ in \eqref{eq:supp:program} is optimal if and only if,
for every feasible $p$ of the same size,
\begin{equation}
 \int w_Z(z)F'_+(z,p_*(z);p(z)-p_*(z))\,dz\le0,
 \label{eq:supp:variational}
\end{equation}
where the derivative is the right derivative at $t=0$ of
$F(z,p_*(z)+t\{p(z)-p_*(z)\})$, using
the appropriate one-sided slope at values corresponding to marginal rejection boundaries. An infinite positive
directional derivative violates this condition.
\end{suppprop}
\begin{proof}
For fixed $p,p_*$, secant slopes along
$p_*+t(p-p_*)$ increase as $t\downarrow0$ by concavity. They are
bounded below by the integrable full-segment difference
$F(z,p)-F(z,p_*)$. Monotone convergence after subtracting this lower
bound interchanges integration and the one-sided derivative.
At an optimum every feasible short segment has nonpositive
derivative, proving necessity. Conversely, a concave function lies
below its first-order directional bound, so
\eqref{eq:supp:variational} bounds every feasible power difference
by zero.
\end{proof}

For the penalised formulation, the same condition holds without a
size restriction after subtracting $k(p-p_*)$ inside the integral.
On an interval where both slope constraints and the upper-bound constraint
are inactive, arbitrary compactly supported perturbations give the boundary equation
$H=k$ on the boundary. Attainment of the upper derivative bound
$p'=\gamma_+I(p)$ gives $h'=-1$ and a vertical boundary segment in
$(X,Y)$; attainment of the lower derivative bound gives $h'=1$ and a horizontal
segment. Optimality also requires the global condition
\eqref{eq:supp:variational}.

\section{Rearrangement in the marginal-test strips and step-up optimality}
\label{supp:rectangles}

The conditional averaging step below is an instance of classical
optimisation under a monotonicity constraint; see
\citet{O01,O02}. Its role here is to preserve the quadrant
decisions while rearranging rejection decisions in the two strips where
only one marginal test rejects.

Write $S=(c,\infty)\times(-\infty,c]$ and
$F=(-\infty,c]\times(c,\infty)$. For fixed target noncentrality parameters
at the worst-case offsets, define
\begin{align*}
 z(x)&=(c-\rho x)/s,\qquad w(x)=\varphi(x)\Phi(z(x)),\\
 C_S(x)&=e^{\mu_S x-\mu_S^2/2}\frac{\Phi(z(x)+e_S)}{\Phi(z(x))},\qquad
 C_F(y)=e^{\mu_F y-\mu_F^2/2}\frac{\Phi(z(y)+e_F)}{\Phi(z(y))}.
\end{align*}
These are conditional null averages of the corresponding likelihood
ratios within the two strips. For example, conditioning on
$X=x,Y\le c$ reduces the average to the identity
\[
 E[e^{-e_SW-e_S^2/2}\mid W\le z]
   =\Phi(z+e_S)/\Phi(z),\qquad W\sim N(0,1).
\]
Both averaged likelihood ratios are strictly increasing. To verify this, the
lower-tail inverse Mills ratio (reverse hazard) $\lambda(z)=\varphi(z)/\Phi(z)$ satisfies
$\lambda'(z)=-\lambda(z)(z+\lambda(z))<0$. For $z<0$ the last
inequality uses $\lambda(z)>-z$, obtained by integrating
$(-u)\varphi(u)$ over $(-\infty,z)$; for $z\ge0$ it is immediate.
Consequently,
\[
 (\log C_S)'(x)=\mu_S+\frac\rho s
  [\lambda(z(x))-\lambda(z(x)+e_S)]\ge \mu_S>0,
\]
and similarly for $C_F$.

\begin{supplemma}[Simultaneous improvement of rejection regions within the two strips]
\label{supp:arms}
For a nonrandomised monotone rule supported on $\mathcal G$, keep its rejection set
in $Q$ fixed. Its rejection decisions in $S$ and $F$ can be replaced by the threshold rules $x>u$
and $y>v$ within those strips, respectively. The replacement preserves null probability
in each strip, remains upper, and increases or preserves every
target power throughout the two subfamilies at the worst-case offsets. This holds for all
finite nonnegative offset bounds, without a determinant condition.
\end{supplemma}
\begin{proof}
Let $L$ be the missed set in $Q$. It is a relative lower set: any
coordinatewise decrease from a point of $L$ that remains in $Q$ also lies
in $L$. Let $X_L$ and $Y_L$ be the suprema of its coordinate projections; for
the empty set put $X_L=Y_L=c$. Infinite suprema are allowed.
If $x<X_L$, rejection at $(x,y)$ with $y\le c$ would force
rejection of a point of $L$ with larger first coordinate. Thus no
$S$ rejection is possible below $X_L$, apart from irrelevant
boundary lines.

Set $p(x)=E_0(\phi\mid X=x,Y\le c)$. Under this conditional distribution,
$\phi$ increases with $Y$ whereas $L_S$ decreases. Their covariance
is nonpositive: with independent copies $Y_1,Y_2$, expand
\[
 E[(\phi(Y_1)-\phi(Y_2))(L_S(Y_1)-L_S(Y_2))]\le0.
\]
It follows that
\[
 E_0(\phi L_S1_S)\le\int_c^\infty p(x)C_S(x)w(x)\,dx.
\]
Choose $u\ge X_L$, possibly $u=\infty$, so that
$\int_u^\infty w=\int_c^\infty pw$. This is possible since
$p=0$ below $X_L$. If this null rejection probability is zero, the replacement is
empty and the power comparison is immediate. Otherwise $u<\infty$.
Subtracting $C_S(u)$ and comparing signs on
either side of $u$ gives
\[
 \int_c^\infty pC_Sw\le\int_u^\infty C_Sw.
\]
The threshold rule within the strip attains equality in the conditional
covariance bound. Its cutoff uses only the null rejection probability in the strip and thus is
independent of the noncentrality parameter; the ordering of $C_S$ holds at every
$\mu_S>0$, proving simultaneous improvement. It does not affect
full-population target power, because $M_F=0$ in $S$. Since $u\ge X_L$, all
quadrant points above the new strip rejection region already belong to the fixed
rejection set. The replacement is upper. Repeat the argument on
$F$.
\end{proof}

Consider a step-up procedure with finite cutoffs and intersection rejection region
\[
 R=\{x>x_0\}\cup\{y>y_0\}\cup Q,\qquad x_0,y_0>c.
\]
Put $K=P_0(S)>0$ and $L_0(t)=\int_c^t w(x)\,dx$.
Its null size is $\alpha+K-L_0(x_0)-L_0(y_0)$, so calibration is
$L_0(x_0)+L_0(y_0)=K$. Its regrets at the worst-case offsets are
\begin{align}
 r_S(\mu_S;x_0)&=\int_c^{x_0}\Phi(z(x)+e_S)\varphi(x-\mu_S)\,dx,
 \nonumber\\
 r_F(\mu_F;y_0)&=\int_c^{y_0}\Phi(z(y)+e_F)\varphi(y-\mu_F)\,dy.
 \label{eq:supp:rectrisk}
\end{align}
The formula follows because the subgroup target is missed exactly when
$c<X\le x_0,Y\le c$, and similarly for the full-population target.

\begin{suppprop}[Unique equal-regret step-up procedure]
\label{supp:equaliser}
There is a unique minimax step-up procedure among calibrated step-up procedures.
It has finite cutoffs above $c$ and equalises the two interval
maximum regrets. Each maximum is attained at a unique target noncentrality parameter.
\end{suppprop}
\begin{proof}
Parameterise calibration by $L_0(x_0)=\lambda K$,
$L_0(y_0)=(1-\lambda)K$, $0\le\lambda\le1$, allowing the
limiting marginal-test rules at the endpoints. Increasing $\lambda$
raises $x_0$ and lowers $y_0$. The positive kernels in
\eqref{eq:supp:rectrisk} make every subgroup-target regret strictly increase
and every full-population-target regret strictly decrease along this allocation.
The interval maxima inherit these strict inequalities, by
evaluating at a maximiser for the smaller risk. They are continuous,
including the endpoints, by dominated convergence uniformly on
the compact noncentrality intervals. At one endpoint the subgroup-target maximum
is zero and the full-population-target maximum positive; at the other the reverse
holds. Their difference crosses zero once, and the larger maximum
decreases before the crossing and increases afterwards.
Uniqueness of each maximising noncentrality parameter follows from Lemma~\ref{supp:shape}.
\end{proof}

For fixed $\mu_S,\mu_F,q$, suppose the conditional likelihood-ratio averages at the
step-up procedure's cutoffs satisfy
\begin{equation}
 qC_S(x_0)=(1-q)C_F(y_0)=k>0.
 \label{eq:supp:fit}
\end{equation}
For a measurable relative lower set $L\subset Q$ put
\begin{align}
 \Psi_R(L)={}&\int_L(H_Q-k)f_0\,dx\,dy\nonumber\\
 &+\int_{x_0}^{\max(x_0,X_L)}[qC_S(x)-k]w(x)\,dx\nonumber\\
 &+\int_{y_0}^{\max(y_0,Y_L)}[(1-q)C_F(y)-k]w(y)\,dy,
 \label{eq:supp:Psi}
\end{align}
where $H_Q=qL_S+(1-q)L_F$ is the continuous extension of the test statistic from
the open quadrant. The two strip terms are nonnegative. Infinite
endpoints cause no integrability problem: all terms are bounded
by integrals of likelihood ratios and constants against $P_0$.

\begin{suppthm}[Integral condition for step-up optimality]
\label{supp:integrated}
Under \eqref{eq:supp:fit}, $R$ maximises the penalised power
$E_0[\phi(H-k)]$ over nonrandomised upper rules if and only if
$\Psi_R(L)\ge0$ for every measurable relative lower set $L\subset Q$.
The same equivalence holds over randomised upper rules.

For a calibrated step-up procedure, let $\mu_{S,R},\mu_{F,R}$ be the target noncentrality
parameters maximising its regrets over their respective intervals and set
\begin{equation}
 q=\frac{C_F(y_0)}{C_S(x_0)+C_F(y_0)},\qquad
 k=\frac{C_S(x_0)C_F(y_0)}{C_S(x_0)+C_F(y_0)}.
 \label{eq:supp:qk}
\end{equation}
The step-up procedure is minimax in $\mathcal U$ if and only if its two
interval maximum regrets are equal and the integrated criterion
holds at \eqref{eq:supp:qk}. Under \eqref{eq:supp:J0}, this is also
equivalent to minimax optimality in the nonrandomised upper class.
Without that condition the criterion remains sufficient for
minimax optimality in the nonrandomised class.
\end{suppthm}
\begin{proof}
First intersect a nonrandomised competitor with $\mathcal G$; this preserves
power, preserves coordinatewise monotonicity and removes only the negative penalty
outside $\mathcal G$. Fix its quadrant missed set $L$. By
Lemma~\ref{supp:arms}, the competitor can be replaced by one with
threshold rejection regions within the strips and no smaller penalised power, since
the null rejection probabilities in the strips are preserved. Their cutoffs satisfy $u\ge X_L$
and $v\ge Y_L$. The $S$ contribution is
$\int_u^\infty[qC_S(x)-k]w(x)\,dx$. Strict increase of $C_S$ and
\eqref{eq:supp:fit} show that the maximising allowed cutoff is
$u=\max(x_0,X_L)$. Similarly $v=\max(y_0,Y_L)$. This best
competitor rejects $Q\setminus L$ and the two resulting arms;
its objective is the objective of $R$ minus exactly $\Psi_R(L)$.
It is upper and attains that value. Thus nonnegativity of every
$\Psi_R(L)$ is both necessary and sufficient. Layer decomposition
of a randomised monotone test function into indicators of upper sets proves
the same penalised equivalence by Fubini's theorem.

For sufficiency in the minimax problem, calibration and $k>0$
turn penalised Bayes optimality into size-constrained Bayes
optimality. Equalisation and the two noncentrality parameters maximising regret make
the prior-average regret equal the step-up procedure's maximum over the full noncentrality intervals.
Every competitor's maximum regret is at least its prior average,
which is no smaller than the step-up procedure's Bayes average.

For necessity in $\mathcal U$, a minimax step-up procedure must be the
unique calibrated equal-regret step-up procedure by
Proposition~\ref{supp:equaliser}. Pair it with a least-favourable
prior from Theorem~\ref{supp:randomsaddle}. Its two support points
are the step-up procedure's own $\mu_{S,R},\mu_{F,R}$, and both weights are positive.
For this prior, the optimal power over randomised upper rules of
size at most $A$ is finite, nondecreasing and concave in $A$.
At the interior size $\alpha$ a finite nonnegative supporting
slope $k$ therefore exists, and gives global penalised optimality
of the step-up procedure. Varying its finite $x_0$ alone, without a size
restriction in this penalised problem, gives
\[
 [k-qC_S(x_0)]w(x_0)=0.
\]
The $y_0$ derivative similarly gives $k=(1-q)C_F(y_0)$.
Positivity forces $k>0$ and the two identities force
\eqref{eq:supp:qk}. The penalised equivalence already proved now
gives every integrated inequality. Under \eqref{eq:supp:J0},
Theorem~\ref{supp:deterministic} equates the nonrandomised and
randomised values, giving the nonrandomised necessity assertion.
\end{proof}

This criterion is an infinite family of inequalities. The simpler
condition $H_Q\ge k$ throughout $Q$ is sufficient because it makes every term of
\eqref{eq:supp:Psi} nonnegative. The corner inequality
$H_Q(c,c)\ge k$ is necessary: otherwise a sufficiently small
corner rectangle has a negative quadrant integral and no arm
cost. Away from the corner, removal of a point with a low statistic value also
forces removal of all points below it, which is why pointwise
positivity need not be necessary.

\section{The unperturbed case and a cubic expansion of excess regret}
\label{supp:nominal}

\begin{suppcor}[Scalar criterion in the unperturbed case]
\label{supp:nominalcriterion}
When $e_S=e_F=0$, a finite calibrated step-up procedure is minimax if and
only if its two interval maximum regrets are equal and
\begin{equation}
 e^{-\mu_{S,R} (x_0-c)}+e^{-\mu_{F,R} (y_0-c)}\ge1.
 \label{eq:supp:nominalfit}
\end{equation}
\end{suppcor}
\begin{proof}
Here $J_0=\underline{\mu}_S\underline{\mu}_F>0$, $C_S(x)=e^{\mu_S x-\mu_S^2/2}$ and similarly for
$C_F$. At the prior specified by the cutoff equations,
\[
 H_Q/k=e^{\mu_{S,R} (x-x_0)}+e^{\mu_{F,R} (y-y_0)}.
\]
Its minimum on the closed quadrant is the left side of
\eqref{eq:supp:nominalfit}. If that value is at least one, the
pointwise sufficient criterion applies. If it is smaller than
one, continuity gives a sufficiently small corner rectangle with
negative $\Psi_R$, since its projections remain below the two
arm cutoffs. Theorem~\ref{supp:integrated} proves both directions.
\end{proof}

Let $t>c$ be the calibrated symmetric step-up cutoff, and let
$\mu_0$ be the global maximiser of its nominal regret. For identical
noncentrality intervals $I$, Corollary~\ref{supp:nominalcriterion}
reduces to
$\operatorname{clip}(\mu_0,I)(t-c)\le\log2$.
Here clipping replaces a value outside the interval by its nearest
endpoint. The following supporting fact removes the unknown maximiser from this
classification at conventional significance levels.

\begin{supplemma}[Bound on the symmetric regret maximiser]
\label{supp:modebound}
If $0<\alpha\le1/2$, then $\mu_0 (t-c)<\log2$.
Consequently, for identical intervals $[\ell,u]$, the symmetric
step-up procedure is minimax if and only if $\ell\le\log2/(t-c)$.
\end{supplemma}
\begin{proof}
The regret is $r(\mu )=\int_c^t w_0(x)\varphi(x-\mu )\,dx$, with
$w_0(x)=\Phi((c-\rho x)/s)$ nonincreasing. At $m=(c+t)/2$,
pairing points equidistant from $m$ gives
\[
 r'(m)=\int_0^{(t-c)/2}u\varphi(u)
       [w_0(m+u)-w_0(m-u)]\,du\le0.
\]
The unimodality lemma gives $\mu_0\le m$.

Put $t_B=\Phi^{-1}(1-\alpha/2)$ and $d_B=t_B-c$.
The null rejection probability using $t_B$ as the upper cutoff is
$S(\rho)=1+\Phi_2(c,c;\rho)-2\Phi_2(t_B,c;\rho)$.
It equals $\alpha$ at $\rho=0$ and tends to $\alpha$ as
$\rho\uparrow1$. Differentiating the bivariate density and
integrating gives
$S'(\rho)=f_\rho(c,c)-2f_\rho(t_B,c)$. The ratio of these two
densities is $e^{J_B(\rho)}$, where
\[
 J_B(\rho)=\frac{d_B^2}{2(1-\rho^2)}+\frac{cd_B}{1+\rho},\qquad
 J_B'(\rho)=\frac{d_B}{(1+\rho)^2}
             \left\{\frac{d_B\rho}{(1-\rho)^2}-c\right\}.
\]
Since $c\ge0$, the normal upper-tail hazard is strictly larger
than its argument on $[c,t_B]$, and therefore
$J_B(0)=(t_B^2-c^2)/2<\log2$. The derivative formula shows
that $J_B$ first decreases and then increases, or increases
immediately if $c=0$, and crosses $\log2$ exactly once before
diverging. Thus $S'$ first is negative and then positive, and its
equal endpoint values imply $S(\rho)\le\alpha$. Strictly
decreasing null size in the upper cutoff yields $t\le t_B$.
Consequently
\[
 \mu_0 (t-c)\le(t^2-c^2)/2\le(t_B^2-c^2)/2<\log2.
\]
Now $\mu_0<\log2/(t-c)$, so clipping $\mu_0$ to $[\ell,u]$
satisfies the scalar criterion exactly when the lower endpoint
does. The density differentiation used above follows directly
from $\partial_\rho f_\rho=\partial_x\partial_yf_\rho$;
Gaussian tails justify integration and remove boundary terms.
\end{proof}

\begin{suppthm}[Cubic expansion of excess regret in the unperturbed case]
\label{supp:cubic}
Fix $\alpha,\rho$ and the symmetric cutoff $t$. Put
$D_t=t-c$, $h_0=\log2/D_t$, and consider a common fixed target
noncentrality parameter $h>h_0$. Define
\[
 d(h)=1-2e^{-hD_t},\quad r(h)=e^{-hD_t},\quad
 k_0(h)=\tfrac12 e^{ht-h^2/2},\quad f_{cc}=f_0(c,c).
\]
The actual excess minimax regret of the symmetric step-up procedure is
\begin{align}
 \Delta(h)
 &=\frac{k_0(h)f_{cc}}{6h^2r(h)^2}\,d(h)^3+O(d(h)^4)
 \nonumber\\
 &=\frac{2k_0(h_0)f_{cc}}{3h_0^2}\,d(h)^3+O(d(h)^4),
 \qquad h\downarrow h_0.
 \label{eq:supp:cubic}
\end{align}
The constants need not be uniform as $\rho\uparrow1$.
\end{suppthm}
\begin{proof}
At the equal prior, the nominal weighted likelihood-ratio statistic is
\[
 H_h=\tfrac12 M_Se^{hX-h^2/2}+\tfrac12 M_Fe^{hY-h^2/2}.
\]
Thresholding at its calibrated level gives a unique monotone Bayes rule: the statistic
is coordinatewise increasing, its positive levels have zero
Gaussian probability, and $P_0(\mathcal G)>\alpha$. The rule is symmetric,
so its two regrets are equal and the saddle-point inequality makes
it minimax. Hence $\Delta$ is its weighted-power improvement
over the symmetric step-up procedure.

At threshold $k_0(h)$ the threshold rule and step-up procedure differ only on
$W_h=\{x>c,y>c,H_h<k_0(h)\}$. Let
$\mathcal B(h)=E_0[(k_0(h)-H_h)1_{W_h}]$. In corner coordinates
$u=x-c,v=y-c$, the normalised shortfall below the threshold is
\[
 d-r(e^{hu}-1)-r(e^{hv}-1).
\]
Since $e^z\ge1+z$, $W_h$ lies in $u,v\ge0$, $hr(u+v)<d$.
Rescale $u=d\xi/(hr)$, $v=d\eta/(hr)$. On the fixed unit
triangle the normalised deficit divided by $d$ equals
$1-\xi-\eta+O(d)$ uniformly. Taking positive parts extends this
uniform expansion across the curved wedge boundary. The density
is $f_{cc}+O(d)$, uniformly there, and the Jacobian is
$d^2/(h^2r^2)$. Since the integral of $1-\xi-\eta$ over the unit
triangle is $1/6$,
\[
 \mathcal B(h)=\frac{k_0(h)f_{cc}}{6h^2r(h)^2}d^3+O(d^4).
\]
The wedge null probability $m(h)$ is $O(d^2)$ by its area bound.

Let $k_*(h)$ be the calibrated statistic threshold. At $k_0(h)$ its
size is $\alpha-m(h)$, so $k_*<k_0$. A uniform positive lower
bound on the density of $H_h$ near $k_0(h)$ follows from the
two strips alone. For a value $z$ of the statistic near $k_0(h)$, set
$x_h(z)=\{\log(2z)+h^2/2\}/h>c$. Their density contribution is
\[
 \frac{2\varphi(x_h(z))}{hz}
       \Phi\{(c-\rho x_h(z))/s\},
\]
continuous and positive near $(h_0,k_0(h_0))$, where $x_h(z)=t$.
Restoring null rejection probability $m(h)$ therefore requires
$0<k_0-k_*=O(d^2)$.

The Lagrange dual function $B_h(k)=k\alpha+E_0(H_h-k)_+$ satisfies
$B_h'(k)=\alpha-P_0(H_h>k)$ at positive thresholds.
The calibrated threshold minimises it, and
$\mathcal B=B_h(k_0)-E_0(1_RH_h)$ whereas
$\Delta=B_h(k_*)-E_0(1_RH_h)$. For $k_*\le k\le k_0$,
$0\le B_h'(k)\le m(h)$, giving
\[
 0\le\mathcal B-\Delta\le m(h)(k_0-k_*)=O(d^4).
\]
This proves the first expansion. Since $h-h_0=O(d)$,
$r(h)=1/2+O(d)$ and $k_0$ is smooth, the second follows.
\end{proof}

If $\alpha\le1/2$ and the identical intervals are $[h,h+w]$ for
fixed finite $w\ge0$, the same expansion holds for $h>h_0$
sufficiently near $h_0$. Indeed, Lemma~\ref{supp:modebound}
places the symmetric step-up procedure's regret maximiser strictly below $h_0$.
The calibrated Bayes rules just constructed converge almost
everywhere to that step-up procedure as $h\downarrow h_0$. Dominated
convergence in the conditional regret derivative, with the rule
held fixed during differentiation, makes their derivative at
noncentrality parameter $h$ strictly negative for all sufficiently close $h$.
Lemma~\ref{supp:shape} then puts each interval maximum at $h$,
so the singleton prior remains least favourable for the intervals.

The cubic order has a geometric interpretation related to margin
arguments for excess risk and excess mass
\citep{RigolletVert2009,AudibertTsybakov2007}. The likelihood-ratio
deficit is linear near the corner and the deleted area is of order $d^2$;
restoring its probability mass moves the critical value by order $d^2$
and costs order $d^4$. The expansion fixes a nonsingular correlation
and is not uniform as $\rho$ approaches one.

\section{Bounds on excess regret and reoptimisation under perturbations}
\label{supp:stability}

\begin{suppprop}[Bound on excess minimax regret]
\label{supp:defect}
For a calibrated step-up procedure choose $q,k$ at the target noncentrality parameters maximising its regrets
by \eqref{eq:supp:qk}. Let its maximum regret be $v_R$, let $v_*$
be the minimax value in either $\mathcal U$ or
$\mathcal U_{\rm det}$, and set
\[
 \epsilon_R=v_R-[q r_S(\mu_{S,R})+(1-q)r_F(\mu_{F,R})],\qquad
 \mathcal B_R=E_0[1_Q(k-H_Q)_+].
\]
Then $0\le v_R-v_*\le\epsilon_R+\mathcal B_R$.
\end{suppprop}
\begin{proof}
For the fitted two-point prior, write
$W_R=E_0[1_RH]$ and
$W^*=\sup_{\phi\in\mathcal U}E_0[\phi H]$.
The fixed-prior lower bound on minimax regret gives
\[
 v_R-v_{\rm ran}\le\epsilon_R+(W^*-W_R).
\]
Let $G=qC_S(x)$ in $S$, $G=(1-q)C_F(y)$ in $F$, $G=H_Q$ in
$Q$, and $G=0$ outside $\mathcal G$. The conditional covariance argument
in Lemma~\ref{supp:arms} gives $E_0(\phi H)\le E_0(\phi G)$
for every upper $\phi$, with equality for a step-up procedure because
its decisions within each strip depend only on the coordinate corresponding
to the rejecting marginal test.
Since $R$ has size $\alpha$ and $k>0$,
\[
 E_0[(\phi-1_R)H]\le E_0[(\phi-1_R)(G-k)]
                         \le\mathcal B_R.
\]
The last integrand is nonpositive outside
$Q\cap\{H_Q<k\}$, where $R$ agrees with the threshold rule determined by its cutoffs;
inside this set it is bounded above by $k-H_Q$.
Thus prior-weighted power can improve by no more than
$\mathcal B_R$. Equivalently, every feasible rule has prior-average regret at
least $q r_S+(1-q)r_F-\mathcal B_R$. Worst regret dominates
prior regret, and the feasible step-up procedure supplies the opposite
bound $v_*\le v_R$.
\end{proof}

For a conservatively calibrated step-up procedure of exact size
$\gamma\le\alpha$, the same proof adds $k(\alpha-\gamma)$.
Numerical quadrature approximates the exact integral $\mathcal B_R$;
a rigorous bound obtained with outward rounding requires validated numerical error bounds.

The stability result below can be compared with robustness regions in
\citet{O11}.

\begin{suppcor}[Stability under a strict unperturbed optimality condition]
\label{supp:persistence}
If the nominal equal-regret step-up procedure satisfies a strict inequality
in \eqref{eq:supp:nominalfit}, its reoptimised equal-regret step-up procedure
remains exactly minimax for all sufficiently small nonnegative
pairs $(e_S,e_F)$.
\end{suppcor}
\begin{proof}
The equalisation difference is jointly continuous in allocation
and offset bounds and has a unique interior root. Opposite strict signs
on either side of the nominal root trap all nearby roots, proving
continuity of the allocation and hence of the finite cutoffs.
The regret functions then converge uniformly on their compact
noncentrality intervals. Uniqueness of the maxima implies continuity
of the regret-maximising noncentrality parameters: any subsequential limit must maximise
the limiting regret. This argument includes endpoints and
singletons.

The values $q,k$ determined by the cutoff equations, the statistic at the corner
and its gradient are consequently continuous. At zero offset bounds, $H_Q(c,c)>k$
by the strict unperturbed optimality condition, and
both corner gradient components are strictly positive. These
inequalities persist for sufficiently small offset bounds. Since $H_Q$
is a positive weighted sum of exponentials of affine functions, it is convex, and its
supporting hyperplane at $(c,c)$ gives $H_Q\ge k$ on the whole
quadrant. The pointwise sufficient version of
Theorem~\ref{supp:integrated}, together with equalisation
within the step-up family, proves the assertion.
\end{proof}

\begin{suppprop}[First-order loss from leaving the cutoffs unchanged]
\label{supp:retuning}
Fix the symmetric cutoff $t$ and a common fixed target noncentrality parameter $h$
with $2e^{-h(t-c)}>1$. Perturb only the subgroup-target subfamily, setting
$e_S=e>0$, $e_F=0$. Define
\[
 J_t=\Phi((c-\rho t)/s)\varphi(t-h),\qquad
 T_h=\int_c^t\varphi((c-\rho x)/s)\varphi(x-h)\,dx>0.
\]
The step-up procedure with unchanged cutoffs has excess robust minimax regret
$T_he/2+O(e^2)$ as $e\downarrow0$. The reoptimised cutoffs are
$x_0=t-T_he/(2J_t)+O(e^2)$ and
$y_0=t+T_he/(2J_t)+O(e^2)$.
\end{suppprop}
\begin{proof}
At the nominal point the derivative of either regret with respect
to its cutoff is $J_t$, while the subgroup-target derivative with respect to the offset bound is $T_h$.
Null calibration gives $dy_0/dx_0=-1$ at $x_0=y_0=t$.
The cutoff derivative of the equalisation difference is therefore
$2J_t>0$. The implicit-function theorem and smoothness of the
finite-cutoff integrals give the displayed cutoff expansions.
Substitution yields a common regret after reoptimisation of
$v_0+T_he/2+O(e^2)$. By
Corollary~\ref{supp:persistence} this is the full minimax value
for sufficiently small $e$. The step-up procedure with unchanged cutoffs instead has
subgroup-target regret $v_0+T_he+O(e^2)$ and unchanged full-population-target regret $v_0$.
Subtracting gives the result.
\end{proof}

\begin{suppthm}[Local cubic expansion under symmetric perturbations]
\label{supp:robustcubic}
Fix equal offset bounds $e_S=e_F=e\ge0$ and fixed target noncentrality parameters
$\mu_S=\mu_F=h$, with the calibrated symmetric cutoff $t$. Put
\[
 D_t=t-c,\quad z_c=c(1-\rho)/s,\quad z_t=(c-\rho t)/s,
 \quad R_t=\Phi(z_t+e)/\Phi(z_t),
\]
\[
 k_0(h)=\tfrac12 e^{ht-h^2/2}R_t,\qquad
 h_0=\frac{\log2-ez_c-e^2/2-\log R_t}{D_t},\qquad
 \eta_0=h_0-(1-\rho)e/s.
\]
Assume $h_0>0$ and $\eta_0>0$. For
$d(h)=1-e^{-(h-h_0)D_t}$, the symmetric step-up procedure's actual
excess robust minimax regret satisfies
\begin{equation}
 \Delta(h)=\frac{2k_0(h_0)f_0(c,c)}{3\eta_0^2}\,d(h)^3
                +O(d(h)^4),\qquad h\downarrow h_0.
 \label{eq:supp:robustcubic}
\end{equation}
This is a local result under the stated positive-gradient
conditions; it does not assert coordinatewise monotonicity of threshold rules
based on the averaged statistic at arbitrary offset bounds or noncentrality parameters.
\end{suppthm}
\begin{proof}
Use the equal prior and conditionally averaged test statistic $G$ from the proof of
Proposition~\ref{supp:defect}. The threshold determined by the cutoffs is $k_0(h)$.
In the quadrant, its normalised value at the corner is $1-d$, and the
two normalised summands are each $(1-d)/2$. Writing
$u=x-c,v=y-c$, the statistic is
\[
 \frac{H_Q(c+u,c+v)}{k_0(h)}
 =\frac{1-d}{2}\left[
 e^{(h+\rho e/s)u-(e/s)v}
 +e^{-(e/s)u+(h+\rho e/s)v}\right].
\]
Both gradient components at the corner converge to $\eta_0/2>0$. Convexity
places the sublevel region $W_h=\{Q:H_Q<k_0(h)\}$ inside
a triangle with side lengths $O(d)$. Taylor expansion of the
two exponentials and the Gaussian density, followed by rescaling
by the two positive gradient components at the corner, gives
\[
 \mathcal B(h)=E_0[1_Q(k_0-H_Q)_+]
   =\frac{k_0(h_0)f_0(c,c)}{6(\eta_0/2)^2}d^3+O(d^4),
 \qquad P_0(W_h)=O(d^2).
\]
The remainder is uniform on that shrinking triangle: its
normalised statistic is $1-d+g_xu+g_yv+O(u^2+v^2)$,
$g_x,g_y=\eta_0/2+O(d)$, and the positive-part map is Lipschitz.

At $k_0$, the threshold rule based on $G$ omits exactly $W_h$ from the
size-$\alpha$ symmetric step-up procedure. The increasing smooth
conditional likelihood-ratio averages within the strips have positive derivatives at their cutoffs
near $t$, and their null strip densities are positive there.
The distribution of $G$ therefore has a uniform positive
density lower bound near $k_0$. The calibrated threshold
$k_*$ satisfies $0<k_0-k_*=O(d^2)$. The same dual calculation
as in Theorem~\ref{supp:cubic}, applied now to $G$, shows that
the calibrated power gain calculated from the averaged statistic differs from
$\mathcal B(h)$ by $O(d^4)$.

It remains to show that this gain is achievable by a valid upper
rule for the actual statistic. The supporting plane at the corner
places every point with $H_Q<k_*$ inside the same $O(d)$
triangle. Both partial derivatives of the test statistic are positive on a
neighbourhood containing that triangle for sufficiently small
$d$, because their limits at the corner are positive. Any
coordinatewise downward path from a point in the sublevel
triangle stays in the triangle, and the test statistic decreases along that
path. Thus the quadrant sublevel set is lower and its complement
is upper. On each strip, $\{G>k_*\}$ is a threshold rejection region
with cutoff tending to $t>c$. The corner sublevel triangle
lies strictly below both arm cutoffs for small $d$, so every
point rejected in a strip remains rejected when the other marginal test
also rejects. The full threshold set is therefore upper.

Within each strip, the rule depends only on the coordinate corresponding
to the rejecting marginal test. This gives equality in $E_0(\phi H)\le E_0(\phi G)$. It maximises
actual weighted power over upper rules because it maximises
the objective based on the averaged statistic over all measurable tests of size $\alpha$.
Its symmetric regrets are equal, so it is minimax for the two
fixed target noncentrality parameters at the worst-case offsets; the offset reduction supplies the full
offset neighbourhoods. Its actual minimax improvement equals
the power gain calculated from the averaged statistic, proving
\eqref{eq:supp:robustcubic}.
\end{proof}

\section{A minimax step-up procedure for which the pointwise sufficient condition fails}
\label{supp:counterexample}

The following construction shows that $H_Q\ge k$ throughout the
quadrant is stronger than the integrated criterion. It uses an
exactly calibrated and equalised step-up procedure. The significance
level makes the structural counterexample transparent.

Set $\rho=0$, $\alpha=1/2$, $c=0$, $\mu_S=1/20$, $e_S=0$ and
$e_F=2$, with singleton noncentrality intervals. Calibration becomes
$\Phi(x_0)+\Phi(y_0)=3/2$. At the prior determined by the cutoff equations, the normalised
quadrant statistic is
\begin{equation}
 \frac{H_Q(x,y)}k
 =e^{\mu_S (x-x_0)}+
   \frac{e^{-\mu_F y_0-2-2x+\mu_F y}}{2\Phi(2)}=:F_{\mu_F}(x,y).
 \label{eq:supp:counter-score}
\end{equation}
It increases in $y$. Define
\begin{align*}
 u&=\log(1+\mu_S/2)/\mu_S,\qquad C=e^{-2}/[2\Phi(2)],\\
 y(x)&=\Phi^{-1}(3/2-\Phi(x)),\qquad
 D_0(x)=\frac{\mu_S}{\mu_S+2}e^{2(x-u)},\qquad
 \mu_{F,*} (x)=\frac{\log[C/D_0(x)]}{y(x)}.
\end{align*}
At $\mu_F=\mu_{F,*} (x_0)>0$, the function $F_{\mu_F}(x,0)$ has its unique
minimum at $x_*=x_0-u$, with minimum exactly one. Its two
exponential terms there are $2/(\mu_S+2)$ and $\mu_S/(\mu_S+2)$; its
first derivative is zero and its second derivative positive.
The step-up procedure's regrets are
\begin{equation}
 r_S(x_0)=\tfrac12[\Phi(x_0-\mu_S)-\Phi(-\mu_S)],\qquad
 r_F(y_0,\mu_F)=\Phi(2)[\Phi(y_0-\mu_F)-\Phi(-\mu_F)].
 \label{eq:supp:counter-risk}
\end{equation}

\begin{suppprop}[Conservativeness of pointwise quadrant positivity]
\label{supp:counterexample-result}
For the preceding family there are positive $\mu_F$ and finite
calibrated, equalised cutoffs for which the step-up procedure is exactly
minimax but $H_Q<k$ on an open subset of $Q$.
\end{suppprop}
\begin{proof}
First choose an equalised step-up procedure at the pointwise transition.
Substituting $y(x)$ and $\mu_{F,*} (x)$ into
\eqref{eq:supp:counter-risk} gives a continuous equalisation
difference. Rigorous interval bounds obtained with outward rounding are
\[
 \begin{array}{cccc}\toprule
 x&y(x)&\mu_{F,*} (x)&r_S-r_F\\\midrule
 .957&(.437,.438)&(.267,.268)&
       (-3.181,-3.180)\times10^{-4}\\
 .959&(.436,.437)&(.259,.260)&
       (4.184,4.185)\times10^{-4}\\\bottomrule
 \end{array}
\]
Thus some $x^\circ\in(.957,.959)$ equalises, with corresponding
$y^\circ,\mu_F^\circ$. The accompanying script
\texttt{verify\_counterexample.py} gives finer enclosures contained in those shown
using 60-digit interval arithmetic. It evaluates $\Phi$ through
the integrated exponential series with an explicit first-omitted
term remainder bound and encloses the inverse cdf by bisection.
For orientation only, the approximate solution is
$(x^\circ,y^\circ,\mu_F^\circ)\simeq(.958,.437,.264)$.

At this transition, put $f(x)=F_{\mu_F^\circ}(x,0)$.
Then $f\ge1$ on $[0,\infty)$ with equality only at one
strictly positive point. Every truncated-normal conditional mean
\[
 \overline f(v)=\frac{\int_0^v f(x)\varphi(x)\,dx}{\Phi(v)-1/2},
 \qquad 0<v<\infty,
\]
is strictly larger than one. Its limits as $v\downarrow0$ and
$v\to\infty$ are respectively $f(0)>1$ and
$2\int_0^\infty f\varphi>1$. It extends continuously to the
compactified interval $[0,\infty]$, so its gap above one has
a strictly positive uniform minimum.

Now increase $\mu_F$ while maintaining equalisation within the step-up family and
calibration. The implicit-function theorem gives differentiable
cutoffs $x_R(\mu_F),y_R(\mu_F)$ near the transition, because
\[
 \partial_{x_0}(r_S-r_F)=r_{S,x}-r_{F,y}y_x>0,
 \qquad y_x=-\varphi(x_0)/\varphi(y_0)<0.
\]
Differentiating the minimum of \eqref{eq:supp:counter-score}
along this path gives
\begin{equation}
 m'(\mu_F^\circ)=\frac{\mu_S}{\mu_S+2}
             [-2x_R'-y^\circ-\mu_F^\circ y_R'].
 \label{eq:supp:counter-derivative}
\end{equation}
The interval bracket implies
$.95<x^\circ<.97$, $.43<y^\circ<.45$ and
$.25<\mu_F^\circ<.28$. Therefore
$\mu_F^\circ>y^\circ-\mu_F^\circ>0$, and
\[
 r_{F,\mu_F}=\Phi(2)[\varphi(\mu_F)-\varphi(y-\mu_F)]<0.
\]
It follows that $x_R'<0$, $y_R'>0$. On these brackets,
$|r_{F,\mu_F}|<.4\cdot.28\cdot.13=1.456\times10^{-2}$, whereas
$r_{S,x}>\varphi(.92)/2>.125$, giving $|x_R'|<.117$.
The bracket in \eqref{eq:supp:counter-derivative} is consequently
less than $2(.117)-.43<0$. For every sufficiently small
increase in $\mu_F$, the minimum is strictly below one and stays
at an interior positive $x$. By continuity, $F_{\mu_F}<1$ then holds
also at points with small positive $y$, and hence on an open
subset of $Q$.

Nevertheless, the positive uniform gap of the truncated-normal conditional means
persists under a sufficiently small parameter change. Near
$v=0$ this follows from uniform continuity of the statistic values;
on a compact middle interval the denominator is bounded away
from zero; and as $v\to\infty$ a common integrable
Gaussian-exponential envelope gives uniform tail control.
Thus
\[
 \int_0^v[F_{\mu_F}(x,0)-1]\varphi(x)\,dx\ge0
 \quad\text{for every }0\le v\le\infty.
\]
Each horizontal section of a lower set $L\subset Q$ is
an interval $(0,v(y))$ beginning at zero. Independence, $F_{\mu_F}(x,y)\ge F_{\mu_F}(x,0)$
and Fubini's theorem imply
$\int_L(H_Q-k)f_0\ge0$. The additional arm costs in
\eqref{eq:supp:Psi} are nonnegative. By
Theorem~\ref{supp:integrated}, the calibrated equalised step-up procedure
is therefore exactly minimax despite failure of the pointwise
quadrant condition. This final argument uses the sufficient
integrated criterion directly; it does not assume a theorem on
elimination of randomisation.
\end{proof}

\section{Compatibility of the efficacy states with target populations}
\label{supp:priority}

Fix a subgroup $S$, its complement $C$, and their union $F$, with
prevalence $0<\pi<1$. Let $\Delta_S,\Delta_C$ denote treatment effects
on an additive benefit scale.

For independent subgroup and complement estimators with variances $v_S,v_C$,
the additive clinical interpretation gives
\begin{equation}\label{eq:covariance}
\Delta_F=\pi\Delta_S+(1-\pi)\Delta_C,\qquad
\rho=\frac{\pi\sqrt{v_S}}{\sqrt{\pi^2v_S+(1-\pi)^2v_C}}.
\end{equation}
Under proportional recruitment and a common per-observation variance,
the estimator variances satisfy $\pi v_S=(1-\pi)v_C$, giving
$\rho=\sqrt\pi$. In this setting, zero offsets represent zero complement effect in
subgroup-target states and homogeneous benefit in full-population-target states. The statistic $(Y-\rho X)/s$ estimates
the standardised complement effect, so negative $\delta$ permits harm.
In full-population-target states, $\eta=(\Delta_S-\Delta_C)/\sqrt{v_S+v_C}$ measures heterogeneity
at fixed overall efficacy; either stratum may benefit more. The additive
representation does not apply directly to log odds or log hazard ratios.

Under proportional recruitment and a common per-observation variance, let
$0<\pi<1$, $\rho=\sqrt\pi$, $s=\sqrt{1-\pi}$ and
$\kappa=\sigma_F=\rho\sigma_S=s\sigma_C$. For additive effects,
the families with their designated target populations imply
\[
\begin{array}{lll}
\theta_S(\mu_S,\delta):&\Delta_S=\kappa \mu_S/\rho,&\Delta_C=\kappa\delta/s,\\
\theta_F(\mu_F,\eta):&\Delta_S=\kappa(\mu_F+s\eta/\rho),&
\Delta_C=\kappa(\mu_F-\rho\eta/s).
\end{array}
\]
For prespecified effect-scale burdens $\tau_S\ge0$ and $\tau_C>0$,
consider the additive values of treating nobody, the subgroup, or the
full population:
\begin{equation}\label{supp:eq:actionvalue}
U_0=0,\qquad
U_S=\pi(\Delta_S-\tau_S),\qquad
U_F=U_S+(1-\pi)(\Delta_C-\tau_C).
\end{equation}
Expanding treatment adds the complement's net benefit:
$U_F-U_S=(1-\pi)(\Delta_C-\tau_C)$. Provided $\Delta_S>\tau_S$,
full-population use uniquely maximises value when $\Delta_C>\tau_C$,
and subgroup use when $\Delta_C<\tau_C$.
This model omits policy-specific fixed costs. We distinguish maximisation
of aggregate value from the additional requirement that every treated
stratum have positive net benefit. Subgroup treatment uniquely maximises value
throughout the subgroup-target states, and full-population treatment throughout the full-population-target states, exactly when
\begin{equation}\label{supp:eq:priority}
\begin{split}
0\le\tau_S&<\kappa \underline{\mu}_S/\rho,\\
\kappa e_S/s<\tau_C&<\kappa(\underline{\mu}_F-\rho e_F/s),\\
\pi\tau_S+(1-\pi)\tau_C&<\kappa \underline{\mu}_F .
\end{split}
\end{equation}
Indeed, unique optimality of subgroup use is equivalent to $U_S>0$ and
$\Delta_C<\tau_C$; unique optimality of full-population use is equivalent to
$U_F>0$ and $\Delta_C>\tau_C$. Taking the appropriate endpoint
of each affine expression proves necessity and sufficiency. Some
threshold pair exists precisely when $e_S+\rho e_F<s\underline{\mu}_F$: choose
$\tau_C$ between its bounds and $\tau_S=0$. A stronger sufficient pair, ensuring positive net benefit
within each stratum throughout the full-population-target states, is
\begin{equation}\label{eq:priorityconditions}
\begin{split}
0\le\tau_S&<\kappa\min\{\underline{\mu}_S/\rho,\ \underline{\mu}_F-se_F/\rho\},\\
\kappa e_S/s<\tau_C&<\kappa(\underline{\mu}_F-\rho e_F/s).
\end{split}
\end{equation}
The full-population value condition then follows automatically. Under these
stronger conditions, $S$ in subgroup-target states and $F$ in full-population-target states are the broadest candidate
populations with positive net benefit in every treated stratum. The clinical planning example uses this stronger requirement for its common-burden intervals.
Failure of that requirement need not preclude the same target designations
under aggregate value maximisation alone.

Table~\ref{tab:priority-compatibility} verifies the stronger pair with
$\tau_S/\kappa=1$ and $\tau_C/\kappa=1.2$;
the minimum subgroup effect over subgroup-target states is
$\min\Delta_S/\kappa\simeq2.372$. These chosen burdens verify mathematical
compatibility under the stronger requirement. Clinical calibration requires justification in original
outcome units before choosing planning ranges. Defining burdens as multiples
of a changing standard error would change clinical valuation with sample size.
The script \texttt{check\_priority\_compatibility.py} reproduces these
checks of extrema of affine functions from the archived settings.

\begingroup
\renewcommand{\thetable}{S\arabic{table}}
\setcounter{table}{0}
\begin{table}[htbp]
\normalsize\setstretch{1}\renewcommand{\arraystretch}{1}
\centering
\caption{Compatibility of the interval examples under the requirement of positive net benefit in every treated stratum, with the illustrative thresholds $\tau_S/\kappa=1$ and $\tau_C/\kappa=1.2$. The settings are $I_S=[1.5,4]$, $I_F=[2,4.5]$, $\pi=0.4$ and $e_S=e_F=e$. The maximum is over subgroup-target states; both minima are over full-population-target states.}
\label{tab:priority-compatibility}
\begin{tabular}{rrrr}
\toprule
$e$ & $\max\Delta_C/\kappa$ & $\min\Delta_C/\kappa$ & $\min\Delta_S/\kappa$ \\
\midrule
0.10 & 0.129 & 1.918 & 1.878 \\
0.25 & 0.323 & 1.796 & 1.694 \\
0.40 & 0.516 & 1.673 & 1.510 \\
0.80 & 1.033 & 1.347 & 1.020 \\
\bottomrule
\end{tabular}
\end{table}
\endgroup

\begin{suppcor}[Per-stratum benefit conditions imply a nonrandomised minimax rule]\label{cor:prioritydirection}
Under proportional recruitment with a common per-observation variance, the benefit--burden
conditions \eqref{eq:priorityconditions} imply \eqref{eq:supp:J0}. Consequently their
minimax solution is nonrandomised and unique without a further directional
assumption. The weaker conditions for maximising total value alone do not imply
\eqref{eq:supp:J0}; a counterexample is given below.
\end{suppcor}
Indeed, positive net subgroup benefit throughout the full-population-target states gives
$\underline{\mu}_F-se_F/\rho>\tau_S/\kappa\ge0$, and
\[
J_0=\underline{\mu}_S (\underline{\mu}_F+\rho e_F/s)+e_S(\rho \underline{\mu}_F/s-e_F)>0.
\]
More generally, either $\rho \underline{\mu}_F\ge se_F$ or $\rho \underline{\mu}_S\ge se_S$
ensures $J_0>0$: a nonnegative non-target mean at the worst-case offset in either family
suffices. The full determinant condition also covers cases in which both
of those means are negative.

The weaker aggregate-value conditions alone do not imply
$J_0>0$: for example, $\rho=0.2$, $\underline{\mu}_S=0.05$, $\underline{\mu}_F=1$,
$e_S=0.5$, $e_F=1$, $\tau_S/\kappa=0.1$ and
$\tau_C/\kappa=0.6$ satisfy \eqref{supp:eq:priority}, but give
$J_0\approx-0.338$. In that example some states in the full-population-target family have
negative subgroup benefit.

Conversely, $J_0>0$ alone does not ensure compatible target designations. Take
$\rho=1/2$, $I_S=I_F=\{1\}$ and $e_S=e_F=1$. Both target families include
states with mean $\mu=(1,1)$ at $\delta=\eta=1/\sqrt3$,
although $J_0=2/\sqrt3>0$.
No fixed value function can make both actions uniquely optimal at this state.

Effect thresholds designate the targets for each efficacy state; the zero-effect nulls remain
unchanged. Their rejection does not establish complement benefit or net
value. Target-power regret is a surrogate: welfare regret also needs
state-dependent value gaps and a specification of the treatment policy following each testing decision.

\section{Removing randomisation from Gaussian tests in arbitrary dimension}
\label{supp:general-purification}

The two-dimensional rearrangement is a consequence of a more general
monotonicity-preserving testing principle. The conditional rearrangement is Gaussian
symmetrisation \citep{Ehrhard1983,CagnettiEtAl2017}; the conditions below
identify when it preserves the original coordinate order and improves
all target powers simultaneously while retaining the fixed rejection requirements.
The optimisation uses one null rejection-probability constraint under $P_0$.
For more than two claims, strong familywise error control additionally requires
the partial-intersection constraints discussed at the end of this section.

Let $P_0=N_n(0,\Sigma)$ with $\Sigma$ positive definite, and let
$G_1,\ldots,G_k$ be Borel upper sets with union $\mathcal G$. Let $\mathcal M$
be a family of alternative means. Suppose there is a vector $\omega$
satisfying
\begin{equation}\label{supp:eq:general-direction}
 \omega\ge0,\qquad d=\Sigma\omega>0,\qquad
 m^{\mathsf T}\omega>0\quad(m\in\mathcal M),
\end{equation}
where the vector inequalities are coordinatewise. Set containment and
decisions are understood up to $P_0$-null sets. Write
$L_m=dN_n(m,\Sigma)/dP_0$. A randomised rejection rule is a coordinatewise nondecreasing function
$0\le\phi\le1_{\mathcal G}$.

\begin{suppthm}[Order-preserving Gaussian removal of randomisation]
\label{supp:general-purification-theorem}
Under \eqref{supp:eq:general-direction}, every randomised rejection rule
$\phi$ has a nonrandomised monotone rejection rule $1_R\le1_{\mathcal G}$ with
$P_0(R)=E_0\phi$ and
\[
 E_0[1_R1_{G_j}L_m]\ge E_0[\phi1_{G_j}L_m]
 \qquad(j=1,\ldots,k,\ m\in\mathcal M).
\]
Its conditional rejection probabilities belong to the convex class
\eqref{supp:eq:general-order} below. For any finite nonnegative mixture
of these likelihood ratios restricted to their respective fixed rejection sets, assigning positive total weight to every region,
and $0<\alpha<P_0(\mathcal G)$, the Bayes-optimal rejection rule subject to
$E_0\phi\le\alpha$ is nonrandomised, has size $\alpha$, and is unique
up to $P_0$-null sets.
\end{suppthm}

\begin{proof}
Put $\nu^2=\omega^{\mathsf T}\Sigma\omega>0$ and use the independent
Gaussian coordinates
\[
 T=\frac{\omega^{\mathsf T}X}{\nu^2},\qquad
 Z=X-dT,\qquad T\sim N(0,\nu^{-2}).
\]
The residual $Z$ lies in the hyperplane
$\mathcal H=\{z:\omega^{\mathsf T}z=0\}$. Independence follows from
$\operatorname{Cov}(X,T)=d/\nu^2$ and
$\operatorname{Var}(T)=1/\nu^2$.
Since $d>0$, an upper set intersects each line $z+\mathbb R d$
in an upper tail $\{t>h(z)\}$, allowing empty and full tails and ignoring
their endpoints. Its conditional rejection probability is
$p(z)=\overline\Phi(\nu h(z))$.

Let $e_i$ be the $i$th coordinate vector and define
\[
 \ell_i=\omega_i/\nu^2,\qquad
 v_i=e_i-d\ell_i,\qquad \gamma_i=\omega_i/\nu.
\]
Increasing $X$ by $u e_i$, $u\ge0$, increases $T$ by $u\ell_i$ and
changes $Z$ by $u v_i$. Thus these conditional upper tails form an upper set exactly
when $h(z+uv_i)\le h(z)+u\ell_i$, equivalently
\begin{equation}\label{supp:eq:general-order}
 p(z+uv_i)\ge G_{\gamma_i u}(p(z)),\qquad
 z\in\mathcal H,\ u\ge0,\ i=1,\ldots,n,
 \quad 0\le p(z)\le p_{\mathcal G}(z),
\end{equation}
where $G_t(p)=\Phi(\Phi^{-1}(p)-t)$ and $p_{\mathcal G}$ is the conditional
probability of $\mathcal G$. The endpoint values are defined continuously.
The maps $G_t$ are convex for $t\ge0$: on $(0,1)$,
$G_t'(p)=\exp\{t\Phi^{-1}(p)-t^2/2\}$ is nondecreasing.
Consequently \eqref{supp:eq:general-order} defines a convex class,
including coordinates with $\omega_i=0$, for which $G_0$ is the identity.

Write $\phi=\int_0^1 1\{\phi>v\}\,dv$. Each nonzero level set is
upper and contained in $\mathcal G$, so its conditional probability $p_v$
satisfies \eqref{supp:eq:general-order}. Jensen's inequality shows that
their average
$p(z)=E_0[\phi(X)\mid Z=z]=\int_0^1p_v(z)\,dv$
satisfies the same constraints. The replacement by a conditional upper tail
$R=\{z+dt:t>\overline\Phi^{-1}(p(z))/\nu\}$ is therefore upper,
contained in $\mathcal G$, and has the same null probability as $\phi$.

Along this line, the likelihood ratio is
\[
 L_m(z+dt)=
 \exp\{m^{\mathsf T}\Sigma^{-1}z
       -\tfrac12m^{\mathsf T}\Sigma^{-1}m
       +t m^{\mathsf T}\omega\}.
\]
It is strictly increasing in $t$, and $1_{G_j}(z+dt)$ is nondecreasing.
Selecting the upper Gaussian tail of probability $p(z)$ consequently
maximises the conditional integral of every likelihood ratio restricted to its fixed rejection set. Integration
over $Z$ proves simultaneous power improvement.

For the final assertion, let $H$ be the specified finite mixture.
It is positive and strictly increasing along each line on
the portion in $\mathcal G$. The maximal conditional integral of $H$ is therefore
a strictly concave function of $p$ on $[0,p_{\mathcal G}(z)]$ whenever
$p_{\mathcal G}(z)>0$. Integrating over $Z$ gives a strictly concave objective
on the convex conditional-probability class, modulo null sets.
An optimum exists: the bounded monotone rejection functions form a weak-star
compact subset of $L^\infty(P_0)$, and the finitely many Gaussian
likelihood ratio integrals are weak-star continuous. Coordinatewise monotonicity is closed,
for example by its nonnegative distributional coordinate derivatives
on compact sets. The constraint $\phi\le1_{\mathcal G}$ is also weak-star closed.
Since $H>0$ on $\mathcal G$ and $\alpha<P_0(\mathcal G)$, mixing a rule whose null rejection probability is below the constraint
with $1_{\mathcal G}$ strictly improves its objective until the constraint is attained.
Eliminating randomisation gives a nonrandomised optimum; strict conditional
concavity gives its unique conditional probability. Strict increase of the likelihood ratios along each line
also excludes a different randomised maximiser.
\end{proof}

For the present two-claim problem, take
$G_1=\{x>c\}$ and $G_2=\{y>c\}$ and the alternative means at worst-case offsets.
The direction in the main paper corresponds to
$\omega=\Sigma_\rho^{-1}d>0$. The endpoint determinant $J_0>0$
provides one such direction throughout both target noncentrality intervals.
The main theorem additionally uses the two-claim marginal-test reduction
argument to obtain strong familywise error control. In dimension
larger than two, elimination of randomisation under one null rejection probability constraint remains
valid, but strong familywise error control would require the relevant
partial-intersection constraints as well.

\section{A computable bound for the continuous problem}\label{supp:continuumcertificate}

The construction below specialises a dual lower bound to monotone rules
satisfying the marginal rejection requirements. Dual bounds for the original continuous multiple-testing problem
were developed by \citet[Section~6.3]{rly2014};
Section~\ref{supp:literaturecomparison} compares the feasible classes and
within-cell arguments.

Fix any $\mu_S\in I_S$, $\mu_F\in I_F$, $0<q<1$ and $k\ge0$. Write $H$ for the
weighted likelihood ratio statistic, including the marginal rejection indicators, at the two means with worst-case offsets, and
$O=q\Phi(\mu_S-c)+(1-q)\Phi(\mu_F-c)$ for its prior-weighted marginal reference power. No optimality of this
prior is required. Partition a finite box $K$ into rectangular cells, with
both marginal critical-value lines included in the partition. For a cell $C$ let $z_C^-$ and
$z_C^+$ denote its lower-left and upper-right vertices, and put
\[
 p_C=\int_C f_0,\qquad m_C=\int_C Hf_0,\qquad W_C=m_C-kp_C.
\]
Choose finite bounds $0\le\ell_C\le H(z)\le u_C$ within each cell. For
Gaussian shift mixtures these follow by bounding each exponential of a
linear function at the appropriate corners; the marginal rejection indicators are
constant within a cell. Define
\begin{equation}\label{eq:certsecant}
 U_C=\begin{cases}
 0,&u_C\le k,\\
 W_C,&\ell_C\ge k,\\
 \displaystyle\frac{u_C-k}{u_C-\ell_C}(m_C-\ell_Cp_C),
       &\ell_C<k<u_C.
 \end{cases}
\end{equation}
When $u_C=\ell_C=k$ either of the first two cases gives zero. Let $\mathcal V$
be the arrays $0\le v_z\le1$ on the grid vertices that are nondecreasing in
both coordinates and vanish when both coordinates are at most $c$. Set
\begin{equation}\label{eq:certvertex}
 M_K(k)=\max_{v\in\mathcal V}\sum_C
 \{W_Cv_{z_C^-}+U_C(v_{z_C^+}-v_{z_C^-})\}.
\end{equation}
Finally, choose $T_K\ge\int_{K^c}Hf_0$. A sum of four marginal Gaussian
tail probabilities for each alternative, weighted by $q$ and $1-q$,
supplies a simple choice.

\begin{suppthm}[Lower bound for the continuous problem]\label{supp:vertexbound}
For every such prior, penalty, partition and tail bound,
\begin{equation}\label{eq:certbracket}
 O-k\alpha-M_K(k)-T_K
 \ \le\ v_{\rm ran}\ \le\ v_{\rm det}
 \ \le\ \sup_{\theta\in\Theta}R(\theta;\widehat\phi)
\end{equation}
for every feasible nonrandomised monotone rule $\widehat\phi$. The vertex
maximum is attained by a coordinatewise nondecreasing array taking values zero and one and is computable by
dynamic programming over upper rejection sets. The lower bound applies at every finite pair of
offset bounds, without the common-direction condition.
\end{suppthm}
\begin{proof}
The positive-part function is bounded above by its secant on
$[\ell_C,u_C]$, so \eqref{eq:certsecant} gives
$U_C\ge\int_C f_0(H-k)_+$. For a coordinatewise nondecreasing rejection function $\phi$, write
$v^-_C=\phi(z_C^-)$, $v^+_C=\phi(z_C^+)$ and
$d_C=v^+_C-v^-_C\ge0$. Pointwise in the cell,
$0\le\phi-v^-_C\le d_C$. Consequently
\[
 \int_C f_0(H-k)\phi
 \le W_Cv^-_C+d_C\int_C f_0(H-k)_+
 \le W_Cv^-_C+U_Cd_C.
\]
Boundary values may be chosen from a monotone version of $\phi$; they
do not change its Gaussian integrals. Summing cells and bounding the
outside-box contribution by $T_K$ yields
\[
 E_0(H\phi)\le k\alpha+M_K(k)+T_K.
\]
Subtracting from $O$ bounds fixed-prior Bayes regret below, and minimax
regret is at least that Bayes regret. A feasible nonrandomised rule gives
the upper endpoint in \eqref{eq:certbracket}.

For each $v\in\mathcal V$, the arrays $1\{v_z>t\}$, $0<t<1$, are binary
upper arrays satisfying the same marginal rejection constraints. Integrating these level-set indicators
expresses the linear objective at $v$ as the average of its objectives on
these binary arrays. Its maximum is therefore attained at one of them.
Combining the two contributions at each vertex reduces
\eqref{eq:certvertex} to the standard finite upper-set dynamic program.
\end{proof}

The result gives a bound valid for all monotone rules in the continuous problem, whereas maximising exact signed
cell masses over cellwise constant monotone rules restricts the optimisation
to those grid rules. Refining the maximal cell width gives convergence. To see this,
first fix the box. A binary vertex array
defines a monotone grid rule with value $v_{z_C^-}$ on each cell. The
difference from \eqref{eq:certvertex} is at most
$\sum_C U_C1\{v_{z_C^-}\ne v_{z_C^+}\}$. The cells in this sum lie on a
monotone staircase; their total area is bounded by
$|K_y|h_x+|K_x|h_y$, where $h_x,h_y$ are maximal side lengths. On a fixed
box the bounded Gaussian mixture densities and bounds on the weighted likelihood ratio statistic make
this difference tend to zero. Extending the monotone grid rule upwards
outside the box costs at most $kP_0(K^c)$ in its penalised objective.
Letting the box grow makes this term and $T_K$ vanish. Thus the relaxation
converges to the maximum penalised power over all monotone rules in the continuous problem.

\paragraph{Integration error.}
The theorem is stated for exact masses. If computed cell masses satisfy
known bounds $\sum_C|\widehat p_C-p_C|\le\delta_0$ and
$\sum_C|\widehat m_C-m_C|\le\delta_1$, then adding
$\delta_1+k\delta_0$ to the computed power bound accounts for these
errors. Indeed, for fixed ordered vertex values the coefficient of $m_C$
in the cell objective lies in $[0,1]$, while the coefficient of $p_C$ lies
in $[-k,0]$. This follows directly from the three cases of
\eqref{eq:certsecant}, and the same Lipschitz bound holds after maximising.
Errors in the bounds on the weighted likelihood ratio statistic, Gaussian tails and reference power require
their own rigorous numerical bounds. Our implementation evaluates the bound with
conditional Gaussian quadrature and compares numerical integration orders to
assess numerical stability. A rigorous numerical error bound requires bounds that include rounding and integration error
for the computed quantities. Near singular correlation, the
implementation explicitly splits narrow conditional-probability transitions before adaptive
integration, since two fixed quadrature orders can miss the same narrow transition.

\paragraph{A conditionally averaged likelihood-ratio comparator.}
This rule uses conditional averaging in the two strips where only one marginal
test rejects. For fixed $\mu_S,\mu_F,q$, let $C_S,C_F$ be the conditional null averages
defined in Section~\ref{supp:rectangles}, and let $H_Q=qL_S+(1-q)L_F$. Define
\[
 G(x,y)=\begin{cases}
 qC_S(x),&x>c,\ y\le c,\\
 (1-q)C_F(y),&x\le c,\ y>c,\\
 H_Q(x,y),&x>c,\ y>c,\\
 0,&x\le c,\ y\le c.
 \end{cases}
\]
Reject the intersection when $G>k_G$, choosing $k_G>0$ by scalar root
search so that $P_0(G>k_G)=\alpha$, and then apply the marginal tests.
Thus the rule thresholds conditional averages within the two strips and the
ordinary mixture likelihood ratio in the quadrant. The averaging inequality
in Lemma~\ref{supp:arms} bounds the prior-weighted power of every monotone
competitor by its integral against $G$. A threshold rule for $G$ attains
that bound when it is itself monotone: its strip decisions depend only on
the corresponding target statistic. Monotonicity therefore needs a separate
check; it is not a general property of $G$.

Here is the sufficient check used in the reported computations. Write
\[
 \begin{gathered}
 A_0=\mu_S+\rho e_S/s,\qquad B_0=\mu_F+\rho e_F/s,\qquad
 u=e_S/s,\qquad v=e_F/s,\\
 \lambda_S=qL_S(c,c),\qquad \lambda_F=(1-q)L_F(c,c).
 \end{gathered}
\]
At distance $(r,t)\ge0$ from the quadrant corner,
\[
 H_Q(c+r,c+t)=\lambda_S e^{A_0r-ut}
                    +\lambda_F e^{-vr+B_0t}.
\]
The implementation first requires strictly positive corner derivatives
$g_x=A_0\lambda_S-v\lambda_F$ and
$g_y=B_0\lambda_F-u\lambda_S$. If
$\Delta=(k_G-\lambda_S-\lambda_F)_+=0$, convexity implies that the
whole quadrant is retained. If $\Delta>0$, put
$U=\Delta/g_x$, $V=\Delta/g_y$, and let $x_G,y_G\ge c$ be the
strip cutoffs. Set $x_G=c$ if $qC_S(c)\ge k_G$; otherwise solve
$qC_S(x_G)=k_G$, and define $y_G$ analogously using $(1-q)C_F$.
The check requires
\[
 U<x_G-c,\qquad V<y_G-c,
\]
\[
 A_0\lambda_Se^{-uV}-v\lambda_Fe^{B_0V}>0,\qquad
 B_0\lambda_Fe^{-vU}-u\lambda_Se^{A_0U}>0.
\]
Convexity places the removed quadrant region inside the triangle
$g_xr+g_yt<\Delta$. The last two inequalities bound the partial
derivatives of $H_Q$ below on the containing rectangle $[0,U]\times[0,V]$,
making the removed region a lower set within the quadrant. The first two
inequalities place it below both strip cutoffs, so no strip rejection is
reversed on entering the quadrant. These conditions ensure a monotone
intersection rejection region. They are sufficient, and failure is inconclusive.
All quantities in this check are evaluated in floating-point arithmetic;
the reported passes are not rigorous numerical enclosures.

For Table~\ref{tab:robust-comparison}, $q,\mu_S,\mu_F$ are the prior weight and
noncentralities returned by the same coarse-grid prior search used for the
staircase comparison, with spacing $0.1$. The equal fixed-noncentrality
case uses $q=1/2$ by symmetry. The interval case at offset bound $0.25$
retains the initial prior after the recorded line-search failure. The
cutoff $k_G$ is then recalibrated for $G$ with those values held fixed.
We evaluate the resulting rule over the full noncentrality intervals;
neither this construction nor the local prior search establishes global
minimaxity of the comparator.

\paragraph{Constructing a feasible nonrandomised rule.}
The clinical computation collects upper staircases encountered while
optimising the prior weight. A finite linear programme mixes them subject
to the null rejection probability constraint and the two target-regret constraints. Each
component's regret derivative is negative at the corresponding lower
target noncentrality endpoint, so the regret-shape lemma identifies these endpoints
as its full-interval maxima. The same remains true for their mixture.
In the reported cases the linear programme uses three staircases and
equalises the two regrets.

Conditional Gaussian rearrangement then converts the mixture into a nonrandomised rule.
In coordinates $X=d_x(T+Z)$, $Y=d_y(T-Z)$, a staircase with upper-quadrant
corners $(x_i,y_i)$ has the explicit conditional threshold
\[
 h_r(z)=\min_i\max\{x_i/d_x-z,\ y_i/d_y+z\}.
\]
If its mixing weight is $\omega_r$, define
\[
 p(z)=\sum_r\omega_r\Phi\{(\beta z-h_r(z))/\sigma\},\qquad
 h_*(z)=\beta z-\sigma\Phi^{-1}\{p(z)\},
\]
where $T\mid Z=z\sim N(\beta z,\sigma^2)$ under the null. Reject the
intersection when $T>h_*(Z)$ and apply the marginal tests. The
rearrangement theorem preserves the null rejection probability and coordinatewise monotonicity and improves
every target power for a common admissible direction. The clinical
examples use $(d_x,d_y)=(1,1)$; the directional likelihood ratio derivatives are
strictly positive at both lower target noncentrality endpoints and hence throughout
the planning intervals. The resulting conditional boundary is fully
specified by the saved component corners and weights.

\paragraph{A reoptimised closed max-$Z$ comparator.}
We also reoptimise the cutoffs of the established closed max-$Z$ rejection region:
\[
 R(k_x,k_y)=\{x>k_x\}\cup\{y>k_y\},\qquad
 k_x,k_y\ge c,\qquad P_0\{R(k_x,k_y)\}=\alpha,
\]
followed by the same marginal tests. Its minimax member is obtained by
equalising the two full-interval worst regrets. This is a global optimum
within the calibrated closed max-$Z$ family. To verify the claim, take
$x_1<x_2$ and $y_1>y_2$ for two equally calibrated members of this family. Their lost
and gained rejection regions are respectively
\[
 L=\{x_1<x\le x_2,\ y\le y_2\},\qquad
 G=\{x\le x_1,\ y_2<y\le y_1\},
\]
with $P_0(L)=P_0(G)$. The subgroup likelihood ratio at the worst-case offset increases in $x$
and decreases in $y$, so it is at least its value at $(x_1,y_2)$
throughout $L$ and at most that value throughout $G$. The marginal subgroup
test rejects everywhere on $L$. Thus subgroup power decreases when
$k_x$ rises. The full-population likelihood ratio and its marginal test give the
reverse comparison, so its power increases. These inequalities hold
for every target noncentrality parameter, hence also for the two interval worst regrets.
Their continuity and the opposite endpoints corresponding to rejection based on only one statistic establish
the optimality of the equal-regret rule.

In the moderate and historical-point clinical scenarios, the reoptimised
closed max-$Z$ rules have cutoffs approximately $(2.210,2.172)$ and
$(2.218,2.165)$, respectively. Their worst regrets are
$6.188\%$ and $5.990\%$, compared with $6.190\%$ and
$5.992\%$ for the nonrandomised candidates. The bounds for the continuous problem use
the reoptimised closed max-$Z$ rules as their feasible upper endpoints. The nonrandomised
rules obtained from mixtures remain constructive demonstrations of the general
rearrangement. Numerical calibration of the closed max-$Z$ rules targets
$\alpha-2\times10^{-12}$, with independent adaptive integration of
size and both worst regrets.

\paragraph{Boundary-search results.}
The initial prior search at common offset bound 0.25 terminated with a line-search
failure. Its calibrated candidate is feasible, and the lower bound for the continuous problem remains valid at that prior. At offset bound 0.40, the original nonrandomised
cutoff interpolation gave a staircase with larger regret than both DT and
closed max-$Z$, despite a favourable discrete objective. These computations
motivated the use of conditional Gaussian rearrangement to remove randomisation from the grid mixture satisfying the null rejection probability constraint,
which preserves its null rejection probability and improves both target powers whenever the
common-direction condition holds. The sufficient monotonicity check for the conditionally averaged likelihood-ratio rule is not satisfied
at the larger offset bounds, so those candidates are omitted from the comparison.

\paragraph{Mesh and monotonicity diagnostics for the numerical illustration.}
For the interval experiment with $\alpha=0.025$, $\rho=\sqrt{0.4}$,
$I_S=[1.5,4]$ and $I_F=[2,4.5]$, the DT rule at common offset bound
$0.10$ has normalised corner excess $0.042$ over its fitted critical value,
and both partial derivatives are positive. Convexity then reduces the
pointwise sufficient condition throughout the quadrant to these corner
inequalities. With asymmetric uncertainty $e_S=0.25$, $e_F=0$, the
reoptimised step-up rule has normalised corner excess $0.013$ and two
positive partial derivatives. These checks use floating-point arithmetic.

The integration box has half-width 10, with base grid spacing 0.05 and
local spacing $5\times10^{-4}$ near active cutoffs, together with explicit
Gaussian tail allowances. At offset bound $0.80$, the last mesh increment
is about 0.005 percentage points, close to the remaining evaluated gap.
Supplementary Table~\ref{tab:continuum-refinement} reports all four meshes;
the refinement leaves the stability of the final displayed digits
unresolved. At offset bound $0.10$, the numerical pointwise sufficient
check supports DT independently of the lower-bound calculation. Its
evaluated gap of less than 0.001 percentage points measures slack in the
bounding construction, subject to that numerical check; arithmetic error
is outside the reported gap.

\paragraph{Complete numerical bounds and cutoff reoptimisation.}
Table~\ref{tab:continuum-certificate-full} retains the full bound and
last-mesh-increment diagnostics. Table~\ref{tab:retuning} gives the
cutoff reoptimisation calculations summarised in the main paper.

\begin{table}[p]
\centering\normalsize\setstretch{1}
\caption{Numerically evaluated regret bounds for the continuous problem, in percentage points.
The lower endpoint uses monotone grid-vertex values to bound all feasible rules;
the upper endpoint is the best evaluated nonrandomised regret.
The main rows use $\rho=\sqrt{0.4}$.
QUEST A and B use complement-effect and stratum-difference bounds
$(0.02,0.08)$ and $(0.005,0.125)$ L, respectively.
Endpoints are rounded to three decimal places; gaps and last increases use
unrounded values in units of $10^{-3}$ percentage points.
Values use floating-point quadrature, with diagnostics in Section~S10.
The last column records the increase in the lower bound as local mesh spacing
falls from $0.001$ to $5\times10^{-4}$; this is a refinement diagnostic, not an error estimate.}
\label{tab:continuum-certificate-full}
\begin{tabular}{lrrrr}
\toprule
Scenario & Lower bound & Feasible regret & \shortstack{Gap\\($10^{-3}$ pp)} & \shortstack{Last increase\\($10^{-3}$ pp)} \\
\midrule
$e=0.1$ & 8.122 & 8.122 & 0.472 & 0.464 \\
$e=0.25$ & 8.469 & 8.471 & 1.324 & 1.277 \\
$e=0.4$ & 8.704 & 8.707 & 2.812 & 2.175 \\
$e=0.8$ & 9.036 & 9.041 & 5.174 & 5.154 \\
QUEST A & 6.181 & 6.188 & 6.438 & 4.943 \\
QUEST B & 5.984 & 5.990 & 5.449 & 5.419 \\
\bottomrule
\end{tabular}
\end{table}

\begin{table}[p]
\normalsize\setstretch{1}\renewcommand{\arraystretch}{1}
\centering
\caption{Reoptimising step-up cutoffs under asymmetric uncertainty, $e_F=0$, using the interval settings in the main paper's numerical illustration. DT retains its original cutoffs. Regrets and gains are percentage points; bold marks the lowest displayed regret.}
\label{tab:retuning}
\begin{tabular}{rrrrrr}
\toprule
$e_S$ & $x_0$ & $y_0$ & DT (fixed) & Reoptimised & Gain \\
\midrule
0.00 & 2.208 & 2.208 & \textbf{7.855} & \textbf{7.855} & 0.000 \\
0.10 & 2.204 & 2.212 & 8.122 & \textbf{7.986} & 0.136 \\
0.25 & 2.198 & 2.218 & 8.479 & \textbf{8.157} & 0.322 \\
\bottomrule
\end{tabular}
\par\smallskip
\begin{minipage}{0.94\textwidth}\normalsize\setstretch{1}
Gains use unrounded values. Each reoptimised rule satisfies the numerically evaluated pointwise sufficient condition. DT's arm cutoff is approximately $2.208$ and $c\simeq1.960$. Numerical evaluations have no rigorous rounding or integration error bounds.
\end{minipage}
\end{table}

\FloatBarrier

\section{Reproduction of the asthma planning example}\label{supp:clinical}
The calculation uses the QUEST protocol's anticipated subgroup proportion
0.42 and the 200 mg versus matched-placebo results in the US prescribing
information, Section 14.2 and Table 26
\citep{BusseEtAl2018,CastroEtAl2018,DupixentLabel2026}. The treatment and control counts
were 631 and 317 overall, and 264 and 148 in the eosinophil subgroup.
Reported mean differences were 0.21 L (95\% confidence interval 0.13--0.29) in the subgroup and 0.14 L (0.08--0.19) overall.
The reported overall interval has width 0.11 L, giving the working standard
error $\kappa\simeq28.1$ mL under a normal-interval approximation. Because
the endpoints were rounded to hundredths, widths from 0.10 to 0.12 L
are consistent with that rounding; their standard errors range from
25.5 to 30.6 mL.
The target noncentrality ranges are $I_S\simeq[3.002,6.697]$ and $I_F\simeq[2.851,6.771]$; their joint coverage is unspecified.
The trial's repeated-measures joint covariance cannot be recovered from these aggregate summaries.

We specify independent stratum estimators for a future experiment with
standard errors $\kappa/\sqrt{0.42}\simeq43.3$ mL and
$\kappa/\sqrt{0.58}\simeq36.8$ mL. Their weighted full-population
estimator has standard error $\kappa$ and correlation $\sqrt{0.42}$ with
the subgroup statistic. At the historical arm counts this precision
corresponds to an equivalent independent-outcome standard deviation
0.408 L. This value expresses the assumed planning precision on an
independent-outcome scale; the trial's repeated-measures residual variance
remains unspecified. The subgroup interval width instead
gives 40.8 mL; our planning standard error is about 6.1\% larger. The
observed subgroup fraction is $412/948\simeq0.435$; proportional recruitment with a common per-observation variance
for planning at that fraction would give correlation approximately 0.659 rather than 0.648.

Let $h_C$ bound the absolute complement effect in subgroup-target states and $h_H$ the absolute difference between stratum effects in full-population-target states, both in litres.
The planning sets in outcome units are
\[
 \begin{aligned}
 \text{Subgroup target:}\quad&
 \Delta_S\in[0.13,0.29],\quad |\Delta_C|\le h_C,\\
 \text{Full-population target:}\quad&
 \Delta_F\in[0.08,0.19],\quad |\Delta_S-\Delta_C|\le h_H.
\end{aligned}
\]
Using the same units for $\kappa$ and the widths gives
\[
 e_S=sh_C/\kappa,\qquad e_F=\rho s h_H/\kappa.
\]
The endpoint ranges use the published marginal intervals as reproducible
ranges for sensitivity analysis. Simultaneous coverage of their Cartesian product
with the analyst-selected widths is unspecified. With
$d=\Delta_S-\Delta_C$, the identities
$\Delta_S=\Delta_F+(1-\pi)d$ and
$\Delta_C=\Delta_F-\pi d$ yield the minima over full-population-target states:
\[
 \min\Delta_S=0.08-0.58h_H,\qquad
 \min\Delta_C=0.08-0.42h_H.
\]
Combining these minima with $\max\Delta_C=h_C$ over subgroup-target states gives the outcome-scale
benefit--burden conditions in \eqref{eq:priorityconditions}, which require
positive net benefit in every treated stratum. In particular,
\begin{equation}\label{eq:clinical-frontier}
0\le\tau_S<\min\{0.13,0.08-0.58h_H\},\qquad
h_C<\tau_C<0.08-0.42h_H.
\end{equation}
Burdens are fixed in outcome units, so this compatibility condition is independent of sample size.
A stratum-specific pair of nonnegative burdens
with strictly positive net benefit in each treated stratum exists exactly
when
\[
 h_C+0.42h_H<0.08,\qquad 0.58h_H<0.08.
\]
Under this requirement, for a common burden the upper endpoint is
$0.08-0.58h_H$, so its nonempty interval is
$(h_C,0.08-0.58h_H)$.
Equivalently, a common burden exists exactly when $h_C+0.58h_H<0.08$.
For example, widths 0.04 and 0.08 L admit
stratum-specific burdens, but no common burden; widths 0.06 and 0.08 L
admit neither under the requirement of positive net benefit in every
treated stratum. These failures do not by themselves rule out the weaker
aggregate-value conditions in Section~S8. All rows satisfying the stronger
conditions have $J_0>0$; the aggregate-value conditions alone do not imply
this inequality.

For an illustrative reconstruction of the future planning estimand,
substituting $\pi=0.42$ into the additive identity gives
\[
\Delta_C^{\rm rec}=\frac{0.14-0.42(0.21)}{0.58}\simeq0.089\ {\rm L},
\qquad d^{\rm rec}\simeq0.121\ {\rm L}.
\]
This reconstruction transforms reported point estimates under the future
planning estimand; its complement effect is a derived quantity. The
overall and subgroup adjusted contrasts need not obey an exact stratum
decomposition across their repeated-measures fits. Full-precision derived
values are retained in the code; Section~S13 varies the planning prevalence.
The protocol's detectable overall effect of 0.15 L was a power-planning
input; a burden and a minimum clinically important difference require
separate clinical specification.

The script \texttt{clinical\_calibration.py} writes
\texttt{results/clinical\_calibration.json} and the table used in the main
paper. It checks the mapping from treatment effects to test-statistic means on every scenario endpoint and
computes regrets at worst-case offsets by one-dimensional quadrature. For the
equalised step-up procedure, the two target-specific regrets agree; their worst target noncentrality parameters
are the lower endpoints of $I_S$ and $I_F$ in every displayed row. Independent adaptive and Gauss--Legendre evaluations, including the reoptimised
closed max-$Z$ rule, differed by at most $8.4\times10^{-17}$. The null rejection probability discrepancy
for the calibrated step-up procedure, symmetric closed max-$Z$ and
Dunnett--Tamhane rules was below $2.1\times10^{-14}$. Each reoptimised closed max-$Z$ rule
uses a null rejection probability constraint of $0.025-2\times10^{-12}$; integrating in either
coordinate gives a discrepancy below $1.5\times10^{-15}$ from that constraint.
Table~2 in the main paper distinguishes the reoptimised and symmetric closed max-$Z$ rules. Table~\ref{tab:clinical-burden-calibration} additionally gives the full common-burden intervals.
Supplementary Section~S10 supplies the continuum comparison.

\subsection{Numerical comparisons and the reconstructed-effect scenario}

Table~2 of the main paper compares established testing families,
with the step-up and closed max-$Z$ procedures reoptimised separately for each row.
At $(h_C,h_H)=(0.02,0.08)$ L, reoptimising closed max-$Z$ reduces worst regret
from 6.760 to 6.188 percentage points. This 0.573-point gain is within
the closed max-$Z$ family. The reoptimised closed max-$Z$ procedure gains
0.279 points over the reoptimised
step-up procedure and 0.997 over DT with its original cutoffs. The gains use distinct baselines
and are interpreted separately.

The reoptimised closed max-$Z$ procedure is the best evaluated rule in this scenario.
The evaluated lower bound for the continuous problem, 6.181,
leaves a 0.006-point gap (Table~1 of the main paper).
Equalising the two worst regrets minimises regret within the calibrated
closed max-$Z$ family (Supplementary Section~S10). A randomised mixture
of staircase rejection regions satisfies the intersection-test size constraint.
Replacing it by the nonrandomised rule obtained through conditional
rearrangement preserves its size and improves both target powers, giving
regret 6.190.

Figure~4 in the main paper shows the corresponding target-specific
power curves at the worst-case offsets. At the lower endpoints of the
noncentrality intervals, the reoptimised closed max-$Z$ procedure gives subgroup
and full-population target powers of
78.949\% and 75.163\%, versus marginal reference powers of 85.137\% and
81.351\%. Each target-specific shortfall for this procedure is 6.188
percentage points at its respective state. Relative to symmetric closed max-$Z$, reoptimisation gains 0.573
points of full-population power and gives up 0.503 points of subgroup
power, shifting allocation towards the alternative family with the largest regret.

Including the reconstructed point
in full-population-target states requires $h_H\ge7/58$ L (approximately 0.121 L) and hence $h_C<0.010$ L for a common
burden satisfying the requirement of positive net benefit in every treated
stratum. The row $(h_C,h_H)=(0.005,0.125)$ L admits common burdens
strictly between 5 and 7.5 mL; its reoptimised closed max-$Z$ procedure
has regret 5.990 points and a gap from the evaluated lower bound of
0.005 points (Figure~\ref{fig:clinical-planning}).

Converting these FEV$_1$
summaries to a clinical burden
requires additional valuation of exacerbations, lung function, adverse
effects and treatment costs.

\begin{table}[p]
\centering\normalsize\setstretch{1}
\setlength{\tabcolsep}{3pt}
\caption{QUEST-informed sensitivity calculations. Bold entries have the smallest regret among the four procedures in each row at the displayed precision. The complement bound limits its absolute effect in subgroup-target states; the stratum-difference bound limits the absolute difference between stratum effects in full-population-target states. Bounds and common-burden intervals are in mL; regret is in percentage points. The step-up and closed max-$Z$ cutoffs are reoptimised separately for each row, with the step-up procedure equalising the two worst regrets. Symmetric closed max-$Z$ uses equal cutoffs; both max-$Z$ rules apply the marginal tests. The widths are planning choices. An empty interval means no common burden preserves both target population designations with positive net benefit in each treated stratum. The last row also fails this requirement with stratum-specific burdens. These feasibility statements use the additional requirement on every treated stratum; aggregate value maximisation alone has weaker conditions.}
\label{tab:clinical-burden-calibration}
\begin{tabular}{rrrrrrr}
\toprule
\shortstack{Complement\\effect\\bound} & \shortstack{Stratum\\difference\\bound} & \shortstack{Common\\burden} & \shortstack{Equal-regret\\step-up} & \multicolumn{2}{c}{Closed max-$Z$} & DT \\
\cmidrule(lr){5-6}
& & & & Reoptimised & Symmetric & \\
\midrule
10 & 40 & $(10,56.8)$ & 6.105 & \textbf{5.983} & 6.561 & 6.798 \\
20 & 80 & $(20,33.6)$ & 6.466 & \textbf{6.188} & 6.760 & 7.185 \\
40 & 80 & $\varnothing$ & 6.686 & \textbf{6.312} & 6.760 & 7.185 \\
5 & 125 & $(5,7.5)$ & 6.173 & \textbf{5.990} & 6.798 & 7.276 \\
60 & 80 & $\varnothing$ & 6.764 & \textbf{6.350} & 6.760 & 7.185 \\
\bottomrule
\end{tabular}
\end{table}

\FloatBarrier

\section{Additional numerical comparisons}\label{supp:additional}
Tables~\ref{tab:robust-comparison} and \ref{tab:grid-refinement} compare
procedures and report calculations on successively finer grids.
Their finite-grid objectives concern optimisation restricted to grid rules;
Table~\ref{tab:continuum-refinement} reports refinement of the lower bound
for the continuous problem proved in Section~S10.
For the robust cubic expansion (Theorem~\ref{supp:robustcubic} and
Figure~\ref{fig:retuning-cubic}(B)), a common offset bound $0.05$ and fixed, equal target noncentrality parameters give
a transition near $2.524$ and coefficient $C\simeq1.156\times10^{-2}$ in probability units.
Here $h-h_0$ is the departure of the common target noncentrality from the
transition, whereas $d=1-\exp\{-(h-h_0)(t-c)\}$ is the normalised corner
likelihood-ratio deficit used in the cubic expansion. As $h-h_0$ decreases
from $0.20$ to $1/80$, the ratio of the actual regret loss to $Cd^3$ increases from $0.795$ to $0.986$;
monotonicity holds throughout.

\renewcommand{\thetable}{S\arabic{table}}
\begin{table}[p]
\normalsize\setstretch{1}\renewcommand{\arraystretch}{1}
\centering
\caption{Worst target regret (percentage points) under common offset bound $e$. The interval experiment uses $I_S=[1.5,4]$ and $I_F=[2,4.5]$; the fixed-noncentrality experiment uses $\mu_S=\mu_F=3.2$. All entries use $\alpha=0.025$ and $\rho=\sqrt{0.4}$. The non-target mean offsets remain uncertain in both experiments.}
\label{tab:robust-comparison}
\begin{tabular}{rrrrr}
\toprule
$e$ & DT & \shortstack{Symmetric\\max-$Z$} & Grid candidate & \shortstack{Averaged\\likelihood ratio} \\
\midrule
\multicolumn{5}{l}{\emph{Intervals}} \\
0.10 & \textbf{8.122} & 8.280 & \textbf{8.122} & \textbf{8.122} \\
0.25 & 8.479 & 8.515 & \textbf{8.471} & \textbf{8.471} \\
0.40 & 8.783 & \textbf{8.707} & 8.810 & --- \\
0.80 & 9.355 & \textbf{9.041} & \textbf{9.041} & --- \\
\addlinespace
\multicolumn{5}{l}{\emph{Fixed target noncentralities, $\mu_S=\mu_F=3.2$}} \\
0.10 & 4.365 & 4.414 & 4.365 & \textbf{4.360} \\
0.25 & 4.557 & 4.541 & \textbf{4.530} & --- \\
0.40 & 4.722 & 4.644 & \textbf{4.643} & --- \\
0.80 & 5.032 & \textbf{4.823} & \textbf{4.823} & --- \\
\bottomrule
\end{tabular}
\par\smallskip
\begin{minipage}{0.97\textwidth}\normalsize\setstretch{1}
Bold entries have the smallest regret in each row at the displayed precision; shared bold entries need not be equal before rounding. Grid candidates are calibrated nonrandomised monotone rules, evaluated over continuous target noncentrality intervals; their optimality in the continuous problem has not been established. The final column uses the conditionally averaged likelihood-ratio rule defined in Section~S10. A dash indicates that its sufficient monotonicity check is not satisfied, so no feasible risk is reported. The optimisation over the prior for the interval experiment at $e=0.25$ has a recorded line-search failure. At $e=0.40$, the calibrated staircase rule has greater regret than symmetric closed max-$Z$. Values and monotonicity checks use floating-point arithmetic. DT denotes the two-test Gaussian Dunnett--Tamhane step-up procedure.
\end{minipage}
\end{table}

\begin{table}[p]
\normalsize\setstretch{1}\renewcommand{\arraystretch}{1}
\centering
\caption{Grid refinement in the interval experiment at common offset bound $0.8$. Regrets are in percentage points. The prior is reoptimised on the finer grid and then held fixed for the final two resolutions.}
\label{tab:grid-refinement}
\begin{tabular}{rrrr}
\toprule
Spacing & Grid Bayes regret & Calibrated worst regret & Null size \\
\midrule
0.025 & 9.046 & 9.041 & 0.025 \\
$1/80$ & 9.042 & 9.041 & 0.025 \\
\bottomrule
\end{tabular}
\par\smallskip
\begin{minipage}{0.94\textwidth}\normalsize\setstretch{1}
Grid Bayes regret is the fixed-prior value when optimisation is restricted to randomised grid rules. It is not a lower bound for minimax regret in the continuous problem. Both calibrated rules coincide with symmetric closed max-$Z$; independent adaptive integration gives worst regret $9.041$ and null size $0.025$. Agreement under grid refinement does not establish optimality in the continuous problem.
\end{minipage}
\end{table}

\begin{table}[p]
\normalsize\setstretch{1}\renewcommand{\arraystretch}{1}
\centering
\caption{Rounded summaries of lower bounds on regret in the continuous problem under local mesh refinement, in
percentage points. Column headings give the local spacing on $[c,2.4]$;
the box is $[-10,10]^2$ with base spacing $0.05$. Entries are rounded to three decimal places. Each entry evaluates
the bound based on monotone grid-vertex values in Section~S10. QUEST A and B are the
clinical scenarios in the clinical calibration table in the main paper.}
\label{tab:continuum-refinement}
\begin{tabular}{lrrrr}
\toprule
Scenario & $0.004$ & $0.002$ & $0.001$ & $5\times10^{-4}$ \\
\midrule
$e=0.1$ & 8.118 & 8.120 & 8.121 & 8.122 \\
$e=0.25$ & 8.461 & 8.466 & 8.468 & 8.469 \\
$e=0.4$ & 8.689 & 8.698 & 8.702 & 8.704 \\
$e=0.8$ & 9.000 & 9.020 & 9.031 & 9.036 \\
QUEST A & 6.147 & 6.167 & 6.176 & 6.181 \\
QUEST B & 5.946 & 5.968 & 5.979 & 5.984 \\
\bottomrule
\end{tabular}
\par\smallskip
\begin{minipage}{0.97\textwidth}
The entries use quadrature order 24. At the finest mesh, changing to order
40 gives a largest observed change of $5.55\times10^{-16}$
to the displayed precision. The largest evaluated bound on alternative probability
outside the box is $8.98\times10^{-13}$; both diagnostics are
in probability units. The order comparison is a numerical check, not a
rigorous numerical error bound.
\end{minipage}
\end{table}

\FloatBarrier
\renewcommand{\thefigure}{S\arabic{figure}}
\begin{figure}[p]
\normalsize\setstretch{1}
\centering
\includegraphics[width=\textwidth]{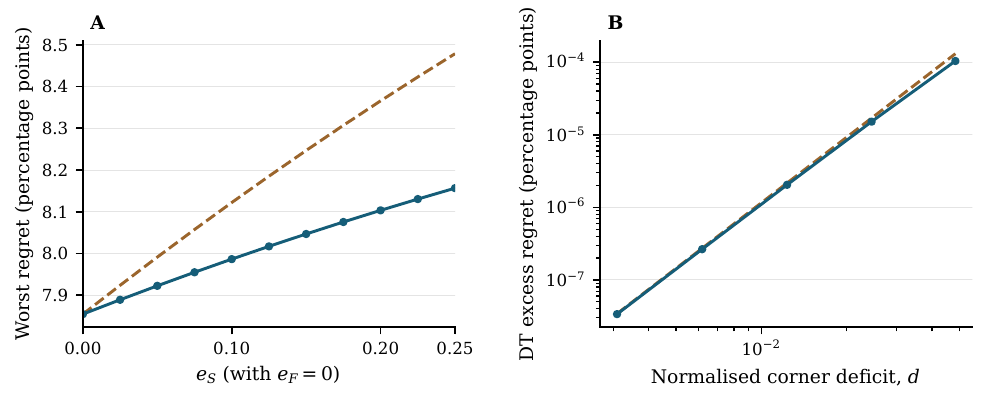}
\caption{Reoptimisation and the local cost of retaining DT, with $\alpha=0.025$ and $\rho=\sqrt{0.4}$. A: worst regret for DT with its original cutoffs (dashed brown), and the step-up procedure with reoptimised cutoffs (solid blue with points), under $e_F=0$ and the target noncentrality intervals $I_S=[1.5,4]$, $I_F=[2,4.5]$. The numerically evaluated pointwise sufficient optimality condition is satisfied at each displayed reoptimised rule. B: DT excess regret (solid blue with points) and the leading term $Cd^3$ (dashed brown), for fixed, equal target noncentrality parameters approaching a transition near $2.524$ from above at fixed common offset bound $0.05$. Here $d$ is the normalised deficit of the weighted likelihood ratio statistic at the corner and $C\simeq1.156\times10^{-2}$ in probability units. Both axes in B are logarithmic. This panel illustrates the local robust cubic result; its numerical evaluations do not include rigorous rounding and integration error bounds.}
\label{fig:retuning-cubic}
\par\smallskip\begin{minipage}{\textwidth}
\textbf{Alt text:} Left: both regret curves rise with uncertainty, with a
slower increase for the reoptimised step-up procedure. Right: on logarithmic scales,
DT's excess regret approaches the cubic approximation as the normalised
deficit of the weighted likelihood ratio statistic at the corner decreases towards zero.
\end{minipage}
\end{figure}

\FloatBarrier

\section{Sensitivity to correlation and prevalence}\label{supp:sensitivity}
\subsection{Rules with fixed cutoffs under a perturbed correlation}
Here the procedures retain their original cutoffs. We retain the design correlation $\rho_0=\sqrt{0.4}$, the target noncentrality
intervals $I_S=[1.5,4]$, $I_F=[2,4.5]$ and the mean neighbourhoods specified
at $\rho_0$. Only the covariance correlation changes to
$r\in\{\rho_0-0.05,\rho_0,\rho_0+0.05\}$; marginal variances remain one.
For each common offset bound $e\in\{0.25,0.80\}$, the means at worst-case offsets are therefore
$(\mu_S,\rho_0\mu_S-s_0e)$ and $(\rho_0\mu_F-s_0e,\mu_F)$, where
$s_0=\sqrt{1-\rho_0^2}$, at every value of $r$.
The marginal critical value and both procedures' cutoffs remain fixed:
$c\simeq1.960$, $t_{\rm DT}\simeq2.208$ and
$t_{\max Z}\simeq2.193$. These displayed approximations are rounded; all calculations
retain the unrounded, originally calibrated cutoffs. The saved reoptimised members of these two families
are symmetric for these scenarios and give the same cutoffs to numerical
precision. We report each rule once.

Coordinatewise monotonicity still places the worst intersection-null rejection probability
at the origin, while the marginal tests bound partial-null false rejection
by $\alpha$. For these two rules the partial-null supremum is $\alpha$,
so their strong FWER supremum is the larger of $\alpha$ and the origin
probability. Table~\ref{tab:correlation-sensitivity} presents the lower,
design and higher correlations side by side: reducing
the true correlation by 0.05 raises the origin probabilities from 2.5\%
to 2.537\% for DT and 2.550\% for max-$Z$.
Increasing it by 0.05 lowers them to 2.457\% and 2.442\%, while
the strong FWER supremum remains 2.5\% for both rules.
Reduced correlation lowers regret while inflating error, so these changes
compare different error levels. DT retains lower
worst regret than max-$Z$ at $e=0.25$, and max-$Z$ retains lower worst
regret at $e=0.80$, across these three correlations.

Coordinatewise monotonicity still makes the lower non-target offset the worst case. The
regret-shape argument requires a new calculation because, on the
subgroup-target family,
\[
 Y\mid X=x\sim
 N\{rx+(\rho_0-r)\mu_S-s_0e,\,1-r^2\}.
\]
We consequently maximise each target-specific regret over its full target noncentrality
interval by locating derivative sign changes and sampled local maxima,
including both endpoints, on grids of 501 and 2001 noncentrality values and
refining the resulting brackets continuously. The maxima agree within
$4.2\times10^{-17}$ in probability units. Independent integration of
rejection power over the other coordinate, followed by a separate
bracketed search, agrees within $8.4\times10^{-17}$; sum--difference
coordinates independently check the null probabilities within
$1.8\times10^{-17}$. These agreements measure floating-point consistency;
rigorous global numerical error bounds have not been computed. The procedures
retain their original calibration throughout.

\begin{table}[htbp]
\centering
\normalsize\setstretch{1}\renewcommand{\arraystretch}{1}
\caption{Correlation sensitivity with all cutoffs and efficacy mean neighbourhoods fixed at their design values, $\rho_0=\sqrt{0.4}$. Bold identifies the lower regret at the design correlation; perturbed-correlation columns have different error levels. Intersection-null sizes are percentages and do not depend on $e$; worst target regrets are percentage points and coincide for the two target families in these examples.}
\label{tab:correlation-sensitivity}
\begin{tabular}{lrrr}
\toprule
Rule & $\rho_0-0.05$ & $\rho_0$ & $\rho_0+0.05$ \\
\midrule
\multicolumn{4}{l}{Intersection-null size (\%)} \\
DT & 2.537 & 2.500 & 2.457 \\
Symmetric closed max-$Z$ & 2.550 & 2.500 & 2.442 \\
\addlinespace
\multicolumn{4}{l}{Worst target regret, $e=0.25$ (percentage points)} \\
DT & 8.366 & \textbf{8.479} & 8.616 \\
Symmetric closed max-$Z$ & 8.420 & 8.515 & 8.626 \\
\addlinespace
\multicolumn{4}{l}{Worst target regret, $e=0.80$ (percentage points)} \\
DT & 9.270 & 9.355 & 9.448 \\
Symmetric closed max-$Z$ & 8.985 & \textbf{9.041} & 9.100 \\
\bottomrule
\end{tabular}
\end{table}

\subsection{Outcome-scale prevalence sensitivity}
This calculation varies the prevalence in the illustrative additive
planning reconstruction, holding the reported subgroup and overall
contrasts at 0.21 and 0.14 L and retaining the outcome-scale planning
ranges. The complement effect is reconstructed algebraically from these
aggregate contrasts.
For $0.40\le\pi\le0.45$, the minimum heterogeneity bound needed to
contain the reconstructed effect estimate is
\[
 h_H^{\min}=\frac{0.21-0.14}{1-\pi}.
\]
Since $\pi<1/2$, the smaller stratum benefit throughout the full-population-target states is bounded
below by $0.08-(1-\pi)h_H$. Requiring positive net benefit in every
treated stratum, in addition to the designated targets maximising
aggregate value, makes the common-burden condition at $h_H=h_H^{\min}$
reduce to
\[
 h_C<0.08-(1-\pi)h_H^{\min}=0.010\ \mathrm{L}.
\]
Thus the 10 mL restriction persists across this prevalence range under
the common-burden model with positive net benefit in every treated
stratum and the same additive planning reconstruction;
using a larger heterogeneity bound makes the restriction stricter.

The fixed illustrative widths $(h_C,h_H)=(5,125)$ mL require a separate
check. At $\pi=0.40$ their strict common-burden interval is empty, and at
$\pi=0.45$ they exclude the reconstructed effect estimate.
Both inclusion of the reconstructed effect estimate and common-burden
feasibility under the additional requirement on every treated stratum hold exactly when
$0.40<\pi\le0.44$ within the examined range.
Table~\ref{tab:prevalence-sensitivity} includes the anticipated
prevalence 0.42 and the count-based fraction $412/948$ from the
200 mg and matched-placebo groups. These prevalence calculations update
the algebraic target conditions while holding the testing designs fixed.
The script \texttt{build\_sensitivity\_analysis.py}
computes the sensitivity quantities and saves the full-precision results in
\texttt{results/sensitivity\_analysis.json}; exact rational arithmetic
also checks every listed prevalence boundary.

\begin{table}[htbp]
\centering
\normalsize\setstretch{1}\renewcommand{\arraystretch}{1}
\caption{Prevalence sensitivity of the additive planning reconstruction. Effects, bounds on between-stratum effect differences and burden endpoints are in mL. The $h_C$ bound uses the minimum $h_H$ in the same row. The last two columns indicate whether the reconstructed estimate is included and whether a common burden preserves the target designations with positive net benefit in every treated stratum at fixed widths $(h_C,h_H)=(5,125)$ mL. The displayed prevalence 0.435 represents the observed proportion $412/948$; the calculations use this unrounded fraction. Parentheses denote an interval with strict endpoints.}
\label{tab:prevalence-sensitivity}
\begin{tabular}{rrrrcc}
\toprule
$\pi$ & $\Delta_C^{\rm rec}$ & Minimum $h_H$ & Bound on $h_C$ & Point included & Common burden \\
\midrule
0.40 & 93.3 & 116.7 & $<10$ & Yes & Empty \\
0.42 & 89.3 & 120.7 & $<10$ & Yes & $(5,7.50)$ \\
0.435 & 86.2 & 123.8 & $<10$ & Yes & $(5,9.32)$ \\
0.44 & 85.0 & 125.0 & $<10$ & Yes & $(5,10.00)$ \\
0.45 & 82.7 & 127.3 & $<10$ & No & $(5,11.25)$ \\
\bottomrule
\end{tabular}
\end{table}

\begin{figure}[p]
\normalsize\setstretch{1}\centering
\includegraphics[width=\textwidth]{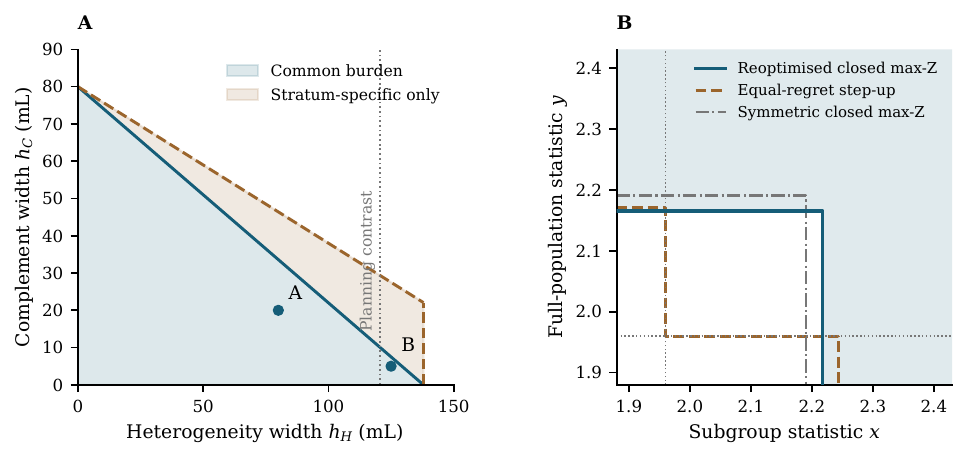}
\caption{QUEST-informed planning. A: uncertainty widths admitting the stated
population targets under the additional requirement of positive net benefit
in every treated stratum; aggregate value maximisation alone imposes weaker
conditions. Blue shading
admits a common burden; brown shading requires stratum-specific burdens.
The dotted line is the illustrative additive planning contrast. Points A and B correspond to uncertainty widths $(20,80)$ and $(5,125)$ mL, respectively.
B: the reoptimised closed max-$Z$ intersection boundary at point B, with its rejection region
shaded, compared with the equal-regret step-up procedure and symmetric closed max-$Z$.
Each elementary hypothesis is rejected only when its marginal test also rejects.}
\label{fig:clinical-planning}
\par\smallskip\begin{minipage}{\textwidth}
\textbf{Alt text:} Under the requirement of positive net benefit in every
treated stratum, the common-burden region narrows as heterogeneity uncertainty increases. The reconstructed planning contrast lies near its
lower-right edge. A second panel compares the reoptimised closed max-$Z$, symmetric
closed max-$Z$ and equal-regret step-up boundaries at scenario B.
\end{minipage}
\end{figure}

\FloatBarrier
\section{Relation to optimal-testing and shortcoming results}
\label{supp:literaturecomparison}

The present Gaussian regret problem brings together optimal-testing,
shortcoming and order-constrained optimisation results. It uses labelled
subgroup and full-population targets, compact positive target noncentrality intervals,
and finite non-target offset bounds in a fixed experiment. Proposition~1
of the main paper equates the minimax value over nonrandomised monotone
decision pairs $\mathcal D_{\rm det}$ with that over the shared-intersection
class $\mathcal U_{\rm det}$ through domination. Allowing the shared
intersection test to take values in $[0,1]$ gives the convex relaxation
$\mathcal U$, with the marginal tests fixed. Under the common-direction
condition, the randomised and nonrandomised minimax values coincide, and the
optimal shared-intersection rule is unique and nonrandomised.

\subsection{Objectives and the meaning of discovery}
\citet[Sections~2.1--2.8]{BelitskayaLevyEtAl2018} study simultaneous inference
for an overall population and a prespecified marker-positive subgroup.
Their formulation assumes ordered treatment effects and adds a safety boundary
for the marker-negative subgroup. Its three possible decisions are rejection
of the subgroup null alone, rejection of both nulls, and rejection of neither.
The COSST-O variant chooses cutoffs to maximise the probability of at least
one rejection at specified alternatives. Our success event is rejection of
the population claim designated by the prespecified planning assignment at
each state. Thus, for a full-population target, subgroup rejection alone receives
zero credit. Regret compares the designated claim's power with its marginal
reference power over the continuous planning ranges. Strong FWER control
here is required for every Gaussian mean pair, including unrestricted nuisance
means under partial nulls. Section~S8 develops the benefit--burden interpretation and compatibility conditions for the planning assignments; the main-paper asthma example reports the resulting testing comparisons.

The state-dependent loss formulation of \citet{rly2014} represents our
regret on the target-labelled state space by taking the loss to be marginal reference
power minus the indicator of rejecting the designated claim. A common
additive constant makes this loss nonnegative and preserves the optimiser.
At a fixed prior, minimising average regret is equivalent to maximising weighted target
power. Under a worst-case objective, the state-dependent reference-power term
sets the benchmark at each efficacy state. The Gaussian
calculation of conditional expected loss then identifies the regret shape and the
states that determine the optimal compromise.

\citet[Section~3.6, equation~(3.8) and Theorem~3.1]{Rosenblum2014UMP}
optimises the probability of rejecting the overall null and at least one
subpopulation null simultaneously. Under the normal known-variance model, at level $0.05$,
the procedure attains the marginal overall-test power while making an
additional subgroup rejection. Section~3.7 also considers the more specific
success event of rejecting a benefiting subgroup while avoiding false
rejections. These criteria illustrate how the discovery event and the
class of procedures determine the power benchmark. Our criterion
credits a particular labelled population, and our decision class requires
coordinatewise monotonicity in the subgroup and overall statistics. The
resulting minimax regret quantifies the shortfall for uniform protection
of those targets.

\subsection{Consonance, monotonicity and stepwise optimality}
Consonance requires rejection of an intersection hypothesis to lead to
rejection of at least one of its constituent elementary hypotheses. Coordinatewise monotonicity requires each decision to remain
nondecreasing when either observed statistic increases. These properties
play separate roles here: the shared intersection rule is supported on the
union of the marginal rejection regions, and its test function is
coordinatewise nondecreasing.
\citet[Theorems~2.1, 3.1 and~4.1]{RomanoShaikhWolf2011} develop closure
representations, consonant versions of closed procedures and maximin results
for the probability of at least one elementary rejection.
\citet{GoemanHemerikSolari2021} place closed testing in a broader admissibility
framework for simultaneous inference, including FWER control. Proposition~1
of the main paper uses the Gaussian extreme-nuisance argument to fix the
marginal rejection requirements, then constructs a dominating monotone
closure. Its converse concerns nonrandomised coordinatewise monotone
decisions. The randomised class $\mathcal U$ retains fixed marginal tests
and a shared monotone intersection decision.

\citet[Propositions~2--3]{BittmanEtAl2009} construct maximin consonant tests,
including optimisation of an intersection test with the marginal tests fixed.
Their common effect direction refers to treatment effects with a common
sign. The common increasing direction in our conditional Gaussian
construction is a direction in observation space along which the relevant
likelihood ratios increase. Their maximin alternative sets and those in
\citet{RomanoShaikhWolf2011} already include continuous ranges of effects.
Our regret-shape argument identifies the states supporting the least-favourable
prior for the labelled target-power shortfall.
\citet{lrs2005} establish stepdown and stepup maximin results under their
monotonicity assumptions. Their decision-rule condition preserves the rejection
set when rejected coordinates increase and unrejected coordinates decrease;
our condition makes each decision nondecreasing in every coordinate.
\citet{ZehetmayerKoenigPosch2024} construct consonant improvements of closed
tests based on symmetric $p$-value combinations, starting with independent
$p$-values and discussing exchangeable dependence. These results provide
context for optimising a shared intersection test and for assessing familiar
procedures under a specified power objective.

\Needspace{10\baselineskip}
\subsection{Result-by-result comparison}
\begin{longtable}{@{}>{\raggedright\arraybackslash}p{.25\textwidth}>{\raggedright\arraybackslash}p{.71\textwidth}@{}}
\caption{Related results and the additional argument in the present model.}
\label{tab:theoremcomparison}\\
\toprule
Present result & Related result and mathematical relationship \\
\midrule
\endfirsthead
\toprule
Present result & Related result and mathematical relationship \\
\midrule
\endhead
\bottomrule
\endfoot
Marginal-test reduction &
\citet[Section~3.2, Lemma~1]{hkr2023} establish the marginal rejection requirements by monotonicity
and extreme alternatives. Closure and consonance have general representations
\citep{RomanoShaikhWolf2011}. Here the extreme-nuisance argument applies to the correlated
Gaussian experiment; a shared monotone intersection test function then reduces strong FWER
to the null rejection probability at the origin.
\\[8pt]
Regret shape and two-point least-favourable prior &
\citet[Theorem~4.1 and p.~670]{VanZwetOosterhoff1967} connect Gaussian
shortcoming shape with a two-point prior in a symmetric global-testing
problem. Our conditional expected loss argument covers randomised monotone rules satisfying the marginal rejection requirements,
a positive limiting power shortfall, asymmetric intervals and offsets, giving
one worst target noncentrality parameter per target population.
\\[8pt]
Conditional rearrangement and uniqueness &
\citet[Definition~1.1 and Proposition~1.2]{Ehrhard1983} supplies Gaussian
symmetrisation; corrected Supplement~S7.2 of \citet{rosset2022} supplies
isotonic optimisation \citep[2024]{RossetCorrection2024}. Under the
common-direction condition, averaging conditional probabilities preserves
the marginal rejection requirements and coordinatewise monotonicity. Reconstruction gives a nonrandomised
rule with unchanged null rejection probability at the origin and at least as much power for every
target; strict concavity gives uniqueness.
\\[8pt]
Integral criterion for step-up optimality &
Stepwise maximin optimality appears in \citet{lrs2005}, and
familiar-rule optimality in \citet[Section~3.3]{hkr2023};
signed-integral order optimisation has classical antecedents \citep{O01}.
Here optimisation of the regions where only one marginal test rejects gives the exact cost
of removing a lower subset of the quadrant. Calibration and equalisation at the procedure's
own worst target noncentrality parameters give the stated minimax equivalence.
\\[8pt]
Continuum lower bound &
\citet[Section~6.3, equation~(17)]{rly2014} use dual variables from a finite linear programme
to bound the original continuous problem. Our bound based on monotone grid-vertex
values uses the same principle while retaining within-cell
monotonicity. The added step is the cell inequality with its convergence
and integration-error analysis.
\\
\end{longtable}

\subsection{Eliminating randomisation}
\citet{DvoretzkyWaldWolfowitz1950,DvoretzkyWaldWolfowitz1951} eliminate
randomisation while matching risks for finite collections of atomless
probability measures and finite action spaces. Under the common-direction
condition, Theorem~2 of the main paper preserves null size, coordinatewise
monotonicity and the marginal rejection requirements while weakly improving
all target powers over the continuous planning sets. The likelihood ratios
increase along common conditional Gaussian lines; reconstruction by upper
tails preserves feasibility and improves power. This construction belongs
to the Gaussian symmetrisation tradition of \citet{Ehrhard1983}. Strict
concavity supplies the Bayes and minimax uniqueness conclusions under the
stated assumptions.

\subsection{Test statistics and monotone optimisation}
Section~3 of \citet{hkr2023} develops the explicit construction of test statistics for
independent global-null $p$-values. Their general framework and discussion
also allow dependence. Under its distributional assumptions, Theorem~2
establishes optimality in the monotone class when the objective
coefficients have the required coordinatewise order. The likelihood ratios at worst-case offsets
in our Gaussian problem satisfy
\[
 \partial_y\log L_S=-e_S/s,\qquad
 \partial_x\log L_F=-e_F/s.
\]
The weighted likelihood ratio statistic, including marginal rejection indicators, can consequently decrease in the other coordinate.
Under the common-direction condition, both likelihood ratios increase along a
direction compatible with the Gaussian covariance. Conditional
probabilities along this direction give a convex representation of monotone
rejection regions satisfying the marginal rejection requirements and yield a nonrandomised rearrangement that improves all target powers.

\citet[Propositions~1--2 and Theorem~2]{rosset2022} provide analytical
optimality, multiplier existence under nonredundancy, and maximin verification
results as well as computational methods. The
corrected Supplement~S7.2 retains the stated distributional assumptions and
symmetry and likelihood-ratio ordering restrictions. Equations~(S7.1)--(S7.2)
incorporate coordinatewise monotonicity through an optimisation subject to monotonicity constraints;
the sufficient maximin optimality condition also applies in that class. Our
labelled problem allows asymmetric target noncentrality intervals and offset bounds.
Section~\ref{supp:purification} develops its conditional-probability
representation and reconstruction, obtaining simultaneous power
improvement and uniqueness under the common-direction condition.

\citet[Theorem~3.1 and Corollary~3.1.1]{khr2025} optimise the final intersection
test with lower-order tests fixed. Their main construction of the test statistic assumes
continuous, coordinatewise decreasing objective coefficients
(Assumption~3) and treats independence, with extensions
discussed for dependence. This construction provides the interpretation
of the shared intersection decision as the remaining design choice.
In the present regret problem, the intersection test and the target
noncentrality parameters supporting its least-favourable prior are selected jointly in the continuous minimax problem.

\subsection{Continuous optimality bounds}
The bound of \citet[Section~6.3, equation~(17)]{rly2014} relaxes the original constraint collection using
nonnegative multipliers obtained from a finite problem, then minimises the
resulting risk over the original measurable decision space. Their sparse linear programming formulation
also incorporates monotone decisions. This dual construction links a
finite computation to a bound on continuous optimality.

In our reduced problem a fixed prior and a single size penalty leave the
penalised power $E_0[(H-k)\phi]$. The upper bound obtained by pointwise maximisation without the monotonicity constraint is
$\int(H-k)_+f_0$. To retain the order constraint,
Section~\ref{supp:continuumcertificate} uses the lower-left and upper-right
vertex values to constrain
$\phi$ throughout each cell, together with a secant bound on $(H-k)_+$.
The resulting maximisation gives an upper bound over all monotone
rules, including those that change inside a cell. Mesh refinement and
expanding boxes recover the maximum penalised power over monotone rules. The cell
inequality, convergence argument and integration-error bounds provide
bounds for the continuous Gaussian regret problem.

\FloatBarrier
\bibliographystyle{oup-abbrvnat}
\bibliography{references}
\end{document}